\documentclass[superscriptaddress,eqsecnum,floats,aps,nofootinbib,showpacs,preprintnumbers]{revtex4}

\usepackage{amsmath,amssymb,amsfonts}
\usepackage{graphicx}
\usepackage{undertilde}

\def\be{\begin{equation}}
\def\ee{\end{equation}}
\def\ba{\begin{eqnarray}}
\def\ea{\end{eqnarray}}
\def\nn{\nonumber}

\newcommand{\abs}[1]{{\left|{#1}\right|}} 

\newcommand{\ftriad}[2]{{}^o\! e^{#1}_{#2}} 
\newcommand{\fcotriad}[2]{{}^o\!\omega_{#1}^{#2}} 
\newcommand{\fq}{{}^o\!q} 
\newcommand{\tE}[2]{{\mbox{$\tilde{E}$}^{#1}}_{#2}} 

\newcommand{\Pl}{\ell_{\rm Pl}} 
\newcommand{\muzero}{\mu^o} 
\newcommand{\mubar}{{\bar \mu}} 
\newcommand{\calK}{{\cal K}}

\newtheorem{remark}{Remark}

\newcommand{\secref}[1]{Sec.~\ref{#1}}

\newcommand{\remref}[1]{Remark~\ref{#1}}
\newcommand{\eqnref}[1]{(\ref{#1})}
\newcommand{\figref}[1]{Fig.~\ref{#1}}
\newcommand{\tabref}[1]{Table~\ref{#1}}
\newcommand{\appref}[1]{Appendix~\ref{#1}}
\newcommand{\footref}[1]{footnote~\ref{#1}}

\begin{document}

\large

\preprint{IGC-08/2-4}

\title{Phenomenological Dynamics of Loop Quantum Cosmology\\ in Kantowski-Sachs Spacetime}
\author{Dah-Wei Chiou}
\email{chiou@gravity.psu.edu}
\affiliation{
Institute for Gravitation and the Cosmos,
Physics Department,
The Pennsylvania State University, University Park, PA 16802, U.S.A.}

\begin{abstract}
The fundamental theory and the semiclassical description of loop quantum cosmology (LQC) have been studied in the Friedmann-Robertson-Walker and Bianchi I models. As an extension to include both anisotropy and intrinsic curvature, this paper investigates the cosmological model of Kantowski-Sachs spacetime with a free massless scalar field at the level of phenomenological dynamics with the LQC discreteness corrections. The LQC corrections are implemented in two different improved quantization schemes. In both schemes, the big bang and big crunch singularities of the classical solution are resolved and replaced by the big bounces when the area or volume scale factor approaches the critical values in the Planck regime measured by the reference of the scalar field momentum. Symmetries of scaling are also noted and suggest that the fundamental spatial scale (area gap) may give rise to a temporal scale. The bouncing scenarios are in an analogous fashion of the Bianchi I model, naturally extending the observations obtained previously.
\end{abstract}


\maketitle



\section{Introduction}
It has been long suggested that the singularities in general relativity signal a breakdown of the classical theory and should be resolved by the quantum effects of gravity. Loop quantum gravity (LQG) is one of such candidate theories of quantum gravity and its application to cosmological models is known as loop quantum cosmology (LQC) (see \cite{Bojowald:2006da} for a review). The comprehensive formulation for LQC has been constructed in detail in the spatially flat and isotropic model with a free massless scalar field \cite{Ashtekar:2006uz,Ashtekar:2006rx,Ashtekar:2006wn}, showing that the quantum evolution is deterministic across the deep Planck regime and the cosmological singularity is replaced by a big bounce for the states which are semiclassical at late times. This construction was then extended to $k=\pm1$ Friedmann-Robertson-Walker (FRW) models to include intrinsic curvature \cite{Ashtekar:2006es,Vandersloot:2006ws} as well as Bianchi I models to include anisotropy \cite{Chiou:2006qq,Chiou:2007dn,Chiou:2007sp,Chiou:2007mg} either in the fundamental theory of LQC or at the level of phenomenological dynamics; the studies in extended models affirm the resolution of cosmological singularities and the occurrence of big bounces.\footnote{However, not all extended models have been developed fully rigorously and in some models the affirmation is only based on the semiclassical treatment, the validity of which has been justified only for the spatially flat and isotropic model with a free massless scalar field. With this caveat, it would be more precise to call the heuristic treatment of semiclassical approach used in this paper ``\emph{phenomenological} dynamics'' instead of ``\emph{effective} dynamics'' as it has not been shown to be so. (The author thanks Martin Bojowald for this comment.)}

To further extend this formulation and enlarge its domain of validity, the next step is to investigate loop quantum geometry of the black hole and to see whether the black hole singularity is also resolved. The simplest step is to consider the interior of a Schwarzschild black hole, in which the temporal and radial coordinates flip roles and thus the metric components are homogeneous with the Kantowski-Sachs symmetry. Because of homogeneity, the loop quantization of the Schwarzschild interior can be formulated in a similar fashion of LQC. This has been developed in \cite{Modesto:2004wm,Ashtekar:2005qt,Modesto:2005zm} and its phenomenological dynamics studied in \cite{Modesto:2006mx} shows that the black hole interior is extended to a white hole interior through the bounce, which resolves the singularity.

The analysis in \cite{Modesto:2006mx} is based on the original quantization strategy (referred to as the ``$\mu_o$-scheme'' in this paper) used in \cite{Ashtekar:2005qt}, which, as a direct transcription of the original LQC construction in \cite{Ashtekar:2006uz}, introduces a fixed parameter to impose fundamental discreteness of quantum geometry. However, it has been argued that the $\mu_o$-scheme quantization in LQC leads to a wrong semiclassical limit in some regimes and should be improved by replacing the discreteness parameters with adaptive variables which depend on the scale factors \cite{Ashtekar:2006wn}. Two improved strategies for loop quantization of the Schwarzschild interior were investigated in \cite{Bohmer:2007wi} at the level of phenomenological dynamics, revealing that the black hole singularity is resolved with the correct semiclassical behaviors in the regime away from the black hole singularity and the event horizon.

However, due to the absence of matter content, the result of the Schwarzschild interior is not easy to be directly compared with the bouncing scenario of LQC. In particular, the occurrence of bounces of triad variables is not indicated by the matter energy density and difficult to be pinpointed; moreover, the interesting symmetry of scaling observed in \cite{Chiou:2007mg} is not notable.

In order to bridge LQC in the isotropic and Bianchi I models with the case of Kantowski-Sahcs symmetry, instead of focusing on the Schwarzschild interior, we consider a cosmological model of Kantowski-Sachs spacetime by introducing a free massless scalar field. The inclusion of the scalar field gives a few advantages. First, in the presence of the scalar field, there is no event horizon and therefore we are working on a ``self-contained'' cosmological model with no need to refer to the ``exterior'' (see \remref{rem:black hole} in \secref{sec:classical solution}). Second, the scalar field serves as emergent time and makes our model directly analogous to those which have been carefully studied. Third, the scaling symmetry noted in \cite{Chiou:2007mg} can be analyzed in a similar manner with the reference of the matter momentum. Furthermore, the Kantowski-Sachs spacetime as a cosmological model possesses both anisotropy and intrinsic curvature; in a sense this is a hybrid of the Bianchi I  and $k=+1$ FRW models (see \remref{rem:Bianchi I} in \secref{sec:classical solution}) and thus sets a new testing ground for the LQC ramifications.

Based on the same semiclassical approach of \cite{Chiou:2007mg} to incorporate the LQC discreteness corrections, the phenomenological dynamics of the cosmological model in Kantowski-Sachs spacetime with a free massless scalar field is investigated in this paper with two improved quantization strategies (called ``$\mubar$-scheme'' and ``$\mubar'$-scheme''). The investigation shows that both the big bang and big crunch singularities of the classical solution are resolved and replaced by big bounces and as a result the evolution of phenomenological dynamics follows (semi)-cyclic patterns.

As a direct analog of the Bianchi I model in \cite{Chiou:2007mg}, the $\mubar$-scheme and $\mubar'$-scheme give rise to different bouncing scenarios, distinct from each other not only in detail but also qualitatively. In the $\mubar$-scheme, the indications of the occurrence of big bounces are ``directional densities'' $\varrho_b$ and $\varrho_c$ and the triad variable $p_b$ is perfectly periodic while the other triad variable $p_c$ only bounces a few times and grows to infinity in the far future and past. By contrast, in the $\mubar'$-scheme, it is the matter energy density $\rho_\phi$ that signals the big bounces and therefore $p_b$ and $p_c$ get bounced roughly around the same moment. Additionally, the $\mubar'$-scheme has the problem that $p_c$ eventually descends into deep Planck regime (while $p_b$ grows huge), signaling a breakdown of the semiclassical description.

As in the Hamiltonian framework for homogeneous models, we have to restrict the spatial integration to a finite sized shell ${\cal I}\times S^2$ to make the Hamiltonian finite. This prescription raises the question whether the resulting dynamics is independent of the choice of the finite interval ${\cal I}$. It can be shown that the phenomenological dynamics in the $\mubar'$-scheme is completely independent of the choice of ${\cal I}$ as is the classical dynamics, while the phenomenological dynamics in the $\mubar$-scheme reacts to the macroscopic scale introduced by the boundary condition of ${\cal I}$. This is an important difference between these two schemes (although the independence of ${\cal I}$ is not required with quantum corrections).

In addition to the issues related to the dependence on ${\cal I}$, the phenomenological dynamics also reveals interesting symmetries of scaling, which are reminiscent of the relational interpretation of quantum mechanics. It is also suggested that the fundamental scale (area gap) imposed for the spatial geometry may gives rise to a fundamental scale in temporal measurement. This observation further supports the speculations made in \cite{Chiou:2007mg}.

This paper follows the steps in \cite{Chiou:2007mg} as closely as possible and uses notations in the similar style.\footnote{Assiduous readers are encouraged to look at \cite{Chiou:2007mg} to see the close parallel (but note that the term ``effective dynamics'' has been rephrased as ``phenomenological dynamics'' in this paper). Also see \remref{rem:Bianchi I} in \secref{sec:classical solution} for some comments.} In \secref{sec:classical dynamics}, the Ashtekar variables of Kantowski-Sachs spacetime are introduced and the classical dynamics with a massless scalar source is solved in Hamiltonian formalism. The phenomenological dynamics with LQC discreteness corrections is constructed and solved in \secref{sec:phenomenological dynamics} for the $\mubar$- and $\mubar'$-schemes, respectively. The scaling symmetry and issues about relational measurements are discussed in \secref{sec:scaling}. Finally, the results are summarized and discussed in \secref{sec:summary}. As a comparison to the $\mubar$- and $\mubar'$-schemes, the phenomenological dynamics in the $\mu_o$-scheme is also included in \appref{sec:muzero dynamics}. The details of heuristic arguments and motivations for the $\mubar$- and $\mubar'$-schemes are given in \appref{sec:mubar schemes}.

\section{Classical dynamics}\label{sec:classical dynamics}
In this section, we first briefly describe the Ashtekar variables in Kantowski-Sachs spacetime \cite{Ashtekar:2005qt}. In the Hamiltonian framework, we then solve the classical solution in terms of Ashtekar variables for the Kantowski-Sachs cosmology with a free massless scalar field.

\subsection{Ashtekar variables in Kantowski-Sachs spacetime}\label{sex:Ashtekar variables}
The metric of homogeneous spacetime with the Kantowski-Sachs symmetry group $\mathbb{R}\times SO(3)$ is given by the line element:
\ba\label{eqn:metric}
ds^2&=&-d\tau^2+g_{xx}(\tau)dx^2+g_{\Omega\Omega}(\tau)d\Omega^2\nn\\
&=&-N(t)^2dt^2+g_{xx}(t)dx^2+g_{\theta\theta}(t)d\theta^2+g_{\phi\phi}(t)d\phi^2,
\ea
where $\tau$ is the proper time, $N(t)$ is the lapse function associated with the arbitrary coordinate time $t$ via $N(t)dt=d\tau$ and $d\Omega^2$ represents the unit 2-sphere given in polar coordinates as
\be
d\Omega^2=d\theta^2+\sin^2\theta\,d\phi^2.
\ee
The topology of the homogeneous spatial slices is $\Sigma=\mathbb{R}\times S^2$, which is coordinatized by $x\in\mathbb{R}$, $\theta\in[0,\pi]$ and $\phi\in[0,2\pi]$.

As in any homogeneous cosmological models, on the homogeneous spacelike slice $\Sigma$, we can choose a fiducial triad field of vectors $\ftriad{a}{i}$ and a fiducial cotriad field of covectors $\fcotriad{a}{i}$ that are left-invariant by the action of the Killing fields of $\Sigma$. (Note $\ftriad{a}{i}\fcotriad{b}{i}=\delta^a_b$.) The \emph{fiducial} 3-metric of $\Sigma$ is given by the cotriad $\fcotriad{a}{i}$:
\be
\fq_{ab}=\fcotriad{a}{i}\,\fcotriad{b}{j}\,\delta_{ij}.
\ee
In the \emph{comoving coordinates} $(x,\theta,\phi)$, we can choose $\fq_{ab}$ to have
\be
\fq_{ab}dx^adx^b=dx^2+d\theta^2+\sin^2\theta\, d\phi^2,
\ee
which gives $\fq:=\det \fq_{ab}=\sin^2\theta$.

In connection dynamics, the canonical pair consists of the Ashtekar variables:
the densitized triads $\tE{a}{i}(\vec{x})$ and connections  ${A_a}^i(\vec{x})$, which satisfy the canonical relation:
\be\label{eqn:Poisson of A nd E}
\{{A_a}^i(\vec{x}),\tE{b}{j}(\vec{x}')\}
=8\pi G\gamma\,\delta^i_j\,\delta_a^b\,\delta^3(\vec{x}-\vec{x}'),
\ee
where $\gamma$ is the Barbero-Immirzi parameter. In the case that connections and triads admit the Kantowski-Sachs symmetry $\mathbb{R}\times SO(3)$, ${A_a}^i$ and $\tE{a}{i}$ after gauge fixing of the Gauss constraint are of the form \cite{Ashtekar:2005qt}:
\ba
\label{eqn:symmetric A}
A={A_a}^i\tau_idx^a&=&\tilde{c}\tau_3dx+\tilde{b}\tau_2d\theta
-\tilde{b}\tau_1\sin\theta d\phi+\tau_3\cos\theta d\phi,\\
\label{eqn:symmetric E}
\tilde{E}=\tE{a}{i}\tau_i\partial_a&=&\tilde{p}_c\tau_3\sin\theta\,\partial_x
+\tilde{p}_b\tau_2\sin\theta\,\partial_\theta-\tilde{p}_b\tau_1\,\partial_\phi,
\ea
where $\tilde{b}$, $\tilde{c}$, $\tilde{p}_b$, $\tilde{p}_c$ are functions of time only and $\tau_i=-i\sigma_i/2$ are $SU(2)$ generators satisfying $[\tau_i,\tau_i]={\epsilon_{ij}}^k\tau_k$ (with $\sigma_i$ being the Pauli matrices.)

The symplectic structure on the symmetry-reduced phase space is given by the complete symplectic structure [as in \eqnref{eqn:Poisson of A nd E}] integrated over the finite sized shell ${\cal I}\times S^2$:
\be\label{eqn:Omega tilde}
\tilde{{\bf \Omega}}=\frac{1}{8\pi G\gamma}\int_{{\cal I}\times S^2}d^3x
\;d{A_a}^i(\vec{x})\wedge d\tE{a}{i}(\vec{x})
=\frac{L}{2G\gamma}
\left(d\tilde{c}\wedge d\tilde{p}_c+2d\tilde{b}\wedge d\tilde{p}_b\right),
\ee
where the integration is over $\theta\in[0,\pi]$, $\phi\in[0,2\pi]$ and restricted to $x\in{\cal I}:=[0,L]$; the finite interval ${\cal I}$ is prescribed to circumvent the problem due to homogeneity that the spatial integration over the whole spatial slice $\mathbb{R}\times S^2$ diverges. [We will see that this prescription does not change the classical dynamics but might have effects on the quantum corrections.]
The reduced symplectic form leads to the canonical relations for the reduced canonical variables:
\be
\{\tilde{b},\tilde{p}_b\}=G\gamma L^{-1},
\qquad
\{\tilde{c},\tilde{p}_c\}=2G\gamma L^{-1}
\ee
and $\{\tilde{b},\tilde{c}\}=\{\tilde{p}_b,\tilde{p}_c\}=0$.
It is convenient to introduce the rescaled variables:
\be
b:=\tilde{b},\qquad
c:=L\tilde{c},\qquad
p_b:=L\tilde{p}_b,\qquad
p_c:=\tilde{p}_c,
\ee
which satisfy the canonical relations:
\be
\{b,p_b\}=G\gamma,
\qquad
\{c,p_c\}=2G\gamma.
\ee

The relation between the densitized triad and the 3-metric is given by
\be\label{eqn:q and E}
qq^{ab}=\delta^{ij}\tE{a}{i}\tE{b}{j},
\ee
which leads to
\be
g_{\Omega\Omega}=g_{\theta\theta}=g_{\phi\phi}\sin^2\theta=p_c,\qquad
g_{xx}=\frac{p_b^2}{L^2p_c}.
\ee
Let $S_{x\phi}$, $S_{x\theta}$ and $S_{\theta\phi}$ be the three surfaces of interest, respectively, bounded by the interval ${\cal I}$ and the equator, ${\cal I}$ and a great circle along a longitude, and the equator and a longitude (so that $S_{\theta\phi}$ forms a quarter of the sphere $S^2$). It follows that the physical areas of $S_{x\phi}$, $S_{x\theta}$ and $S_{\theta\phi}$ are given by
\be\label{eqn:p and A}
{\bf A}_{x\phi}={\bf A}_{x\theta}=2\pi L\sqrt{g_{xx}g_{\Omega\Omega}}=2\pi p_b,
\qquad
{\bf A}_{\theta\phi}=\pi g_{\Omega\Omega}=\pi p_c,
\ee
and the physical volume of ${\cal I}\times S^2$ is
\be
{\bf V}=4\pi L \sqrt{g_{xx}}\,g_{\Omega\Omega}=4\pi p_b \sqrt{p_c}.
\ee
This gives the physical meanings of the triad variables $p_b$ and $p_c$.\footnote{More precisely, in \eqnref{eqn:p and A} $p_b$ and $p_c$ should be $\abs{p_b}$ and $\abs{p_c}$ \cite{Ashtekar:2005qt}. With the gauge fixing $p_b>0$, the opposite sign of $p_c$ corresponds to the inverse spatial orientation, which we do not need to consider in this paper.}

The gravitational part of the Hamiltonian constraint of Kantowski-Sachs spacetime is given in terms of Ashtekar variables as
\be\label{eqn:cl H grav}
H_{\rm grav}
=-\frac{N}{2G\gamma^2}\left[
2bc\sqrt{p_c}+(b^2+\gamma^2)\frac{p_b}{\sqrt{p_c}}
\right].
\ee
This can be derived from the Hamiltonian constraint of the full (unreduced) theory. (See the text toward \eqnref{eqn:H grav} in \appref{sec:mubar schemes} or \cite{Ashtekar:2005qt}.)

\subsection{Classical solution}\label{sec:classical solution}


The vacuum solution of Kantowski-Sachs spacetime is identified with the interior of the Schwarzschild black hole \cite{Ashtekar:2005qt}.\footnote{In the standard Schwarzschild solution, the metric of the black hole interior is of the form of \eqnref{eqn:metric}:
\be
ds^2=-N(t)^2dt^2+g_{xx}(t)dx^2+g_{\Omega\Omega}(t)d\Omega^2
=-\left(\frac{2GM}{t}-1\right)^{-1}dt^2
+\left(\frac{2GM}{t}-1\right)dx^2+t^2d\Omega^2,\nn
\ee
where $t\in[0,2GM]$, $x\in \mathbb{R}$ and $M$ is the mass of the black hole. The black hole singularity corresponds to $t=0$ and the event horizon corresponds to $t=2GM$.
}
In this paper, in order to extend the results studied in \cite{Chiou:2007mg} for the Bianchi I model to the Kantowski-Sachs spacetime, instead of the vacuum solution, we investigate the cosmological model of Kantowski-Sachs spacetime with the inclusion of a homogeneous massless scalar field $\phi(\vec{x},t)=\phi(t)$ without introducing any potential of $\phi$ (i.e. $\phi$ is free).\footnote{Do not confuse the matter field $\phi(t)$ with the polar coordinate $\phi$ in \eqnref{eqn:metric}.}

The total Hamiltonian constraint is given by the gravitational part $H_{\rm grav}$ plus the scalar field part $H_\phi$; that is
\ba\label{eqn:cl Hamiltonian}
H&=&H_{\rm grav}+H_\phi=H_{\rm grav}+N{\bf V}\rho_\phi\nn\\
&=&-\frac{N}{2G\gamma^2}\left[
2bc\sqrt{p_c}+(b^2+\gamma^2)\frac{p_b}{\sqrt{p_c}}
\right]
+\frac{Np_\phi^2}{8\pi p_b\sqrt{p_c}},
\ea
where the matter energy density $\rho_\phi$ is given by
\be
\rho_\phi=\frac{p_\phi^2}{2{\bf V}^2}=\frac{p_\phi^2}{32\pi^2 p_b^2p_c}
\ee
with $p_\phi$ being the scalar field momentum:
\be\label{eqn:p-phi and phi dot}
p_\phi={\bf V}\dot{\phi}:={\bf V}\frac{d\phi}{d\tau}.
\ee

To solve the classical solution, we can simplify the Hamiltonian by choosing the lapse function $N=\gamma p_b\sqrt{p_c}$
corresponding to the new time variable $dt'=(\gamma p_b\sqrt{p_c})^{-1}d\tau$.
The rescaled Hamiltonian is given by
\be\label{eqn:cl rescaled Hamiltonian}
H'=-\frac{1}{2G\gamma}\left[2bcp_bp_c+(b^2+\gamma^2)p_b^2\right]
+\gamma\frac{p_\phi^2}{8\pi}.\\
\ee
The equations of motion are governed by the Hamilton's equations:
\ba
\label{eqn:cl eom 1}
\frac{dp_\phi}{dt'}&=&\{p_\phi,H\}=0\quad\Rightarrow\quad p_\phi\ \text{is constant}\\
\label{eqn:cl eom 2}
\frac{d\phi}{dt'}&=&\{\phi,H\}=\frac{\gamma}{4\pi}p_\phi,\\
\label{eqn:cl eom 3}
\frac{dc}{dt'}&=&\{c,H'\}=2G\gamma\,\frac{\partial\, H'}{\partial p_c}=-2c b p_b,\\
\label{eqn:cl eom 4}
\frac{dp_c}{dt'}&=&\{p_c,H'\}=-2G\gamma\,\frac{\partial\, H'}{\partial c}=2p_c b p_b,\\
\label{eqn:cl eom 5}
\frac{db}{dt'}&=&\{b,H'\}=G\gamma\,\frac{\partial\, H'}{\partial p_b}=-b\left(bp_b+cp_c\right)-\gamma^2p_b,\\
\label{eqn:cl eom 6}
\frac{dp_b}{dt'}&=&\{p_b,H'\}=-G\gamma\,\frac{\partial\, H'}{\partial b}= p_b\left(bp_b+cp_c\right),
\ea
as well as the constraint that the Hamiltonian must vanish:
\be\label{eqn:cl eom 7}
H'=0\quad\Rightarrow\quad
\frac{G\gamma^2p_\phi^2}{4\pi}=2bcp_bp_c+\left(b^2+\gamma^2\right)p_b^2.
\ee
Notice that substituting \eqnref{eqn:p and A} into \eqnref{eqn:cl eom 4} and \eqnref{eqn:cl eom 6} gives us
\ba
\label{eqn:cl b}
b&=&\frac{\gamma}{2\sqrt{p_c}}\frac{dp_c}{d\tau}
=\gamma\frac{d}{d\tau}\sqrt{g_{\Omega\Omega}}\,,\\
\label{eqn:cl c}
c&=&\frac{\gamma}{p_c^{1/2}}\frac{dp_b}{d\tau}-\frac{\gamma p_b}{2p_c^{3/2}}\frac{dp_c}{d\tau}
=\gamma\frac{d}{d\tau}\left(\frac{p_b}{\sqrt{p_c}}\right)
=\gamma\frac{d}{d\tau}\left(L\sqrt{g_{xx}}\right),
\ea
which tells that, \emph{classically}, the connection variable $b$ is the time change rate of square root of the physical area of $S^2$ (up to constant $(4\pi)^{-1}\gamma$) and $c$ is the time change rate of the physical length of ${\cal I}$ (up to constant $\gamma$).

To solve the equations of motion, first note that combining \eqnref{eqn:cl eom 3} and \eqnref{eqn:cl eom 4} gives
\ba\label{eqn:Kc}
\frac{d}{dt'}(p_c c)=0\quad\Rightarrow\quad
p_c c=\gamma K_c\ \text{is constant},
\ea
and on the other hand, \eqnref{eqn:cl eom 5} and \eqnref{eqn:cl eom 6} yield
\ba\label{eqn:Kb}
\frac{d}{dt'}(K_b)=-\gamma p_b^2\qquad
\text{with}\ p_bb=:\gamma K_b(t').
\ea
The Hamiltonian constraint \eqnref{eqn:cl eom 7} then reads as
\be\label{eqn:Kphi Kb Kc}
K_\phi^2=2K_bK_c+K_b^2+p_b^2
\ee
if we define
\be
K_\phi^2:=\frac{G p_\phi^2}{4\pi}.
\ee
By \eqnref{eqn:Kb} and \eqnref{eqn:Kphi Kb Kc}, we have
\be\label{eqn:cl diff eq for Kb}
\gamma^{-1}\frac{dK_b}{dt'}=\frac{K_\phi}{\sqrt{4\pi G}}\frac{dK_b}{d\phi}
=2K_bK_c+K_b^2-K_\phi^2,
\ee
where \eqnref{eqn:cl eom 2} has been used. It follows from \eqnref{eqn:cl eom 2} that $\phi$ is a monotonic function of time and therefore can be regarded as \emph{emergent time}. In terms of $\phi$, the solution to \eqnref{eqn:cl diff eq for Kb} is given by
\be\label{eqn:cl sol of Kb}
K_b(\phi)=-K_c-\sqrt{K_c^2+K_\phi^2}
\,\tanh\left(\frac{\sqrt{4\pi G\left(K_c^2+K_\phi^2\right)}}{K_\phi}
\left(\phi-\phi_0\right)-\alpha\right)
\ee
with $\alpha$ being the constant specified by the initial state:
\be
\alpha:=\tanh^{-1}\left(\frac{K_b(\phi_0)+K_c}{\sqrt{K_c^2+K_\phi^2}}\right).
\ee
In terms of $K_b(\phi)$ and the constants $K_c$ and $K_\phi$, \eqnref{eqn:cl eom 4} and \eqnref{eqn:cl eom 6} read as
\ba
\label{eqn:cl diff eq for pc}
\frac{1}{p_c}\frac{dp_c}{d\phi}&=&2\frac{\sqrt{4\pi G}}{K_\phi}K_b(\phi),\\
\label{eqn:cl diff eq for pb}
\frac{1}{p_b}\frac{dp_b}{d\phi}&=&\frac{\sqrt{4\pi G}}{K_\phi}
\left[K_b(\phi)+K_c\right],
\ea
the solutions to which are given by
\be\label{eqn:cl sol of pb}
p_b(\phi)=p_b(\phi_0)\cosh\alpha \left[\cosh\left(\frac{\sqrt{4\pi G\left(K_c^2+K_\phi^2\right)}}{K_\phi}
\left(\phi-\phi_0\right)-\alpha\right)\right]^{-1}
\ee
and
\ba\label{eqn:cl sol of pc}
p_c(\phi)&=& g_{\Omega\Omega}(\phi)\\
&=&p_c(\phi_0)\left(\cosh\alpha\right)^2
e^{-\frac{4\sqrt{\pi G}K_c}{K_\phi}\left(\phi-\phi_0\right)}
\left[\cosh\left(\frac{\sqrt{4\pi G\left(K_c^2+K_\phi^2\right)}}{K_\phi}
\left(\phi-\phi_0\right)-\alpha\right)\right]^{-2}.\nn
\ea
Consequently, we have
\be
g_{xx}(\phi)=\frac{p_b^2}{L^2p_c}=\frac{p_b(\phi_0)^2}{L^2p_c(\phi_0)}
\ e^{\frac{4\sqrt{\pi G}K_c}{K_\phi}\left(\phi-\phi_0\right)}
\ee
and
\ba
{\bf V}(\phi)&=&4\pi p_b(\phi_0)\sqrt{p_c(\phi_0)}
\left(\cosh\alpha\right)^2
e^{-\frac{2\sqrt{\pi G}K_c}{K_\phi}\left(\phi-\phi_0\right)}\nn\\
&&\quad\times
\left[\cosh\left(\frac{\sqrt{4\pi G\left(K_c^2+K_\phi^2\right)}}{K_\phi}
\left(\phi-\phi_0\right)-\alpha\right)\right]^{-2}.
\ea
Furthermore, from \eqnref{eqn:cl sol of Kb} and \eqnref{eqn:cl sol of pb}, we have
\be
K_b(\phi)=
\left\{
\begin{array}{ccl}
K_{b\uparrow}:=-K_c-\sqrt{K_c^2+K_\phi^2-p_b^2} & & \quad\text{for }\phi>\phi_{\rm max},\\
K_{b\downarrow}:=-K_c+\sqrt{K_c^2+K_\phi^2-p_b^2} & & \quad\text{for }\phi<\phi_{\rm max},
\end{array}
\right.
\ee
with the convention $K_\phi>0$.\footnote{We get the equivalent solution if the signs of $K_\phi$, $K_c$ and $K_b$ are flipped simultaneously. Thus, we fix the convention $K_\phi>0$.} That is, $K_b=K_{b\uparrow}$ in the \emph{collapsing phase} (i.e. $\phi>\phi_{\rm max}$) and $K_b=K_{b\downarrow}$ in the \emph{expanding phase} (i.e. $\phi<\phi_{\rm max}$), while $p_b$ reaches the maximal value
\be
p_{b,\,{\rm max}}=\sqrt{K_c^2+K_\phi^2}
\ee
at the epoch $\phi=\phi_{\rm max}$.
[Note that the two branches $K_{b\uparrow\downarrow}$ are the two solutions of the quadratic equation of $K_b$ given by \eqnref{eqn:Kphi Kb Kc}.]
In particular, when $p_b^2\ll K_c^2+K_\phi^2$, $K_b$ approaches the constants $K_{b\pm}$:
\be\label{eqn:Kb asymp}
K_b\rightarrow
\left\{
\begin{array}{ccl}
K_{b+}(K_c):=-K_c-\sqrt{K_c^2+K_\phi^2} & & \quad\text{as }\phi\gg\phi_{\rm max},\\
K_{b-}(K_c):=-K_c+\sqrt{K_c^2+K_\phi^2} & & \quad\text{as }\phi\ll\phi_{\rm max},
\end{array}
\right.
\ee
and we call the periods with $p_b^2\ll K_c^2+K_\phi^2$ the ``$K_{b\pm}$-asymptotic phases''.\footnote{We will show that the loop quantum corrections (in both the $\mubar$- and $\mubar'$-schemes) take effect only in the $K_{b\pm}$-asymptotic phases, provided that the semiclassicality condition is well retained. See footnotes \ref{footnote:approx}, \ref{footnote:approx2} and \ref{footnote:approx3}.}

Additionally, the stress tensor of the scalar field is given by $T_{ab}=\nabla_a\phi\nabla_b\phi-\frac{1}{2}g_{ab}\nabla_c\phi\nabla_c\phi$ and thus the Einstein's equation $R_{ab}-\frac{1}{2}Rg_{ab}=8\pi G T_{ab}$ gives the scalar curvature
\be
R=-8\pi G T\equiv-8\pi G {T^a}_a=8\pi G \dot{\phi}^2
=8\pi G\frac{p_\phi^2}{{\bf V}^2}.
\ee

\begin{figure}
\begin{picture}(500,270)(0,0)

\put(10,260){\textbf{(a)}}
\put(270,260){\textbf{(b)}}
\put(10,125){\textbf{(c)}}
\put(270,125){\textbf{(d)}}

\put(110,126){\text{$\phi-\phi_0$ ($G^{-1/2}$)}}
\put(370,126){\text{$\phi-\phi_0$ ($G^{-1/2}$)}}
\put(13,175){\rotatebox{90}{\text{$p_b$, $p_c$ ($\Pl^2$)}}}
\put(273,170){\rotatebox{90}{\text{$g_{xx}$ ($\Pl^2/L^2$)}}}
\put(13,50){\rotatebox{90}{\text{${\bf V}$ ($\Pl^3$)}}}
\put(268,50){\rotatebox{90}{\text{$K_b$ ($\Pl^2$)}}}

\put(26,135.5)
{
\resizebox{0.394\textwidth}{!}{\includegraphics{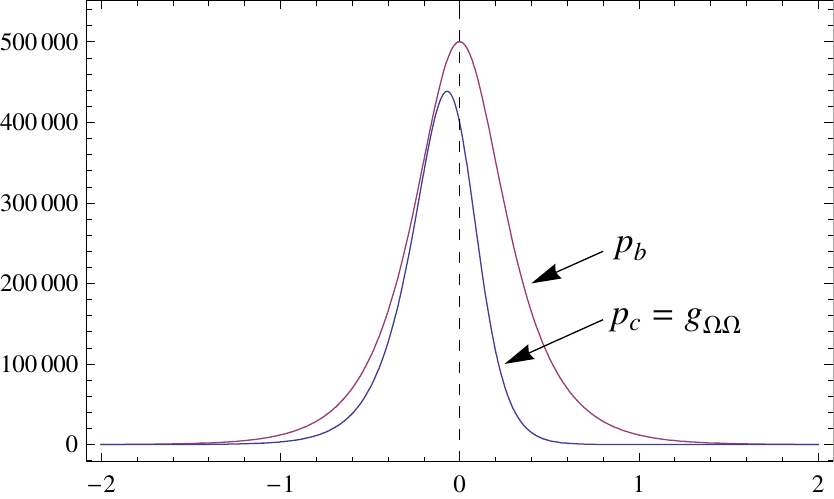}}
}

\put(283,135)
{
\resizebox{0.4\textwidth}{!}{\includegraphics{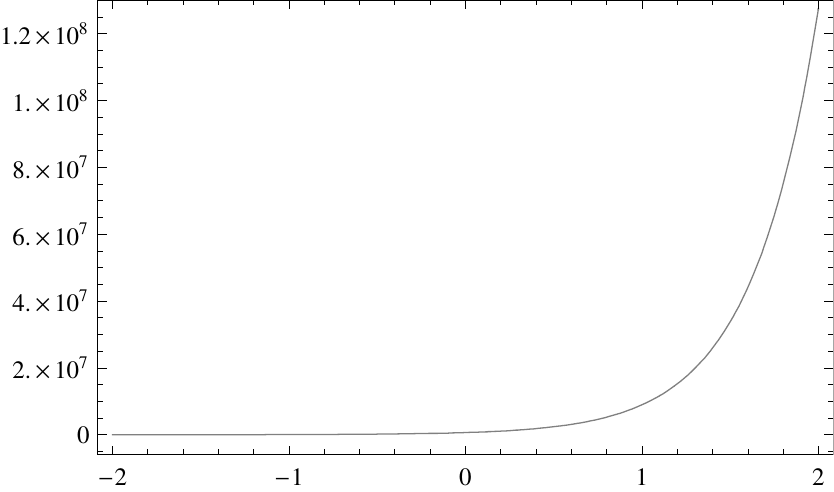}}
}

\put(26,0)
{
\resizebox{0.394\textwidth}{!}{\includegraphics{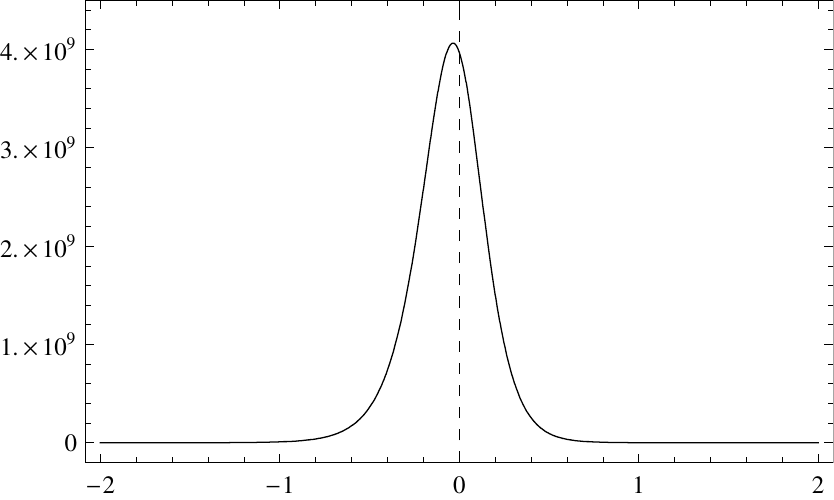}}
}

\put(281.5,-0.4)
{
\resizebox{0.403\textwidth}{!}{\includegraphics{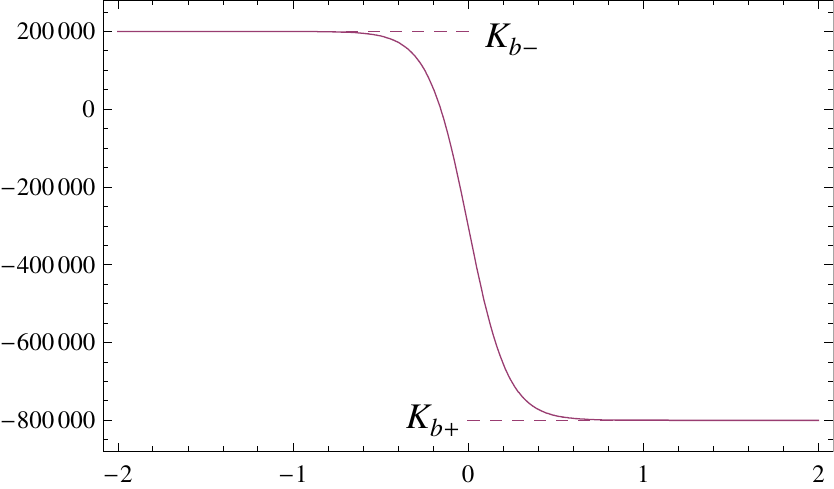}}
}

\end{picture}
\caption{\textbf{Classical solution.} The initial condition is given at $\phi_0=\phi_{\rm max}$ with $p_b(\phi_0)=p_{b,{\rm max}}=\sqrt{K_c^2+K_\phi^2}=5.\times10^5\Pl^2$ ($\Pl:=\sqrt{G\hbar}$\,), $p_c(\phi_0)=4.\times10^5\Pl^2$, $K_c=3.\times10^5\Pl^2$ and $K_\phi=4.\times10^5\Pl^2$ (i.e. $p_\phi=4.\times10^5\hbar\sqrt{4\pi G}$\,). \textbf{(a)} $p_b(\phi)$ and $p_c(\phi)=g_{\Omega\Omega}(\phi)$. \textbf{(b)} $g_{xx}(\phi)$. \textbf{(c)} ${\bf V}(\phi)$. \textbf{(d)} $K_b(\phi)$, which asymptotically approaches $K_{b\pm}$ as indicated by dashed lines.}\label{fig:classical solution}
\end{figure}

In the future and past, the classical solution eventually approaches singularities with the asymptotic behaviors:
\be
p_b\rightarrow 0,\quad p_c=g_{\Omega\Omega}\rightarrow 0,\quad {\bf V}\rightarrow 0,
\quad R\rightarrow\infty
\qquad\text{as }\phi\rightarrow \pm\infty
\ee
and
\be
g_{xx}\rightarrow
\left\{
\begin{array}{ccl}
\infty & & \quad\text{as }K_c\phi\rightarrow\infty,\\
0 & & \quad\text{as }K_c\phi\rightarrow-\infty.
\end{array}
\right.
\ee
The finite sized shell ${\cal I}\times S^2$ collapses to a point at $K_c\phi\rightarrow-\infty$ while to an infinite line at $K_c\phi\rightarrow\infty$. Regardless the different signatures, we always call the singularity in the future ($\phi\rightarrow\infty$) the \emph{big crunch singularity} and the other in the past ($\phi\rightarrow-\infty$) the \emph{big bang singularity}. Although the singularities correspond to $\pm\infty$ in $\phi$, the universe actually takes \emph{finite} proper time to reach either of the singularities (in forward or backward evolution), which can be verified by showing that \eqnref{eqn:proper time} is finite even when the upper limit of the integral is taken to be $\phi=\pm\infty$.
The behaviors of the classical solution are depicted in \figref{fig:classical solution}.

It should be noted that, if we flip the sign of $K_c$ (while fix $K_\phi$), we have $K_b(\phi)\longrightarrow-K_b(-\phi)$. Equations \eqnref{eqn:cl sol of pb} and \eqnref{eqn:cl sol of pc} then tell us that $K_c\longrightarrow-K_c$ corresponds to the time reversal.

Notice that, by \eqnref{eqn:p and A}, \eqnref{eqn:p-phi and phi dot} \eqnref{eqn:cl b} and \eqnref{eqn:cl c}, $p_\phi$, $p_b$ and $c$ depend on the choice of the interval ${\cal I}$ and scale as $p_\phi,p_b,c\propto L$, while $p_c$ and $b$ are independent of ${\cal I}$. As a result, the constants of motion $K_\phi$ and $K_c$ (as well as the function $K_b$) all scale as $\propto L$. The ratios $K_c/K_\phi$ and $K_b/K_\phi$ are nevertheless independent of ${\cal I}$ and hence the classical solutions of $p_b/p_b(\phi_0)$ and $p_c/p_c(\phi_0)$ given by \eqnref{eqn:cl sol of pb} and \eqnref{eqn:cl sol of pc} do \emph{not} depend on the choice of ${\cal I}$. Furthermore, once $p_b(\phi)$ and $p_b(\phi)$ are solved, to know the solutions $p_b(\tau)$ and $p_b(\tau)$ as functions of $\tau$, we only need to convert $\phi$ back to $\tau$ via
\ba\label{eqn:proper time}
\tau-\tau_0=\int_{\tau_0}^\tau d\tau=\int_{\phi_0}^\phi\frac{{\bf V}}{{\bf V}\dot{\phi}}\,d\phi
=\int_{\phi_0}^\phi\frac{4\pi p_b\sqrt{p_c}}{p_\phi}\,d\phi,
\ea
where, again, the dependence of $L$ is gone.
Therefore, whether in terms of the proper time $\tau$ or the emergent time $\phi$, the classical dynamics is \emph{completely independent} of the finite interval ${\cal I}$ we choose to make sense of the Hamiltonian formalism. [The independence of the choice of ${\cal I}$ is not necessarily retained when quantum corrections are taken into account. In any case, however, the dynamics is independent of ${\cal I}$ in terms of $\tau$ if and only if it is so in terms of $\phi$.]


\begin{remark}\label{rem:Bianchi I}
\textbf{Comparison with the Bianchi I model}

It is instructive to note that the Hamiltonian formulation with the Kantowski-Sachs symmetry features some characteristics reminiscent of the Bianchi I model. Exploiting these resemblances can help us solve and understand the equations of motions for both classical and phenomenological dynamics.

Recall that in \cite{Chiou:2007mg} the classical Hamiltonian constraint in the Bianchi I model is given by
\be\label{eqn:H in Bianchi I}
H=-\frac{c_2p_2c_3p_3+c_1p_1c_3p_3+c_1p_1c_2p_2}{8\pi G\gamma^2\sqrt{p_1p_2p_3}}
+\frac{p_\phi^2}{2\sqrt{p_1p_2p_3}}.
\ee
If we formally make the identifications:
\ba
p_1,p_2&\longrightarrow& 4\pi p_b,\qquad
c_1,c_2\longrightarrow b,\nn\\
p_3&\longrightarrow& p_c,\qquad\quad
c_3\longrightarrow 4\pi c,
\ea
then \eqnref{eqn:H in Bianchi I} becomes the same as \eqnref{eqn:cl Hamiltonian} (with $N=1$), except that the term $\gamma^2p_bp_c^{-1/2}$ inside the bracket of \eqnref{eqn:cl Hamiltonian} is missing. Since the term $\gamma^2p_bp_c^{-1/2}$ corresponds to the intrinsic curvature of $S^2$ (see \appref{sec:mubar schemes}), it suggests that if we ignore the curvature of $S^2$, the Kantowski-Sachs spacetime is the same as the Bianchi I model as $I=1,2$ directions are identified with the spherical directions $\partial_\theta$ and $\partial_\phi$ and $I=3$ with the ``radial'' direction $\partial_x$.

The presence of the curvature of $S^2$ complicates the equations of motion and makes the evolution look somewhat familiar with the $k=+1$ FRW model, which has both expanding and collapsing phases and also possesses both the big bang and big crunch. Nevertheless, in the $K_{b\pm}$-asymptotic phases, the $S^2$ curvature is negligible and the strategy used to solve the dynamics of Bianchi I models as in \cite{Chiou:2007mg} can be carried over. In particular, we exploit the fact that in classical dynamics, $p_cc$ is constant and so is $p_bb$ approximately in the $K_{b\pm}$-asymptotic phases.\footnote{Cf. The slightly different notations are used in \cite{Chiou:2007mg}, in which $c_Ip_I=:8\pi G\gamma\hbar\calK_I$ and $p_\phi=:\hbar\sqrt{8\pi G}\calK_\phi$. The constants $\calK_I\equiv\calK\kappa_I$ and $\calK_\phi\equiv\calK\kappa_\phi$ are to be identified with $K_b$, $K_c$ and $K_\phi$ used in this paper up to overall dimensionful constants.}
Since the quantum effect is expected to take effect only in the $K_{b\pm}$-asymptotic phases (in which $p_b$ is small enough), this strategy can be easily adopted to deal with the phenomenological dynamics in both the $\mubar$- and $\mubar'$-schemes in the same fashion as in \cite{Chiou:2007mg}. In \secref{sec:phenomenological dynamics}, we will follow closely the treatment used in \cite{Chiou:2007mg} to analyze the phenomenological dynamics.

It should also be noted that, as discussed in \cite{Chiou:2007sp} for $\calK_I$ in the Bianchi I model, $K_c$ and $K_b$ characterize the anisotropy (between the spherical directions and the radial direction.) The Hamiltonian constraint \eqnref{eqn:Kphi Kb Kc} can be understood as the relation which relates matter energy with anisotropy and spherical curvature.
\end{remark}


\begin{remark}\label{rem:black hole}
\textbf{Comparison with the Schwarzschild interior}

If we did not include any matter content, the vacuum solution with Kantowski-Sachs symmetry would represent the interior of a Schwarzschild black hole \cite{Ashtekar:2005qt}, which encounters the black hole singularity (with $p_b,p_c\rightarrow0$) in the future and the event horizon (with $p_b\rightarrow0$, $p_c\rightarrow(2GM)^2$) in the past. The horizon is however not a singularity but is to be extended to the exterior of the black hole; thus, there might be no quantum corrections near the event horizon. (The analysis in \cite{Bohmer:2007wi} shows that the quantum corrections for the event horizon are present only in the $\mubar'$-scheme phenomenological dynamics but absent in the $\mubar$-scheme.)

With the presence of a free massless scalar field, the situation is a bit different. The Kantowski-Sachs spacetime with a massless scalar possesses the big bang singularity in the past and the big crunch singularity in the future but no event horizon anywhere. Therefore, this is a ``self-contained'' cosmological model and the appeal to the ``exterior'' is not needed.\footnote{The reader might think the Kantowski-Sachs spacetime with inclusion of matter describes the interior of a collapsing black hole. This is not the case, since what we include is a \emph{homogeneous} matter filed, whereas the collapsing black hold requires inhomogeneous (but spherically symmetric) distribution of matter density.} Both singularities are genuine singularities ($p_b,p_c\rightarrow0$) and consequently we shall expect both of them are resolved and replaced by the big bounces if the loop quantum corrections are taken into account. However, these two singularities have different signatures: one of them yields $g_{xx}\rightarrow\infty$ while the other gives $g_{xx}\rightarrow0$.
\end{remark}

\section{Phenomenological dynamics with LQC discreteness corrections}\label{sec:phenomenological dynamics}
In the fundamental loop quantum theory of Kantowski-Sachs spacetime, the connection variables $b$ and $c$ do not exist and should be replaced by holonomies \cite{Ashtekar:2005qt}. At the level of phenomenological theory, to reflect the quantum corrections on the states which are semiclassical when the universe is large, following the procedures used for the isotropic cosmology \cite{Singh:2005xg} and the Bianchi I model \cite{Chiou:2007mg}, we take the prescription to replace $b$, $c$ with
\be\label{eqn:sin prescription}
b\longrightarrow\frac{\sin(\mubar_bb)}{\mubar_b},
\qquad
c\longrightarrow\frac{\sin(\mubar_cc)}{\mubar_c},
\ee
introducing the variables $\mubar_b$ and $\mubar_c$ to impose the fundamental discreteness of quantum geometry.\footnote{This prescription is sometimes referred to as ``polymerization'' or ``holonomization'' in the literature.}
The heuristic argument starting from the full theory of LQG for this prescription is presented in \appref{sec:mubar schemes}

The discreteness of quantum geometry also modifies the cotriad component $\omega_c:= p_b/\sqrt{p_c}=L\sqrt{g_{xx}}$. The eigenvalues of $\hat{\omega}_c$ are very close to the classical expectations far away from the classical singularities but become significantly different from the classical values close to the singularity at which $p_b/\sqrt{p_c}$ diverges \cite{Ashtekar:2005qt}. In the semiclassical description, however, this modification on the cotriad $\omega_c$ is less important and we will ignore it by simply taking the classical function $p_b/\sqrt{p_c}$ for the cotriad $\omega_c$. [We will see that for the solutions which are semiclassical far away from the singularities, the big bounces take place when $p_c$ is still much larger than the square of Planck length and the discreteness correction on $\omega_c$ is yet to be considerable. It is the ``nonlocality'' effect (i.e., using the holonomies) that accounts for the occurrence of big bounces.]\footnote{However, as will be seen in \secref{sec:mubar' dynamics}, in the $\mubar'$-scheme, $p_c$ eventually descends into the deep Planck regime at some point and there we can no longer trust the phenomenological theory without taking into account the modification on the cotriad $\omega_c$.}

As a result, with the prescription of \eqnref{eqn:sin prescription} adopted and the cotriad component $\omega_c$ unchanged, by choosing $N=\gamma p_b\sqrt{p_c}$ and $dt'=(\gamma p_b\sqrt{p_c})^{-1}d\tau$, the (rescaled) classical Hamiltonian \eqnref{eqn:cl rescaled Hamiltonian} is modified to serve as the effective Hamiltonian for the semiclassical theory:
\ba\label{eqn:qm Hamiltonian}
H'_\mubar&=&
-\frac{1}{2G\gamma}
\left\{
2\frac{\sin(\mubar_bb)}{\mubar_b}\frac{\sin(\mubar_cc)}{\mubar_c}\,p_bp_c
+\left(\frac{\sin(\mubar_bb)}{\mubar_b}\right)^2 p_b^2
+\gamma^2p_b^2
\right\}
+
\gamma\frac{p_\phi^2}{8\pi}.
\ea
The phenomenological theory prescribed here is only heuristic and questionable; a more rigorous understanding of the quantum dynamics would require more sophisticated refinement. Nevertheless, the fact that this phenomenological theory could provide an accurate approximation (for the case that the back-reaction is negligible) has been evidenced in the isotropic cosmology \cite{Ashtekar:2006wn,Singh:2005xg,Bojowald:2006gr,Taveras} and also affirmed in the Bianchi I model \cite{in progress}.

As for imposing the fundamental discreteness of LQG on the formulation of LQC, the original construction ($\mu_o$-scheme) is to take $\mubar_b$ and $\mubar_c$ as constants (referred to as $\muzero_b, \muzero_c$ in \appref{sec:muzero dynamics} and $\delta$ in \cite{Ashtekar:2005qt,Modesto:2006mx,Bohmer:2007wi}). However, it has been shown in both isotropic and Bianchi I models that the $\mu_o$-scheme can lead to the wrong semiclassical limit and should be improved by a more sophisticated construction ($\mubar$-scheme) in which the value of discreteness parameters depends adaptively on the scale factors (e.g. $\mubar\propto1/\sqrt{{p}}$ is used in \cite{Ashtekar:2006wn}) and thus implements the underlying physics of quantum geometry of LQG more directly \cite{Ashtekar:2006wn,Chiou:2007mg}.

To impose the discreteness in the Kantowski-Sachs spacetime, there is a variety of possibilities in the improved ($\mubar$-) scheme. Among them, two well-motivated constructions (referred to as the ``$\mubar$-scheme'' and ``$\mubar'$-scheme'') are focused in this paper:
\begin{itemize}
\item
$\mubar$-scheme:
\be\label{eqn:mubar}
\mubar_b=\sqrt{\frac{\Delta}{{p_b}}}\,,\qquad
\mubar_c=\sqrt{\frac{\Delta}{{p_c}}}\,,
\ee
\item
$\mubar'$-scheme:
\be\label{eqn:mubar'}
\mubar'_b=\sqrt{\frac{\Delta}{p_c}}\,,\qquad
\mubar'_c=\frac{\sqrt{p_c\Delta}}{p_b}\,.
\ee
\end{itemize}
Here $\Delta=\frac{\sqrt{3}}{2}(4\pi\gamma\Pl^2)$ is the \emph{area gap} in the full theory of LQG with $\Pl=\sqrt{G\hbar}$ being the Planck length.

Either scheme of them has its own advantages and disadvantages and until more detailed physics is investigated it is arguable which one makes better sense. In particular, the $\mubar$-scheme (in the version for the Bianchi I model) is suggested in \cite{Chiou:2006qq}, since in the construction of the fundamental LQC the Hamiltonian constraint in the $\mubar$-scheme gives a difference equation in terms of affine variables and therefore the well-developed framework of the isotropic LQC can be straightforwardly adopted. (However, it is argued in \cite{Bojowald:2007ra} that the $\mubar$-scheme may lead to an unstable difference equation.) By contrast, the $\mubar'$-scheme does not admit the required affine variables and the fundamental LQC of it is difficult to construct. On the other hand, the $\mubar'$-scheme has the virtue that its phenomenological dynamics is independent of the choice of ${\cal I}$ as will be seen (although this virtue is not necessarily required when quantum corrections are taken into account). (More details of the heuristic arguments and motivations for both schemes as well as their comparison are presented in \appref{sec:mubar schemes}.) To explore their virtues and differences, we study both the $\mubar$-scheme and $\mubar'$-scheme in the context of phenomenological dynamics in \secref{sec:mubar dynamics} and \secref{sec:mubar' dynamics} respectively. (For comparison, the phenomenological dynamics in the $\mu_o$-scheme is also presented in \appref{sec:muzero dynamics}, where we see the insensible behavior that $p_b$, $p_c$ at the bounce can be made arbitrarily big.)

\subsection{Phenomenological dynamics in the $\mubar$-scheme}\label{sec:mubar dynamics}
The phenomenological dynamics in the $\mubar$-scheme is specified by the Hamiltonian \eqnref{eqn:qm Hamiltonian} with $\mubar_b$, $\mubar_c$ given by \eqnref{eqn:mubar}.
Again, the equations of motion are governed by the Hamilton's equations and the constraint that the Hamiltonian must vanish; these are
\ba
\label{eqn:qm eom 1}
\frac{dp_\phi}{dt'}&=&\{p_\phi,H'_\mubar\}=0\quad\Rightarrow\
p_\phi\ \text{is constant},\\
\label{eqn:qm eom 2}
\frac{d\phi}{dt'}&=&\{\phi,H'_\mubar\}=\frac{\gamma}{4\pi}p_\phi,\\
\label{eqn:qm eom 3}
\frac{dc}{dt'}&=&\{c,H'_\mubar\}=2G\gamma\,\frac{\partial\, H'_\mubar}{\partial p_c}
=-2\left[\frac{3\sin(\mubar_cc)}{2\mubar_c}-\frac{c\cos(\mubar_cc)}{2}\right]
\left[\frac{\sin(\mubar_bb)}{\mubar_b}p_b\right],\\
\label{eqn:qm eom 4}
\frac{dp_c}{dt'}&=&\{p_c,H'_\mubar\}=-2G\gamma\,
\frac{\partial\, H'_\mubar}{\partial c}
=2p_c\cos(\mubar_cc)
\left[\frac{\sin(\mubar_bb)}{\mubar_b}p_b\right],\\
\label{eqn:qm eom 5}
\frac{db}{dt'}&=&\{b,H'_\mubar\}=G\gamma\,\frac{\partial\, H'_\mubar}{\partial p_b}
=-\left[\frac{3\sin(\mubar_bb)}{2\mubar_b}-\frac{b\cos(\mubar_bb)}{2}\right]
\left[\frac{\sin(\mubar_bb)}{\mubar_b}p_b+\frac{\sin(\mubar_cc)}{\mubar_c}p_c\right]\nn\\
&&\qquad\qquad\qquad\qquad\qquad-\gamma^2p_b,\\
\label{eqn:qm eom 6}
\frac{dp_b}{dt'}&=&\{p_b,H'_\mubar\}=-G\gamma\,
\frac{\partial\, H'_\mubar}{\partial b}
=p_b\cos(\mubar_bb)
\left[\frac{\sin(\mubar_bb)}{\mubar_b}p_b+\frac{\sin(\mubar_cc)}{\mubar_c}p_c\right],
\ea
as well as
\be\label{eqn:qm eom 7}
H'_\mubar=0
\quad\Rightarrow\quad
\frac{G\gamma^2p_\phi^2}{4\pi}\equiv K_\phi^2
=2\frac{\sin(\mubar_bb)}{\mubar_b}\frac{\sin(\mubar_cc)}{\mubar_c}\,p_bp_c
+\left[\left(\frac{\sin(\mubar_bb)}{\mubar_b}\right)^2+\gamma^2\right]p_b^2.
\ee
[Note that in the classical limit $\mubar_bb,\mubar_cc\rightarrow0$, we have
$\sin(\mubar_bb)/\mubar_b\rightarrow b$, $\sin(\mubar_cc)/\mubar_c\rightarrow c$ and
$\cos(\mubar_bb),\,\cos(\mubar_cc)\rightarrow1$. By inspection, it follows that
\eqnref{eqn:qm eom 3}--\eqnref{eqn:qm eom 7} reduce to their
classical counterparts \eqnref{eqn:cl eom 3}--\eqnref{eqn:cl eom 7} in the classical limit.]
Also notice that \eqnref{eqn:qm eom 4} and \eqnref{eqn:qm eom 6} lead to
\ba
\label{eqn:qm b}
\frac{\sin(\mubar_bb)}{\mubar_b}&=&\frac{1}{\cos(\mubar_cc)}
\frac{\gamma}{2p_c^{1/2}}\frac{dp_c}{d\tau},\\
\label{eqn:qm c}
\frac{\sin(\mubar_cc)}{\mubar_c}&=&\frac{1}{\cos(\mubar_bb)}
\frac{\gamma}{p_c^{1/2}}\frac{dp_b}{d\tau}
-\frac{1}{\cos(\mubar_cc)}\frac{\gamma p_b}{2p_c^{3/2}}\frac{dp_c}{d\tau},
\ea
which are the modifications of \eqnref{eqn:cl b} and \eqnref{eqn:cl c} with quantum corrections.

Combining \eqnref{eqn:qm eom 3} and \eqnref{eqn:qm eom 4}, we have
\be\label{eqn:qm dKc/dt'}
\left[\frac{3\sin(\mubar_cc)}{2\mubar_c}
-\frac{c\cos(\mubar_cc)}{2}\right]\frac{dp_c}{dt'}
+p_c\cos(\mubar_cc)\frac{dc}{dt'}
=\frac{d}{dt'}\left[p_c\frac{\sin(\mubar_cc)}{\mubar_c}\right]=0,
\ee
which, in accordance with the classical counterpart \eqnref{eqn:Kc}, yields the constant of motion:
\be\label{eqn:qm Kc}
p_c\frac{\sin(\mubar_cc)}{\mubar_c}
=\gamma K_c.
\ee
Similarly, \eqnref{eqn:qm eom 5} and \eqnref{eqn:qm eom 6} lead to
\be\label{eqn:qm dKb/dt'}
\left[\frac{3\sin(\mubar_bb)}{2\mubar_b}
-\frac{b\cos(\mubar_bb)}{2}\right]\frac{dp_b}{dt'}
+p_b\cos(\mubar_bb)\frac{db}{dt'}
=\frac{d}{dt'}\left[p_b\frac{\sin(\mubar_bb)}{\mubar_b}\right]
=-\gamma^2p_b^2\cos(\mubar_bb).
\ee
In accordance with the classical counterpart \eqnref{eqn:Kb}, we define
\be\label{eqn:qm Kb}
p_b\frac{\sin(\mubar_bb)}{\mubar_b}=:\gamma \bar{K}_b(t').
\ee
The Hamiltonian constraint \eqnref{eqn:qm eom 7} now read as
\be
K_\phi^2=2\bar{K}_bK_c+\bar{K}_b^2+p_b^2
\ee
and $\bar{K}_b$ satisfies the differential equation:
\be\label{eqn:qm diff eq for Kb}
\gamma^{-1}\frac{d\bar{K}_b}{dt'}=\frac{K_\phi}{\sqrt{4\pi G}}\frac{d\bar{K}_b}{d\phi}
=\cos(\mubar_bb)\left(2\bar{K}_bK_c+\bar{K}_b^2-K_\phi^2\right).
\ee
Substituting \eqnref{eqn:qm Kc} and \eqnref{eqn:qm Kb} into \eqnref{eqn:qm eom 4} and \eqnref{eqn:qm eom 6} yields
\ba
\label{eqn:qm diff eq for pc}
\frac{1}{p_c}\frac{dp_c}{d\phi}&=&
2\frac{\sqrt{4\pi G}}{K_\phi}\,\cos(\mubar_cc)\bar{K}_b,\\
\label{eqn:qm diff eq for pb}
\frac{1}{p_b}\frac{dp_b}{d\phi}&=&
\frac{\sqrt{4\pi G}}{K_\phi}\,\cos(\mubar_bb)\left[\bar{K}_b+K_c\right].
\ea
Here, as in the classical dynamics, $\phi$ is regarded as the emergent time via \eqnref{eqn:qm eom 2}.
[Also note that, as in the classical dynamics, it follows from \eqnref{eqn:qm diff eq for Kb} that the flipping $K_c\longrightarrow-K_c$ gives rise to $\bar{K}_b(\phi)\longrightarrow-\bar{K}_b(-\phi)$ and thus corresponds to the time reversal according to \eqnref{eqn:qm diff eq for pc} and \eqnref{eqn:qm diff eq for pb}.]

Equations \eqnref{eqn:qm diff eq for pc} and \eqnref{eqn:qm diff eq for pb} are the modifications of their classical counterparts \eqnref{eqn:cl diff eq for pc} and \eqnref{eqn:cl diff eq for pb}. Notice that the presence of the $\cos(\cdots)$ terms gives rise to the \emph{repulsive} behavior of gravity as the classical solution approaches singularities at Planckian energy density. More precisely, in the $\mubar$-scheme phenomenological dynamics, $p_c$ and $p_b$ get bounced whenever $\cos(\mubar_cc)$ or $\cos(\mubar_bb)$ flips signs, respectively. To find out the exact moment of occurrence of the bounces, we investigate $\cos(\mubar_cc)$ and $\cos(\mubar_bb)$ in more detail.

First, by \eqnref{eqn:qm Kc}, we have
\be\label{eqn:qm cosc}
\cos(\mubar_cc)=\pm\left[1-\sin^2\mubar_cc\right]^{1/2}
=\pm\left[1-\frac{\gamma^2K_c^2\Delta}{p_c^3}\right]^{1/2}
=\pm\left[1-\frac{\varrho_c}{\varrho_{c,\,{\rm crit}}}\right]^{1/2},
\ee
where we define the \emph{directional density} for the ``$p_c$-direction'' as
\be
\varrho_c:=\frac{p_\phi^2}{32\pi^2p_c^3}
\ee
and its critical value is given by the \emph{Planckian density}
$\rho_{\rm Pl}$ times a numerical factor $K_\phi^2/K_c^2$:
\be\label{eqn:qm crit density c}
\varrho_{c,\,{\rm crit}}:=\frac{K_\phi^2}{K_c^2}\,\rho_{\rm Pl},
\ee
with
\be
\rho_{\rm Pl}:=(8\pi G \gamma^2\Delta)^{-1}.
\ee
[Note that the directional density is of the same dimension as the matter density $\rho_\phi:=p_\phi^2/(32\pi^2p_b^2p_c)$ and thus the name.]
Therefore, the bounce in $p_c$ occurs whenever $\varrho_c$ approaches $\varrho_{c,\,{\rm crit}}$.

Similarly, \eqnref{eqn:qm Kb} gives
\be
\cos(\mubar_bb)=\pm\left[1-\sin^2\mubar_bb\right]^{1/2}
=\pm\left[1-\frac{\gamma^2\bar{K}_b^2\Delta}{p_b^3}\right]^{1/2}.
\ee
Assuming that the bounce in $p_b$ takes place in the $K_{b\pm}$-asymptotic phases (i.e. $p_b^2\ll K_c^2+K_\phi^2$) so that $\bar{K}_b=-K_c\pm\left[K_c^2+K_\phi^2-p_b^2\right]^{1/2}\approx K_{b\mp}:=
-K_c\pm\left[K_c^2+K_\phi^2\right]^{1/2}$ (see \footref{footnote:approx}), we then have
\be\label{eqn:qm cosb}
\cos(\mubar_bb)
\approx \pm\left[1-\frac{\gamma^2K_{b\pm}^2\Delta}{p_b^3}\right]^{1/2}
=\pm\left[1-\frac{\varrho_b}{\varrho_{b\pm,\,{\rm crit}}}\right]^{1/2},
\ee
where the \emph{directional density} for the ``$p_b$-direction'' is defined as
\be
\varrho_b:=\frac{p_\phi^2}{32\pi^2p_b^3}
\ee
and its critical values are given by the \emph{Planckian density}
$\rho_{\rm Pl}$ times numerical factors $K_{b\pm}^2/K_c^2$:
\be\label{eqn:qm crit density b}
\varrho_{b\pm,\,{\rm crit}}:=\frac{K_\phi^2}{K_{b\pm}^2}\,\rho_{\rm Pl}.
\ee
Therefore, the bounce in $p_b$ occurs whenever $\varrho_b$ approaches $\varrho_{b+,\,{\rm crit}}$ in the forward evolution (i.e. the bounce resolves the big crunch singularity) and $\varrho_{b-,\,{\rm crit}}$ in the backward evolution (i.e. the bounce resolves the big bang singularity).\footnote{\label{footnote:approx}The big bounce in $p_b$ occurs as $p_b^3\approx \gamma^2K_{b\pm}^2\Delta$, which in turn justifies the approximation $p_b^2\ll K_c^2+K_\phi^2$ we have used, provided that we have the semiclassical condition $K_\phi\sim K_c \sim K_{b\pm}\gg\gamma^2\Delta$.}

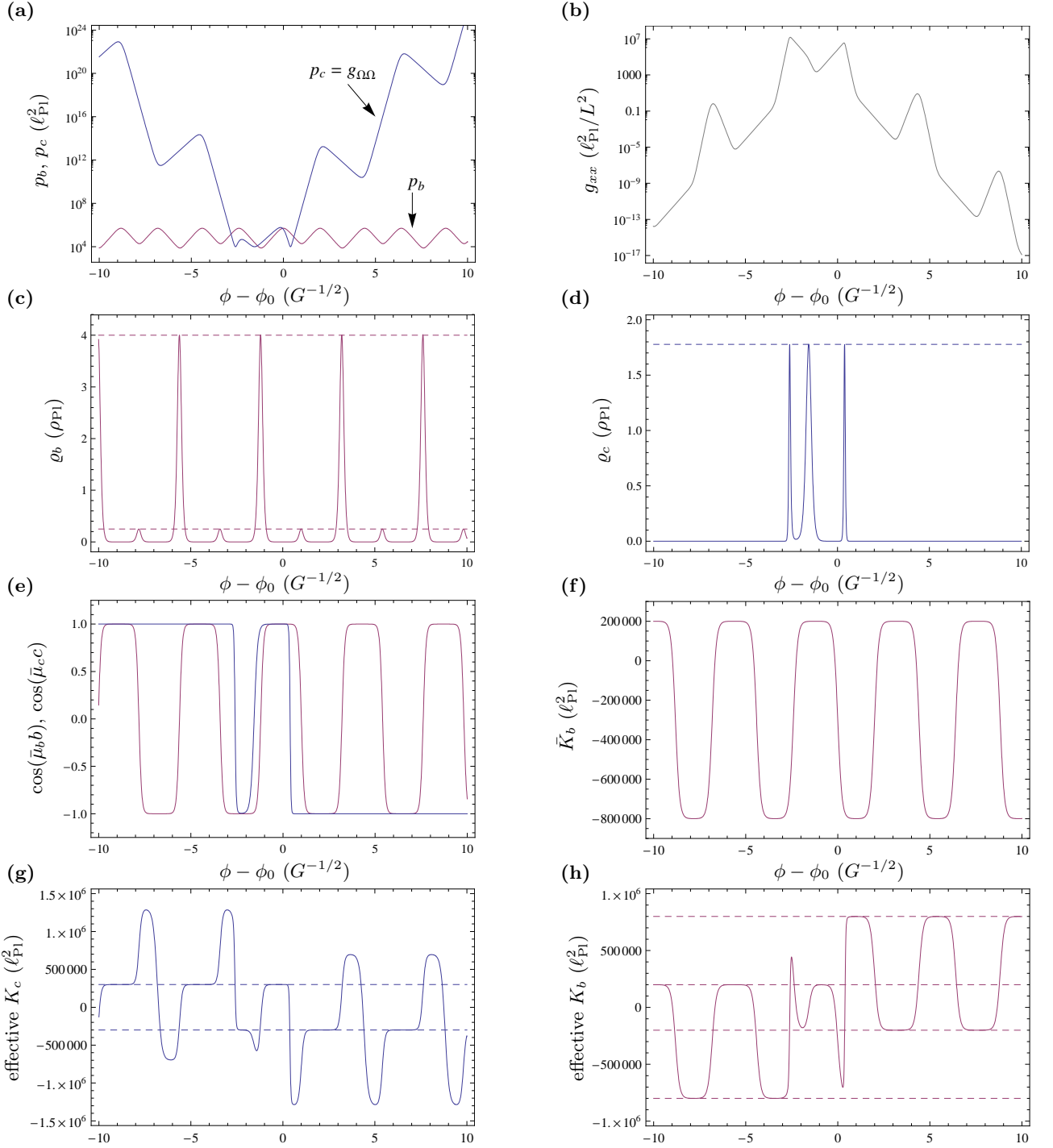
\begin{figure}[Page of floats!]
\begin{picture}(500,540)(0,0)

\put(10,530){\textbf{(a)}}
\put(270,530){\textbf{(b)}}
\put(10,395){\textbf{(c)}}
\put(270,395){\textbf{(d)}}
\put(10,260){\textbf{(e)}}
\put(270,260){\textbf{(f)}}
\put(10,125){\textbf{(g)}}
\put(270,125){\textbf{(h)}}

\put(110,126){\small \text{$\phi-\phi_0$ ($G^{-1/2}$)}}
\put(110,261){\small \text{$\phi-\phi_0$ ($G^{-1/2}$)}}
\put(110,396){\small \text{$\phi-\phi_0$ ($G^{-1/2}$)}}
\put(370,126){\small \text{$\phi-\phi_0$ ($G^{-1/2}$)}}
\put(370,261){\small \text{$\phi-\phi_0$ ($G^{-1/2}$)}}
\put(370,396){\small \text{$\phi-\phi_0$ ($G^{-1/2}$)}}
\put(20,450){\rotatebox{90}{\text{$p_b$, $p_c$ ($\Pl^2$)}}}
\put(278,447){\rotatebox{90}{\text{$g_{xx}$ ($\Pl^2/L^2$)}}}
\put(28,320){\rotatebox{90}{\text{$\varrho_b$ ($\rho_{\rm Pl}$)}}}
\put(285,320){\rotatebox{90}{\text{$\varrho_c$ ($\rho_{\rm Pl}$)}}}
\put(20,165){\rotatebox{90}{\text{$\cos(\mubar_bb)$, $\cos(\mubar_cc)$}}}
\put(268,185){\rotatebox{90}{\text{$\bar{K}_b$ ($\Pl^2$)}}}
\put(8,30){\rotatebox{90}{\text{effective $K_c$ ($\Pl^2$)}}}
\put(273,30){\rotatebox{90}{\text{effective $K_b$ ($\Pl^2$)}}}

\put(34.5,406)
{
\resizebox{0.384\textwidth}{!}{\includegraphics{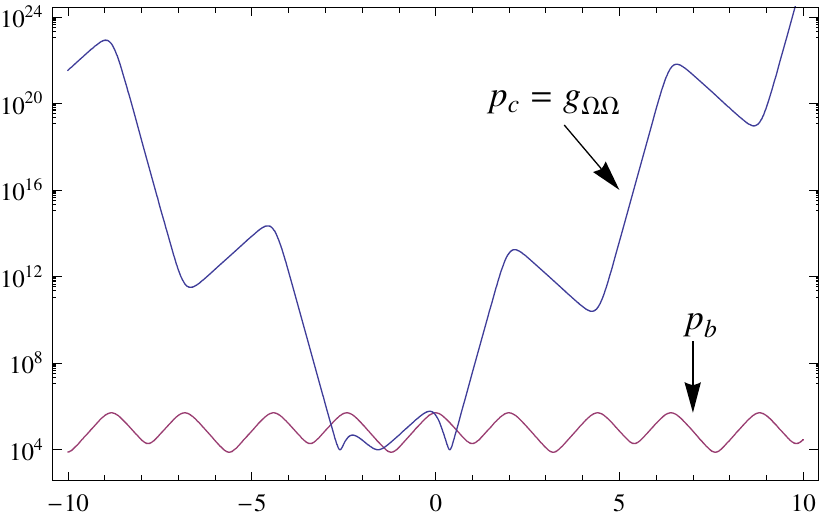}}
}

\put(291.5,405.5)
{
\resizebox{0.390\textwidth}{!}{\includegraphics{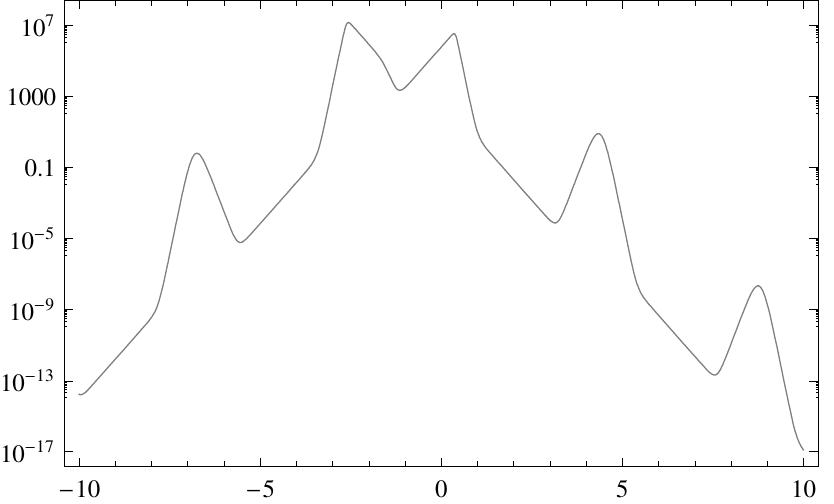}}
}

\put(42,270.5)
{
\resizebox{0.362\textwidth}{!}{\includegraphics{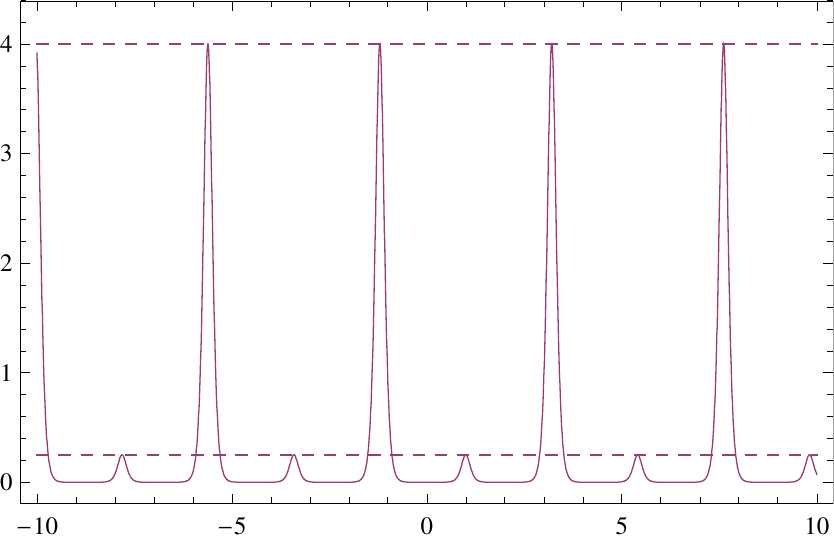}}
}

\put(298,270.5)
{
\resizebox{0.370\textwidth}{!}{\includegraphics{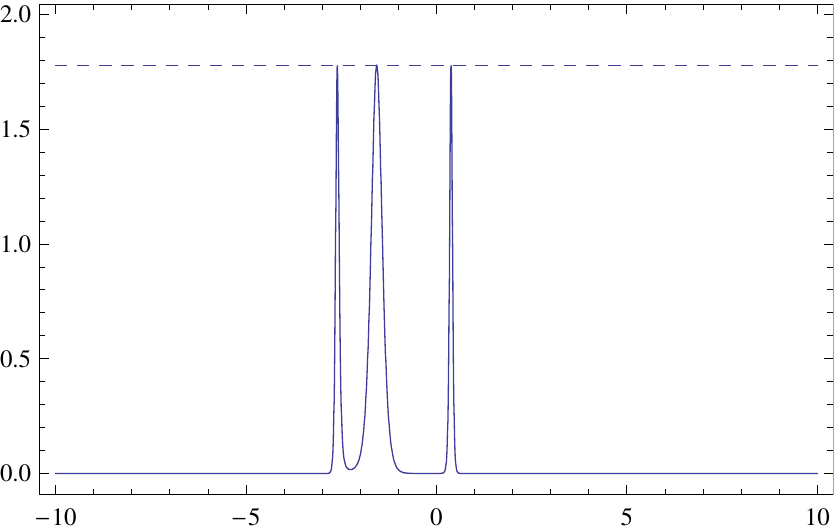}}
}

\put(33.5,136)
{
\resizebox{0.379\textwidth}{!}{\includegraphics{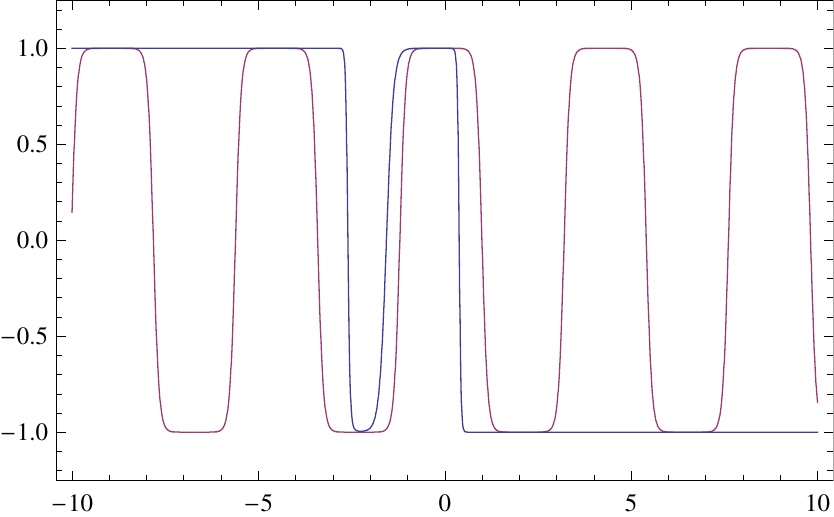}}
}

\put(281.5,135)
{
\resizebox{0.403\textwidth}{!}{\includegraphics{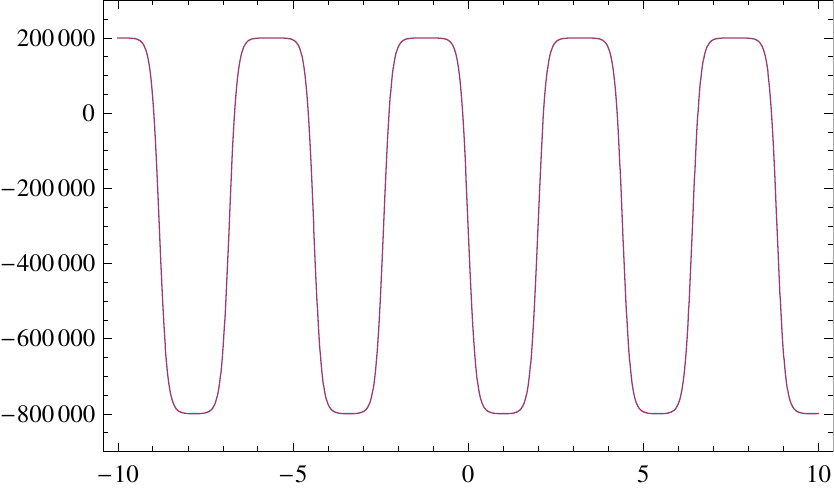}}
}

\put(18,0)
{
\resizebox{0.409\textwidth}{!}{\includegraphics{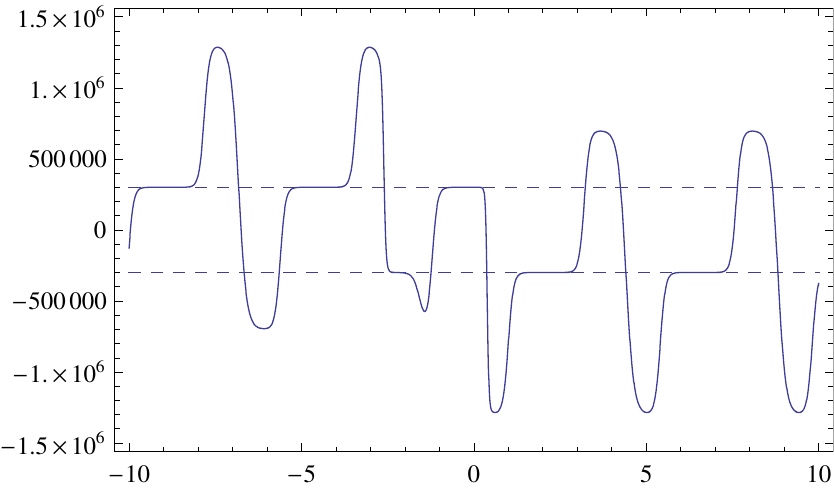}}
}

\put(281.5,0)
{
\resizebox{0.403\textwidth}{!}{\includegraphics{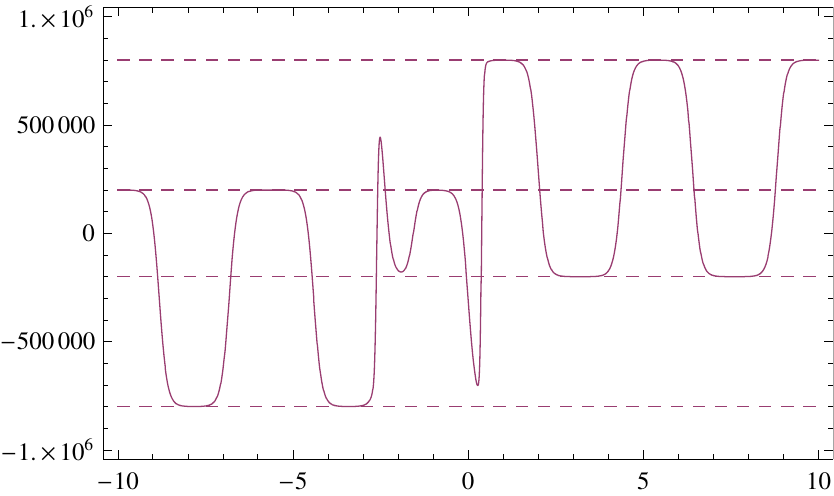}}
}

\end{picture}
\caption{\textbf{Solution in the $\mubar$-scheme phenomenological dynamics.} With the same initial condition as given in \figref{fig:classical solution} (and the Barbero-Immirzi parameter set to $\gamma=1$). \textbf{(a)} $p_b(\phi)$ and $p_c(\phi)=g_{\Omega\Omega}(\phi)$. $p_b$ is periodic while $p_c$ only bounces 3 times and grows up (with wiggled motion) toward infinity in the far future and past. \textbf{(b)} $g_{xx}(\phi)$. \textbf{(c)} $\varrho_b(\phi)$, which signals the occurrence of big bounces of $p_b$ when approaching $\varrho_{b\pm,\,{\rm crit}}$ given by \eqnref{eqn:qm crit density b} and indicated by dashed lines. \textbf{(d)} $\varrho_c(\phi)$, which signals the occurrence of big bounces of $p_c$ when approaching $\varrho_{c,\,{\rm crit}}$ given by \eqnref{eqn:qm crit density c} and indicated by the dashed line. \textbf{(e)} $\cos(\mubar_bb)$: the periodic curve; $\cos(\mubar_cc)$: the curve which flips signs only 3 times (for 3 bounces) and eventually remains $-1$ in the future and $+1$ in the past. \textbf{(f)} $\bar{K}_b(\phi)$, which becomes flat whenever $p_b$ undergoes the bounce. \textbf{(g)} Effective $K_c$ with the constants $\pm K_c$ indicated by dashed lines. (See \eqnref{eqn:qm conjoined Kc}.) \textbf{(h)} Effective $K_b$ with the constants $K_{b\pm}(\pm K_c)$ indicated by dashed lines. (See \eqnref{eqn:qm conjoined Kb}.)}\label{fig:mubar solution}
\end{figure}

In summary, both the big bang and big crunch singularities are \emph{replaced by the big bounces}, which take place in both $p_c$ and $p_b$, whenever $\varrho_c$ or $\varrho_b$ approaches its critical values at Planckian energy density (and thus $\cos(\mubar_cc)$ or $\cos(\mubar_bb)$ flips signs in \eqnref{eqn:qm diff eq for pc} and \eqnref{eqn:qm diff eq for pb} respectively).
Furthermore, the differential equations \eqnref{eqn:qm diff eq for Kb} and \eqnref{eqn:qm diff eq for pb} are independent of $p_c$ and $c$ (the dependence on $p_c$ and $c$ is only through the constant $K_c$).  [Also note that, with the $\cos(\mubar_bb)$ term in \eqnref{eqn:qm diff eq for Kb}, $\bar{K}_b$ becomes flat ($d\bar{K}_b/d\phi=0$) exactly at the same time when $p_b$ get bounced.] As a result, the evolution of $p_b$ is unaffected by the varying of $p_c$ and is expected to be periodic (with respect to $\phi$): i.e. each ``cycle'' of classical evolution of $p_b$ is connected through the quantum bridge with the next/previous classical cycle.\footnote{This periodic behavior is very similar to that of loop quantum cosmology in $k=+1$ FRW model \cite{Ashtekar:2006es}. Also see \remref{rem:Bianchi I} in \secref{sec:classical solution}.} [On the other hand, the numerical analysis shows that the big bounce of $p_c$ occurs only a few times (only 3 times in \figref{fig:mubar solution}).]

Notice that the constants $K_\phi$ and $K_c$ remain the same throughout the evolution. However, this does not mean that the parameters used to parametrize the classical evolutions in different classical cycles remain unchanged, since the physical meanings of $b$ and $c$ are altered before and after the big bounce according to \eqnref{eqn:qm b} and \eqnref{eqn:qm c}.
In order to characterize the classical behaviors of the universe in different classical periods, we define the ``effective $K_c$'' as
\ba\label{eqn:qm effective Kc}
\text{effective }K_c&:=&\gamma^{-1}p_c \left(\frac{\gamma}{p_c^{1/2}}\frac{dp_b}{d\tau}-\frac{\gamma p_b}{2p_c^{3/2}}\frac{dp_c}{d\tau}\right)\nn\\
&=&\gamma^{-1}\cos(\mubar_bb)p_c\frac{\sin(\mubar_cc)}{\mubar_c}
+\gamma^{-1}\left[\cos(\mubar_bb)-\cos(\mubar_cc)\right]
p_b\frac{\sin(\mubar_bb)}{\mubar_b}\nn\\
&=&\cos(\mubar_bb)K_c+\left[\cos(\mubar_bb)-\cos(\mubar_cc)\right]\bar{K}_b
\ea
and similarly the ``effective $K_b$'' as
\be\label{eqn:qm effective Kb}
\text{effective }K_b:=\gamma^{-1}p_b\left(\frac{\gamma}{2p_c^{1/2}}\frac{dp_c}{d\tau}\right)
=\gamma^{-1}\cos(\mubar_cc) p_b \frac{\sin(\mubar_bb)}{\mubar_b}
=\cos(\mubar_cc)\bar{K}_b.\quad
\ee
In the classical period, $\cos(\mubar_bb)\approx\cos(\mubar_cc)\approx\pm1$ and consequently the effective $K_c$ becomes
\be\label{eqn:qm conjoined Kc}
\text{effective }K_c\;\longrightarrow\;\pm K_c.
\ee
That is, the effective $K_c$ in the classical regime can be either $K_c$ or $-K_c$.
Likewise, the effective $K_b$ in the classical period reaches the constant
\be\label{eqn:qm conjoined Kb}
\text{effective }K_b\;\longrightarrow \pm K_{b\pm}(\pm K_c)
\ee
in the $K_{b\pm}$-asymptotic phases.
Furthermore, the evolution also admits a new phase for some periods of time when $\cos(\mubar_bb)\approx-\cos(\mubar_cc)\approx\pm1$. In these ``meta-classical'' periods, we have
\be
\text{effective }K_c\;\longrightarrow\;
\pm\left(K_c+2\bar{K}_b\right),
\ee
which is nonconstant. The new phase is ``meta-classical'' in the sense that $p_b$ evolves classically and $p_c$ is far away from the Planck regime, yet $p_c$ does not follow the classical trajectory. [The occurrence of the meta-classical phases seems to suggest that the $\mubar$-scheme dynamics is problematic, giving wrong semiclassical behavior in some periods, but this may not be the case after all since the Kantowski-Sachs symmetry could be altered by the bounce if the bounce is of $p_b$ alone. We defer this issue for further investigation.]

For given initial conditions, the equations of motion can be solved numerically.\footnote{\label{footnote:numerical method}The common numerical methods (e.g. Runge-Kutta method) encounter numerical instability at some point if we directly solve the coupled differential equations \eqnref{eqn:qm eom 2}--\eqnref{eqn:qm eom 6}. To bypass this problem, which is only a numerical artifact, we solve the reduced coupled equations: \eqnref{eqn:qm diff eq for Kb}, \eqnref{eqn:qm diff eq for pc} and \eqnref{eqn:qm diff eq for pb} for three variables: $\bar{K}_b$, $p_c$ and $p_b$. The variables $b$ and $c$ can be obtained afterwards via \eqnref{eqn:qm Kc} and \eqnref{eqn:qm Kb}.}
The numerical solution is depicted in \figref{fig:mubar solution}.
Note that the bounces occur at the moments exactly when $\varrho_b$ or $\varrho_c$ approaches their critical values. Also notice that $p_b$ and $\bar{K}_b$ are perfectly periodic while $p_c$ only bounces a few times and grows up to infinity in the distant future and past.

It is noteworthy that the directional densities $\varrho_c$ and $\varrho_b$ are the indications of the bounces but their quantities are \emph{not} independent of the choice of ${\cal I}$, as we know
\ba
\varrho_b&=&\frac{p_\phi^2}{32\pi^2p_b^3}=\frac{p_c}{2p_b}\,\dot{\phi}^2
=\frac{{\bf A}_{\theta\phi}}{2{\bf A}_{x\theta}}\,\dot{\phi}^2\propto L^{-1},\nn\\
\varrho_c&=&\frac{p_\phi^2}{32\pi^2p_c^3}=\frac{p_b^2}{2p_c^2}\,\dot{\phi}^2
=\frac{{\bf A}_{x\theta}^2}{2{\bf A}_{\theta\phi}^2}\,\dot{\phi}^2\propto L^2.
\ea
Therefore, contrary to the classical dynamics, the phenomenological dynamics in the $\mubar$-scheme is \emph{dependent} on the choice of the finite sized interval ${\cal I}$. Another subtler dependence on ${\cal I}$ comes from the fact that the constant of motion $K_\phi$ scale as $\propto L$ but $K_c$ and $\bar{K}_b$ scale as $\propto L$ only approximately [see \eqnref{eqn:qm b}, \eqnref{eqn:qm c}, \eqnref{eqn:qm Kc} and \eqnref{eqn:qm Kb} and notice that \eqnref{eqn:qm b} and \eqnref{eqn:qm c} involve the quantum modification with $\cos(\cdots)$ terms]. As a result, the ratios $K_\phi^2/K_c^2$  and $K_\phi^2/K_{b\pm}^2$ are slightly dependent on ${\cal I}$. (Nevertheless, in the classical regime, the dependence on ${\cal I}$ through these ratios is negligible.)

The problem not to be invariant under different choice of ${\cal I}$ is absent in the $\mubar'$-scheme as will be seen in \secref{sec:mubar' dynamics}. However, we should not dismiss the $\mubar$-scheme immediately as it is a common phenomenon that a quantum system reacts to macroscopic scales introduced by boundary conditions (for instance, the well-known ``conformal anomaly'' as a ``soft'' breaking of conformal symmetry). If we have good physical input to tell what exactly the space is to be enclosed as ${\cal I}$ (such as in the compactified Kantowski-Sachs spacetime in which $\Sigma=S^1\times S^2$, instead of $\mathbb{R}\times S^2$, or in the lattice refining model of \cite{Bojowald:2007ra}), the dependence on ${\cal I}$ could be rather meritorious than problematic and the bounce occurrence conditions ($\varrho_c=\varrho_{c,\,{\rm crit}}$, $\varrho_b\approx\varrho_{b\pm,\,{\rm crit}}$) can be understood as: The physical areas ${\bf A}_{\theta\phi}$ and ${\bf A}_{x\theta}={\bf A}_{x\phi}$ get bounced when any of them undergo the Planck regime (times a numerical factor $K_c^2/K_\phi^2$ or $K_{b\pm}^2/K_\phi^2$) measured by the reference of the scalar field momentum $p_\phi$.

\subsection{Phenomenological dynamics in the $\mubar'$-scheme}\label{sec:mubar' dynamics}
The phenomenological dynamics in the $\mubar'$-scheme is specified by the Hamiltonian \eqnref{eqn:qm Hamiltonian} with $\mubar_b$, $\mubar_c$ replaced by $\mubar'_b$, $\mubar'_c$ given in \eqnref{eqn:mubar'}. To simplify the equations of motion, we choose a different lapse function $N=(p_b\sqrt{p_c})^{-1}$ associated with the new time variable $dt''=p_b\sqrt{p_c}\,d\tau$. With the new lapse, the Hamiltonian \eqnref{eqn:qm Hamiltonian} is further rescaled to the simpler form:
\ba\label{eqn:qm' Hamiltonian}
H''_{\mubar'}=
-\frac{1}{2G \gamma^2 \Delta}
\left\{
2\sin(\mubar'_bb)\sin(\mubar'_cc)+\left(\sin(\mubar_bb)\right)^2
+\Delta\frac{\gamma^2}{p_c}
\right\}
+\frac{p_\phi^2}{8\pi p_b^2p_c}.
\ea
Because $\abs{\sin(\mubar'_Ic_I)}\leq 1$, the vanishing of the Hamiltonian constraint $H''_{\mubar'}=0$ immediately implies
\be\label{eqn:rho upper bound}
\rho_\phi=\frac{p_\phi^2}{32\pi^2p_b^2p_c}
\leq\frac{3}{8\pi G \gamma^2 \Delta}+\frac{1}{8\pi G p_c}
<4\rho_{\rm Pl}
\qquad
\text{if }p_c>\gamma^2\Delta.
\ee
This suggests that the matter density $\rho_\phi$ is bounded above and thus the big bounces are expected to occur when the matter density approaches Planckian density (provided that $p_c$ remains large enough).

To know the detailed dynamics for each individual $p_b$ and $p_c$, in addition to the Hamiltonian constraint, we study the Hamilton's equations:
\ba
\label{eqn:qm' eom 1}
\frac{dp_\phi}{dt''}&=&\{p_\phi,H''_{\mubar'}\}=0\quad\Rightarrow\
p_\phi\ \text{is constant},\\
\label{eqn:qm' eom 2}
\frac{d\phi}{dt''}&=&\{\phi,H''_{\mubar'}\}=\frac{p_\phi}{4\pi p_b^2p_c},\\
\label{eqn:qm' eom 3}
\frac{dc}{dt''}&=&\{c,H''_{\mubar'}\}=2 G\gamma\,\frac{\partial\, H''_{\mubar'}}{\partial p_c}\\
&=&
-\frac{c\mubar'_c\cos(\mubar'_cc)\sin(\mubar'_bb)}{\gamma\Delta\,p_c}
+\frac{b\mubar'_b\cos(\mubar'_bb)
\left[\sin(\mubar'_bb)+\sin(\mubar'_cc)\right]}{\gamma\Delta\, p_c}
+\frac{\gamma}{p_c^2}-\frac{G\gamma p_\phi^2}{4\pi p_b^2p_c^2},\nn\\
\label{eqn:qm' eom 4}
\frac{dp_c}{dt''}&=&\{p_c,H''_{\mubar'}\}=-2G\gamma\,\frac{\partial\, H''_{\mubar'}}{\partial c}=
\frac{2\mubar'_c\cos(\mubar'_cc)\sin(\mubar'_bb)}{\gamma\Delta},\\
\label{eqn:qm' eom 5}
\frac{db}{dt''}&=&\{b,H''_{\mubar'}\}=G\gamma\,\frac{\partial\, H''_{\mubar'}}{\partial p_b}=
\frac{c\mubar'_c\cos(\mubar'_cc)\sin(\mubar'_bb)}{\gamma\Delta\,p_b}
-\frac{G\gamma p_\phi^2}{4\pi p_b^3p_c},\\
\label{eqn:qm' eom 6}
\frac{dp_b}{dt''}&=&\{p_b,H''_{\mubar'}\}=-G\gamma\,\frac{\partial\, H''_{\mubar'}}{\partial b}=
\frac{\mubar'_b\cos(\mubar'_bb)
\left[\sin(\mubar'_bb)+\sin(\mubar'_cc)\right]}{\gamma\Delta}.
\ea
Note that \eqnref{eqn:qm' eom 4} and \eqnref{eqn:qm' eom 6} give us
\ba
\label{eqn:qm' b}
\frac{\sin(\mubar'_bb)}{\mubar'_b}&=&\frac{1}{\cos(\mubar'_cc)}
\frac{\gamma}{2p_c^{1/2}}\frac{dp_c}{d\tau},\\
\label{eqn:qm' c}
\frac{\sin(\mubar'_cc)}{\mubar'_c}&=&\frac{1}{\cos(\mubar'_bb)}
\frac{\gamma}{p_c^{1/2}}\frac{dp_b}{d\tau}
-\frac{1}{\cos(\mubar'_cc)}\frac{\gamma p_b}{2p_c^{3/2}}\frac{dp_c}{d\tau},
\ea
which are the modifications of \eqnref{eqn:cl b} and \eqnref{eqn:cl c} with quantum corrections.

Inspecting \eqnref{eqn:qm' eom 3}--\eqnref{eqn:qm' eom 6}, we have
\be
\frac{d}{dt''}\left(p_cc-p_bb\right)=\frac{\gamma}{p_c}.
\ee
In accordance with the constant $K_c$ and the function $K_b(t')$ used for classical solutions in \eqnref{eqn:Kc} and \eqnref{eqn:Kb}, introducing the time-varying function $f(t'')$, we set
\be\label{eqn:qm' Kc}
p_cc=\gamma\left(K_c+f(t'')\right)
\ee
and
\be\label{eqn:qm' Kb}
p_bb=:\gamma\left(\bar{K}'_b(t'')+f(t'')\right),
\ee
where $\bar{K}'_b$ satisfies
\be\label{eqn:qm' diff eq for Kb}
p_b^2p_c\frac{d\bar{K}'_b}{dt''}=
\gamma^{-1}\frac{d\bar{K}'_b}{dt'}=\frac{K_\phi}{\sqrt{4\pi G}}\frac{d\bar{K}'_b}{d\phi}
=-p_b^2,
\ee
which is to be compared with the classical counterpart \eqnref{eqn:Kb}.
To start with a classical regime, we set $K_c\approx\gamma^{-1}p_cc$ and $f\approx0$.

Taking \eqnref{eqn:qm' Kc} and \eqnref{eqn:qm' Kb} into \eqnref{eqn:qm' Hamiltonian}, we have the complicated expression for the Hamiltonian constraint $H''_{\mubar'}=0$:
\ba\label{eqn:qm' eom 7}
\frac{\Delta\gamma^2K_\phi^2}{p_b^2p_c}&=&
2\sin\left(\sqrt{\frac{\Delta\gamma^2}{p_b^2p_c}}\,(\bar{K}'_b+f)\right)
\sin\left(\sqrt{\frac{\Delta\gamma^2}{p_b^2p_c}}\,(K_c+f)\right)\nn\\
&&+\sin^2\left(\sqrt{\frac{\Delta\gamma^2}{p_b^2p_c}}\,(\bar{K}'_b+f)\right)
+\frac{\Delta\gamma^2}{p_c},
\ea
which reduces to
\be\label{eqn:qm' Kph and K}
K_\phi^2=2\bar{K}'_b K_c+\bar{K}_b^{\prime\,2}+p_b^2.
\ee
in the classical limit as $p_b^2p_c\gg\Delta\gamma^2K_\phi^2\sim \Delta\gamma^2K_c^2\sim \Delta\gamma^2K_{b,\pm}^2$ and with $f\approx0$.

As in the $\mubar$-scheme, it is expected that both big bang and big crunch singularities are resolved and replaced by big bounces, which bridge one cycle of classical evolution with the next/previous classical cycle.
One might think that in any classical cycles $p_cc$ becomes constant and so should $f$. This is however not necessarily true (but in fact, $f\approx0$ only in the particular classical cycle in which the initial condition is specified).\footnote{By contrast, in the $\mubar'$-scheme phenomenological dynamics of the Bianchi I model, $f$ are indeed always constant ($f\approx 0$ or $f\approx 2\calK/3$) in classical regimes \cite{Chiou:2007mg}.} The reason is that, due to the quantum modifications in \eqnref{eqn:qm' b} and \eqnref{eqn:qm' c}, $p_bb$ and $p_cc$ have different physical meanings before and after the big bounce and thus $p_cc$ may no longer be constant in the consecutive classical cycle. Instead of $p_cc$, what becomes constant in any classical regimes is the ``effective $K_c$''. Similar to \eqnref{eqn:qm effective Kc} and \eqnref{eqn:qm effective Kb}, we define
\ba\label{eqn:qm' effective Kc}
\text{effective }K_c&:=&\gamma^{-1}p_c \left(\frac{\gamma}{p_c^{1/2}}\frac{dp_b}{d\tau}-\frac{\gamma p_b}{2p_c^{3/2}}\frac{dp_c}{d\tau}\right)\nn\\
&=&\gamma^{-1}\cos(\mubar'_bb)p_c\frac{\sin(\mubar'_cc)}{\mubar'_c}
+\gamma^{-1}\left[\cos(\mubar'_bb)
-\cos(\mubar'_cc)\right]p_b\frac{\sin(\mubar'_bb)}{\mubar'_b}
\ea
and
\be\label{eqn:qm' effective Kb}
\text{effective }K_b:=\gamma^{-1}p_b\left(\frac{\gamma}{2p_c^{1/2}}\frac{dp_c}{d\tau}\right)
=\gamma^{-1}\cos(\mubar'_cc) p_b \frac{\sin(\mubar'_bb)}{\mubar'_b}.
\qquad\qquad\qquad\quad
\ee

Starting with $f\approx0$ and $p_cc\approx \gamma K_c$ in a given cycle of classical phase, in the consecutive classical cycle across the big bounce, rather than constant, $f$ turns out to be a widely time-varying function given as
\be\label{eqn:f}
f\approx\pm\pi\sqrt{\frac{p_b^2p_c}{\gamma^2\Delta}}\,+\delta
\ee
with some constant $\delta$ to be determined.
This follows
\ba
\cos(\mubar'_cc)&\approx&\cos\left(\pm\pi+\sqrt{\frac{\gamma^2\Delta}{p_b^2p_c}}
\,(K_c+\delta)\right)\approx-1,\\
\cos(\mubar'_bb)&\approx&\cos\left(\pm\pi+\sqrt{\frac{\gamma^2\Delta}{p_b^2p_c}}
\,(\bar{K}'_b+\delta)\right)\approx-1,
\ea
and
\ba
\gamma^{-1}p_c\frac{\sin(\mubar'_cc)}{\mubar'_c}&\approx&
\sqrt{\frac{p_b^2p_c}{\gamma^2\Delta}}\
\sin\left(\pm\pi+\sqrt{\frac{\gamma^2\Delta}{p_b^2p_c}}
\,(K_c+\delta)\right)\approx-(K_c+\delta),\\
\gamma^{-1}p_b\frac{\sin(\mubar'_bb)}{\mubar'_b}&\approx&
\sqrt{\frac{p_b^2p_c}{\gamma^2\Delta}}\
\sin\left(\pm\pi+\sqrt{\frac{\gamma^2\Delta}{p_b^2p_c}}
\,(\bar{K}'_b+\delta)\right)\approx-(\bar{K}'_b+\delta)
\ea
provided the classical limit is held: $p_b^2p_c\gg \Delta\gamma^2(K_c+\delta)^2\sim \Delta\gamma^2(K_{b,\pm}+\delta)^2$.
Consequently, by \eqnref{eqn:qm' effective Kc} and \eqnref{eqn:qm' effective Kb}, we obtain that in the consecutive cycle of classical evolution
\ba
\label{eqn:new Kc}
\text{new effective }K_c&=:&\widetilde{K}_c\text{ or } \utilde{K}_c \approx K_c+\delta,\\
\label{eqn:new Kb}
\text{new effective }K_b&=:&\widetilde{K}_b\text{ or } \utilde{K}_b \approx K_b+\delta,
\ea
where $\widetilde{K}_c$ (resp. $\utilde{K}_c$) and $\widetilde{K}_b$ (resp. $\utilde{K}_b$) denote the new effective $K_c$ and $K_b$ in the next (resp. previous) classical cycle across the big bounce. Note that \eqnref{eqn:f} indeed gives a new constant effective $K_c$.

As in the $\mubar$-scheme analysis (see \footref{footnote:approx}), we assume that the bounce in $p_b$ takes place in the $K_{b\pm}$-asymptotic phases (i.e. $p_b^2\ll K_c^2+K_\phi^2$). Furthermore, notice that \eqnref{eqn:Kb} and \eqnref{eqn:qm' diff eq for Kb} are formally identical. As a result, once the universe enters the $K_{b\pm}$-asymptotic phase, $\bar{K}_b'$ remains almost constant ($\bar{K}_b'\approx K_{b\pm}$) even when quantum corrections take effect later. (In the bouncing period, quantum effect varies $f$ dramatically but modifies $\bar{K}_b'$ only slightly.)\footnote{Do not confuse $\bar{K}_b'$ with the \emph{effective} $K_b$. The former remains constant through the big bounce while the latter is offset by $\delta$ as \eqnref{eqn:new Kb} suggests.} Exploiting this fact and using \eqnref{eqn:new Kc} and \eqnref{eqn:new Kb}, we have
\ba
&&K_{b-}(\widetilde{K}_c)=K_{b+}(K_c)+\delta\nn\\
&&\qquad\Leftrightarrow\quad
-(K_c+\delta)+\sqrt{(K_c+\delta)^2+K_\phi^2}=-K_c-\sqrt{K_c^2+K_\phi^2}\,+\delta,\\
&&K_{b+}(\utilde{K}_c)=K_{b-}(K_c)+\delta\nn\\
&&\qquad\Leftrightarrow\quad
-(K_c+\delta)-\sqrt{(K_c+\delta)^2+K_\phi^2}=-K_c+\sqrt{K_c^2+K_\phi^2}\,+\delta,
\ea
which yields $\delta=2K_c/3\pm4(K_c^2+K_\phi^2)^{1/2}/3$ and gives the new effective $K_c$:
\ba
\label{eqn:K tilde}
\widetilde{K}_c&=&\frac{5}{3}K_c+\frac{4}{3}\sqrt{K_c^2+K_\phi^2}\,,\\
\label{eqn:K undertilde}
\utilde{K}_c&=&\frac{5}{3}K_c-\frac{4}{3}\sqrt{K_c^2+K_\phi^2}\,.
\ea
Note that $\utilde{\widetilde{K}_c}=\widetilde{\utilde{K}_c}=K_c$.

In summary, the effective $K_c$ in one cycle of classical regime is shifted to $\widetilde{K}_c$ in the next classical cycle across the putative big crunch and to $\utilde{K}_c$ in the previous classical cycle across the putative big bang.
Schematically, we have [cf. \eqnref{eqn:qm conjoined Kc}]
\be\label{eqn:qm' conjoined Kc}
\cdots
\quad
\utilde{\utilde{K}}_c\;
{
\underrightarrow{\ \mbox{\tiny big crunch}\ }\\
\atop
\overleftarrow{\ \;\mbox{\tiny big bang}\ \;}
}\;
\utilde{K}_c\;
{
\underrightarrow{\ \mbox{\tiny big crunch}\ }\\
\atop
\overleftarrow{\ \;\mbox{\tiny big bang}\ \;}
}\;
K_c\;
{
\underrightarrow{\ \mbox{\tiny big crunch}\ }\\
\atop
\overleftarrow{\ \;\mbox{\tiny big bang}\ \;}
}\;
\widetilde{K}_c\;
{
\underrightarrow{\ \mbox{\tiny big crunch}\ }\\
\atop
\overleftarrow{\ \;\mbox{\tiny big bang}\ \;}
}\;
\widetilde{\widetilde{K}}_c
\quad
\cdots
\ee
and [cf. \eqnref{eqn:qm conjoined Kb}]
\be
K_{b+}(\utilde{K}_c)=-\utilde{K}_c-\sqrt{\utilde{K}_c^2+K_\phi^2}
\quad
{
\underrightarrow{\ \mbox{\tiny big crunch}\ }\\
\atop
\overleftarrow{\ \;\mbox{\tiny big bang}\ \;}
}
\quad
K_{b-}(K_c)=-K_c+\sqrt{K_c^2+K_\phi^2}\,,\nn
\ee
\be\label{eqn:qm' conjoined Kb}
K_{b+}(K_c)=-K_c-\sqrt{K_c^2+K_\phi^2}
\quad
{
\underrightarrow{\ \mbox{\tiny big crunch}\ }\\
\atop
\overleftarrow{\ \;\mbox{\tiny big bang}\ \;}
}
\quad
K_{b-}(\widetilde{K}_c)=-\widetilde{K}_c+\sqrt{\widetilde{K}_c^2+K_\phi^2}\,.
\ee

To find out the exact condition for the occurrence of big bounces, by substituting \eqnref{eqn:qm' Kc} and \eqnref{eqn:qm' Kb} into \eqnref{eqn:qm' eom 4} and \eqnref{eqn:qm' eom 6} and regarding $\phi$ as the emergent time via \eqnref{eqn:qm' eom 2}, we study the differential equations:
\ba
\label{eqn:qm' diff eq for pc}
\frac{1}{p_c}\frac{dp_c}{d\phi}
&=&\frac{2\sqrt{4\pi G}}{K_\phi}\sqrt{\frac{p_b^2p_c}{\gamma^2\Delta}}
\ \cos\left(\sqrt{\frac{\gamma^2\Delta}{p_b^2p_c}}\,(K_c+f)\right)
\sin\left(\sqrt{\frac{\gamma^2\Delta}{p_b^2p_c}}\,(\bar{K}_b'+f)\right),\\
\label{eqn:qm' diff eq for pb}
\frac{1}{p_b}\frac{dp_b}{d\phi}
&=&\frac{\sqrt{4\pi G}}{K_\phi}\sqrt{\frac{p_b^2p_c}{\gamma^2\Delta}}
\ \cos\left(\sqrt{\frac{\gamma^2\Delta}{p_b^2p_c}}\,(\bar{K}_b'+f)\right)\nn\\
&&\quad\times\left[
\sin\left(\sqrt{\frac{\gamma^2\Delta}{p_b^2p_c}}\,(\bar{K}_b'+f)\right)
+\sin\left(\sqrt{\frac{\gamma^2\Delta}{p_b^2p_c}}\,(K_c+f)\right)
\right].
\ea
These are the modifications of the classical counterparts \eqnref{eqn:cl diff eq for pc} and \eqnref{eqn:cl diff eq for pb}. [Also note that, as in the classical and $\mubar$-scheme dynamics, the flipping $K_c\longrightarrow-K_c$ together with $\bar{K}'_b(\phi),\,f(\phi)\longrightarrow-\bar{K}_b'(-\phi),\,-f(-\phi)$ gives rise to the time reversal according to \eqnref{eqn:qm' diff eq for pc} and \eqnref{eqn:qm' diff eq for pb}.]

Similar to the case of \eqnref{eqn:qm diff eq for pc} in the $\mubar$-scheme, $p_c$ gets bounced once the ``$\cos(\cdots)$'' term in \eqnref{eqn:qm' diff eq for pc} flips signs. This happens when
\be\label{eqn:qm' condition 1}
\cos\left(\sqrt{\frac{\gamma^2\Delta}{p_b^2p_c}}\,(K_c+f)\right)=0
\quad\Rightarrow\quad
K_c+f=\frac{\pi}{2}\sqrt{\frac{p_b^2p_c}{\gamma^2\Delta}}.
\ee
Assuming $p_b$ also gets bounced roughly around the same moment,\footnote{This is because \eqnref{eqn:qm' diff eq for pc} and \eqnref{eqn:qm' diff eq for pb} are coupled through $\rho_\phi$. We will see that this is indeed the case in the numerical solution.} at which \eqnref{eqn:qm' condition 1} is satisfied, we have the approximation:
\ba\label{eqn:qm' condition 2}
\sin\left(\sqrt{\frac{\gamma^2\Delta}{p_b^2p_c}}\,(\bar{K}_b'+f)\right)
=\sin\left(\frac{\pi}{2}+\sqrt{\frac{\gamma^2\Delta}{p_b^2p_c}}\,(\bar{K}_b'-K_c)\right)\nn\\
=\cos\left(\sqrt{\frac{\gamma^2\Delta}{p_b^2p_c}}\,(\bar{K}_b'-K_c)\right)
\approx 1-\frac{\gamma^2\Delta}{2p_b^2p_c}(\bar{K}_b'-K_c)^2+\cdots.
\ea
Taking \eqnref{eqn:qm' condition 1} and \eqnref{eqn:qm' condition 2} into \eqnref{eqn:qm' eom 7}, we have
\be
K_\phi^2 \approx 2\frac{p_b^2p_c}{\gamma^2\Delta}
\left[1-\frac{\gamma^2\Delta}{2p_b^2p_c}(\bar{K}_b'-K_c)^2\right]
+\frac{p_b^2p_c}{\gamma^2\Delta}
\left[1-\frac{\gamma^2\Delta}{2p_b^2p_c}(\bar{K}_b'-K_c)^2\right]^2+p_b^2+\cdots,
\ee
which, provided that $K_\phi\gg p_b,p_c\gg \gamma^2\Delta$ when the bounce occurs,\footnote{\label{footnote:approx2}We will see that this is true until $p_c$ eventually descends into deep Planck regime in the far future and pase. (Cf. also see \footref{footnote:approx} for the case of the $\mubar$-scheme.)} leads to the condition for the occurrence of the bounce:
\ba\label{eqn:qm' bounce for pc}
\frac{p_b^2p_c}{\gamma^2\Delta}&\approx&\frac{1}{6}
\left[
2(\bar{K}_b'-K_c)^2+K_\phi^2
+\sqrt{\left(2(\bar{K}_b'-K_c)^2+K_\phi^2\right)^2
-3(\bar{K}_b'-K_c)^4}
\,\right]\nn\\
&=:&F(K_c,\bar{K}_b').
\ea
Since the Taylor series of $\cos x=1-x^2/2+\cdots$ converges very rapidly, the approximation made above is fairly accurate if
\be
\abs{x}=\sqrt{\frac{\gamma^2\Delta}{p_b^2p_c}}\ \abs{\bar{K}_b'-K_c}<\pi
\quad\Rightarrow\quad
p_b^2p_c>\frac{\gamma^2\Delta}{\pi^2}\,(\bar{K}_b'-K_c)^2
\ee
at the epoch when the bounce takes place.\footnote{\label{footnote:approx3}Again, this can be justified in the end (cf. \footref{footnote:approx}).}

Similarly, $p_b$ gets bounced once the ``$\cos(\cdots)$'' term in \eqnref{eqn:qm' diff eq for pb} flips signs. Following the same argument above, we conclude that the big bounce of $p_b$ happens when
\be
K_\phi^2 \approx 2\frac{p_b^2p_c}{\gamma^2\Delta}
\left[1-\frac{\gamma^2\Delta}{2p_b^2p_c}(\bar{K}_b'-K_c)^2\right]
+\frac{p_b^2p_c}{\gamma^2\Delta}
+p_b^2+\cdots,
\ee
which leads to
\be\label{eqn:qm' bounce for pb}
\frac{p_b^2p_c}{\gamma^2\Delta}\approx
\frac{1}{3}(\bar{K}_b'-K_c)^2+\frac{K_\phi^2}{3}=:G(K_c,\bar{K}_b').
\ee

As mentioned earlier, we have assumed that the big bounces take place in the $K_{b\pm}$-asymptotic phases; therefore, in \eqnref{eqn:qm' bounce for pc} and \eqnref{eqn:qm' bounce for pb}, we can replace $\bar{K}_b'$ with $K_{b\pm}$.
To summarize, we conclude that the big bounces of $p_c$ and $p_b$ take place when the matter density $\rho_\phi$ approaches the critical values $\rho_{c\pm,{\rm crit}}$ and $\rho_{b\pm,{\rm crit}}$, respectively, given by the Planckian density $\rho_{\rm Pl}:=(8\pi G \gamma^2\Delta)^{-1}$ times numerical factors:
\ba
\label{eqn:qm' crit density c}
\rho_{c\pm,\,{\rm crit}}
&\approx&\frac{K_\phi^2}{F(K_c,K_{b\pm})}\,\rho_{\rm Pl},\\
\label{eqn:qm' crit density b}
\rho_{b\pm,\,{\rm crit}}
&\approx&\frac{K_\phi^2}{G(K_c,K_{b\pm})}\,\rho_{\rm Pl},
\ea
where ``$+$'' is for the bounce which resolves the big crunch singularity and ``$-$'' is for the bounce which resolves the big bang singularity.

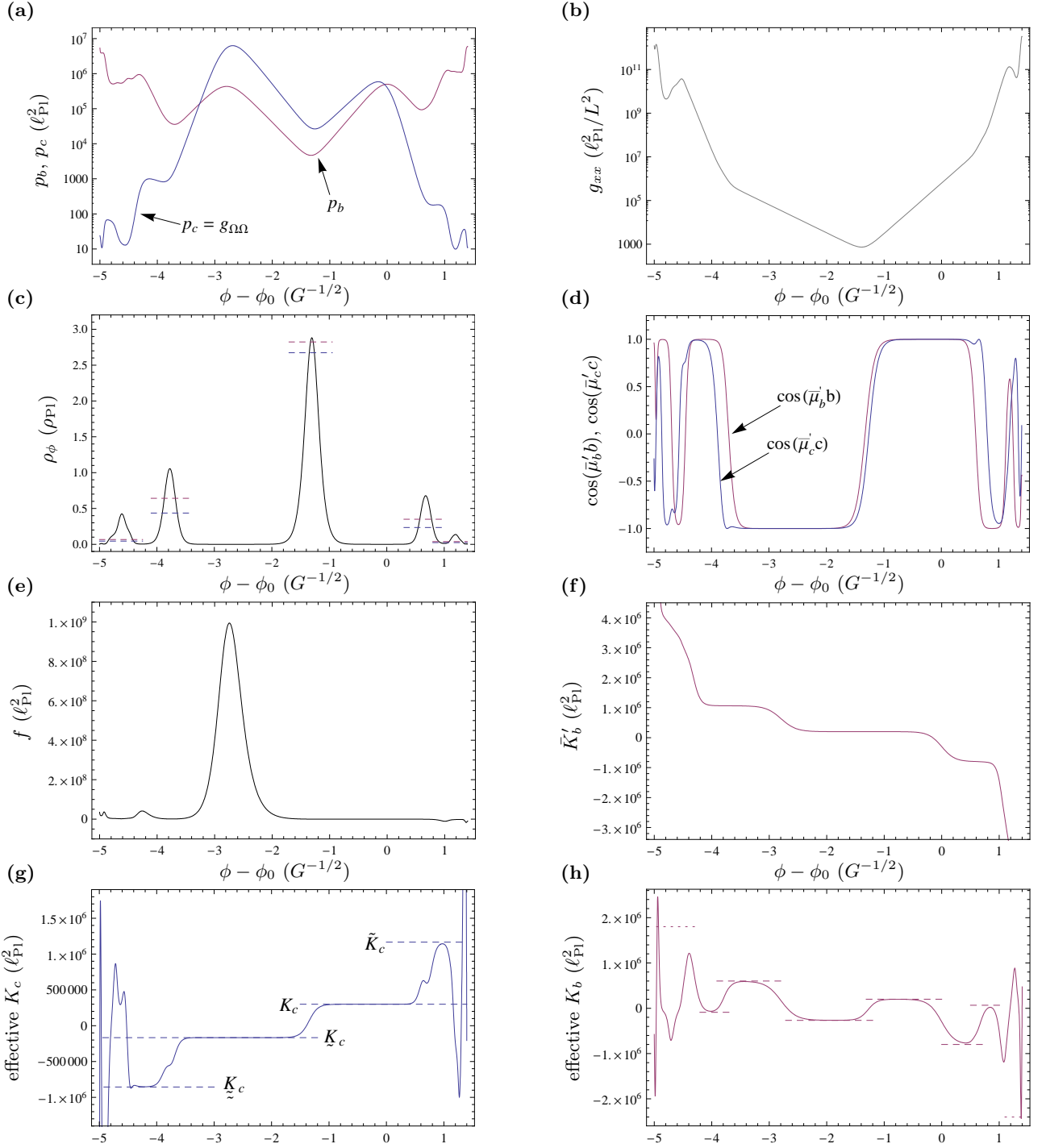
\begin{figure}[Page of floats!]
\begin{picture}(500,540)(0,0)

\put(10,530){\textbf{(a)}}
\put(270,530){\textbf{(b)}}
\put(10,395){\textbf{(c)}}
\put(270,395){\textbf{(d)}}
\put(10,260){\textbf{(e)}}
\put(270,260){\textbf{(f)}}
\put(10,125){\textbf{(g)}}
\put(270,125){\textbf{(h)}}

\put(110,126){\text{\small $\phi-\phi_0$ ($G^{-1/2}$)}}
\put(110,261){\text{\small $\phi-\phi_0$ ($G^{-1/2}$)}}
\put(110,396){\text{\small $\phi-\phi_0$ ($G^{-1/2}$)}}
\put(370,126){\text{\small $\phi-\phi_0$ ($G^{-1/2}$)}}
\put(370,261){\text{\small $\phi-\phi_0$ ($G^{-1/2}$)}}
\put(370,396){\text{\small $\phi-\phi_0$ ($G^{-1/2}$)}}
\put(20,450){\rotatebox{90}{\text{$p_b$, $p_c$ ($\Pl^2$)}}}
\put(280,447){\rotatebox{90}{\text{$g_{xx}$ ($\Pl^2/L^2$)}}}
\put(25,320){\rotatebox{90}{\text{$\rho_\phi$ ($\rho_{\rm Pl}$)}}}
\put(280,300){\rotatebox{90}{\text{$\cos(\mubar'_bb)$, $\cos(\mubar'_cc)$}}}
\put(12,190){\rotatebox{90}{\text{$f$ ($\Pl^2$)}}}
\put(270,185){\rotatebox{90}{\text{$\bar{K}'_b$ ($\Pl^2$)}}}
\put(8,30){\rotatebox{90}{\text{effective $K_c$ ($\Pl^2$)}}}
\put(271,30){\rotatebox{90}{\text{effective $K_b$ ($\Pl^2$)}}}

\put(33.5,405)
{
\resizebox{0.386\textwidth}{!}{\includegraphics{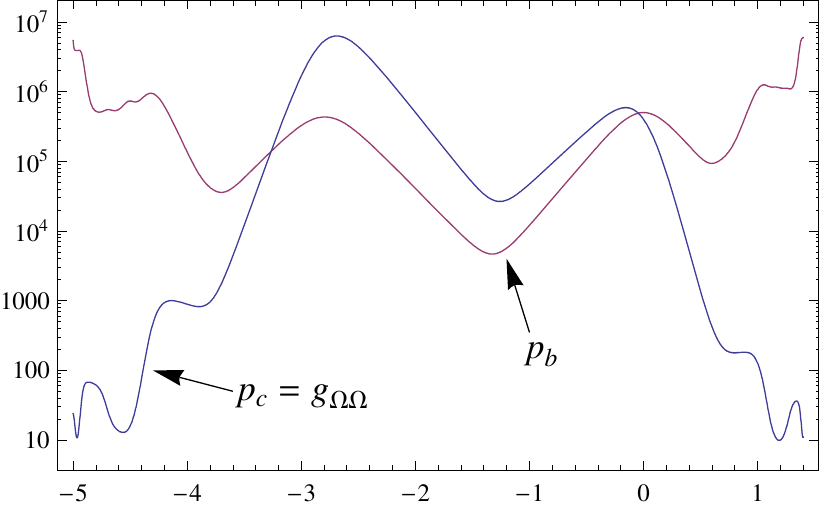}}
}

\put(293.5,405)
{
\resizebox{0.386\textwidth}{!}{\includegraphics{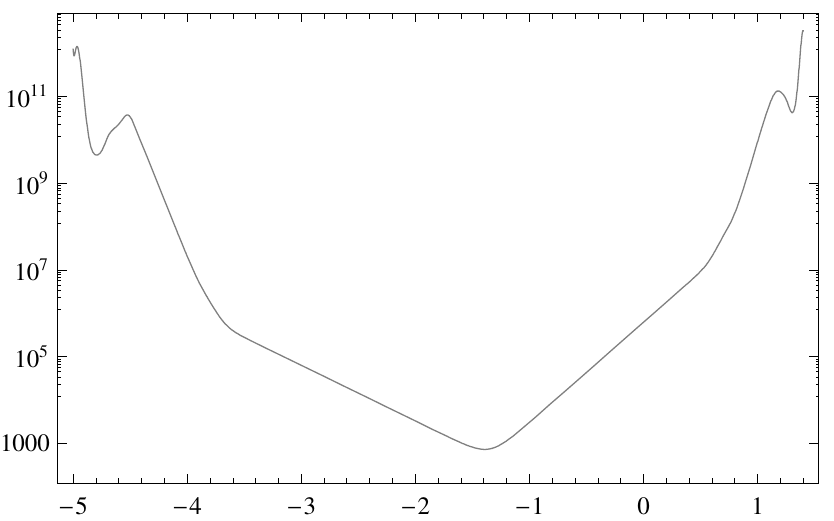}}
}

\put(38,270.5)
{
\resizebox{0.370\textwidth}{!}{\includegraphics{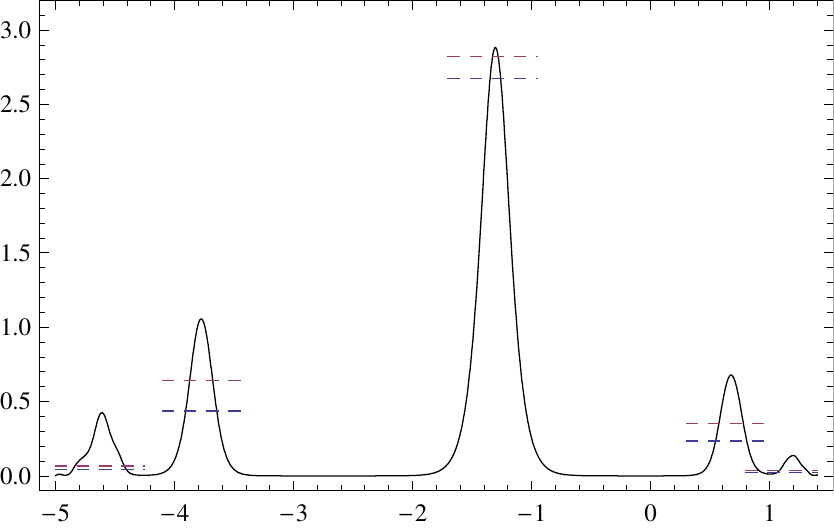}}
}

\put(294,270)
{
\resizebox{0.378\textwidth}{!}{\includegraphics{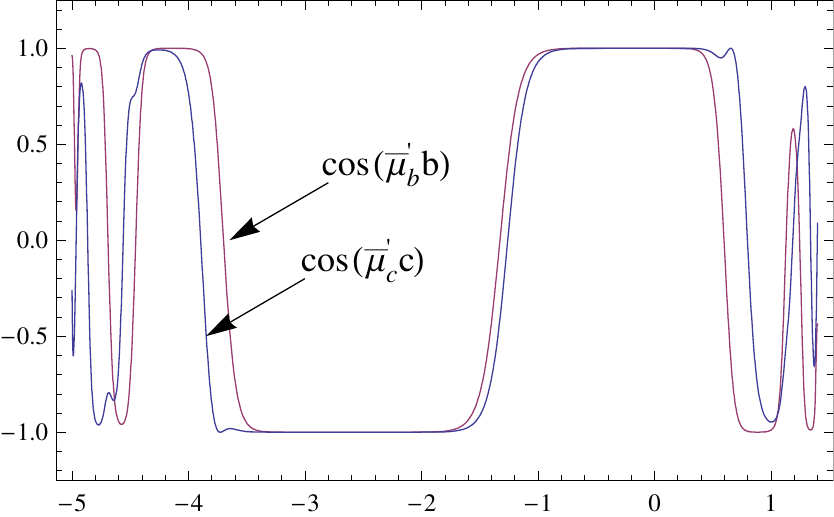}}
}

\put(26.5,135)
{
\resizebox{0.393\textwidth}{!}{\includegraphics{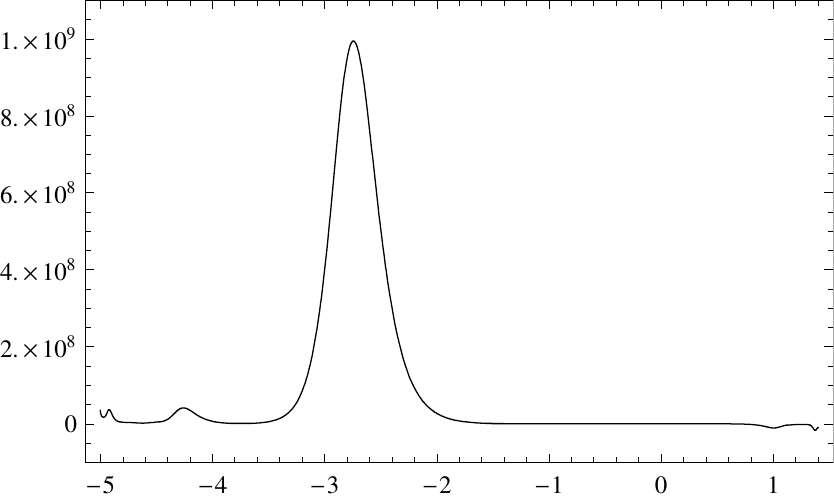}}
}

\put(282,134.5)
{
\resizebox{0.402\textwidth}{!}{\includegraphics{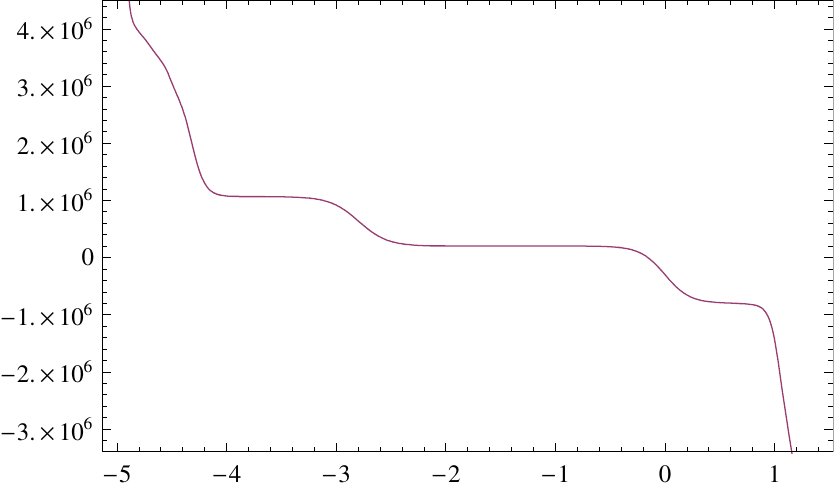}}
}

\put(21.5,0)
{
\resizebox{0.402\textwidth}{!}{\includegraphics{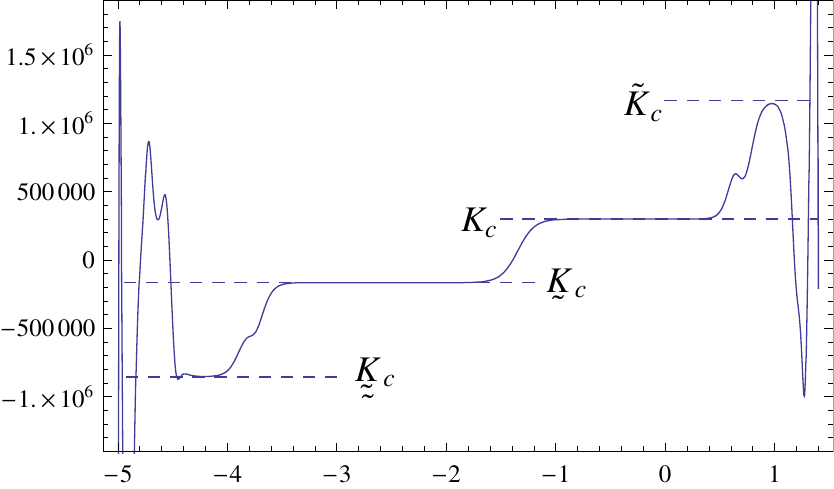}}
}

\put(282,0)
{
\resizebox{0.402\textwidth}{!}{\includegraphics{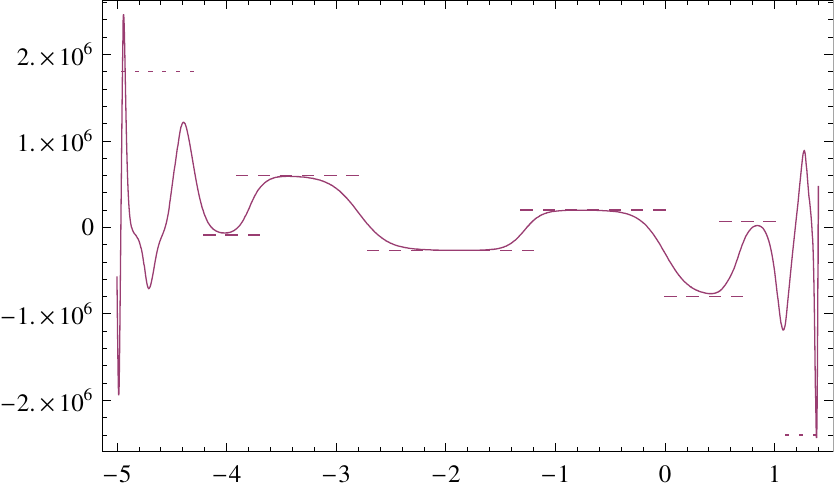}}
}

\end{picture}
\caption{\textbf{Solution in the $\mubar'$-scheme phenomenological dynamics.} Same initial condition as given in \figref{fig:classical solution} (and $\gamma=1$).
\textbf{(a)} $p_b(\phi)$ and $p_c(\phi)=g_{\Omega\Omega}(\phi)$. The bounces of $p_b$ and $p_c$ occur around the same moments. $p_b$ and $p_c$ both oscillate (more and more rapidly) toward the future and past: $p_b$ grows up while $p_c$ descends into deep Planck regime, at which the semiclassical description can no longer be trusted.
\textbf{(b)} $g_{xx}(\phi)$.
\textbf{(c)} $\rho_\phi(\phi)$, which signals the occurrence of big bounces. The critical values given by \eqnref{eqn:qm' crit density c} and \eqnref{eqn:qm' crit density b} with different effective $K_c$ are indicated by dashed lines. The epochs of big bounces are very close to the moments when the curve $\rho_\phi$ intersects one of the dashed lines.
\textbf{(d)} $\cos(\mubar'_bb)$ and $\cos(\mubar'_cc)$, fairly close to each other. \textbf{(e)} $f(\phi)$. $f\approx0$ in the classical cycle around $\phi=\phi_0$ but $f$ yields the time-varying signature given by \eqnref{eqn:f} in the adjacent classical cycles.
\textbf{(f)} $\bar{K}'_b(\phi)$, which remains almost constant across big bounces. \textbf{(g)} Effective $K_c$ with the constants given by \eqnref{eqn:qm' conjoined Kc} indicated by dashed lines.
\textbf{(h)} Effective $K_b$ with the asymptotic values given by \eqnref{eqn:qm' conjoined Kb} indicated by dashed lines.}\label{fig:mubar' solution}
\end{figure}

The differential equations \eqnref{eqn:qm' eom 2}--\eqnref{eqn:qm' eom 6} can be solved numerically for given initial conditions.\footnote{Unlike the case of the $\mubar$-scheme (see \footref{footnote:numerical method}), the numerical method for the $\mubar'$-scheme phenomenological dynamics encounters no problems until the solution of $p_c$ descends into the deep Planck regime.} The numerical solution is shown in \figref{fig:mubar' solution}. The occurrence of big bounces is indicated by the matter density $\rho_\phi$: The bounces of $p_b$ and $p_c$ take place when $\rho_\phi$ is close to $\rho_{b\pm,\,{\rm crit}}$ and  $\rho_{c\pm,\,{\rm crit}}$. Contrary to the $\mubar$-scheme, the epochs of bounces in $p_b$ and $p_c$ are roughly around the same time. Also notice that both $p_b$ and $p_c$ oscillate more and more rapidly toward the future and past; eventually $p_b$ grows very huge (and $\rho_\phi$ subsides) while $p_c$ descends into the deep Planck regime, in which the quantum fluctuations becomes significant (in particular the cotriad component $\omega_c=L\sqrt{g_{xx}}$ grows huge and the quantum corrections on it have to be taken into account) and therefore the analysis of phenomenological dynamics can no longer be trusted.\footnote{As $p_b$ and $p_c$ oscillate too fast, the numerical method also fails to give an accurate solution.} This can be understood by the fact that the absolute value of the effective $K_c$ in the classical cycles becomes larger and larger toward the future and past and as a result the semiclassicality is less and less established.

Contrary to the $\mubar$-scheme dynamics, in which the directional densities $\varrho_b$ and $\varrho_c$ are the indications of bounces, it is the ordinary matter density $\rho_\phi$ that signals the occurrence of bounces in the $\mubar'$-scheme. Unlike $\varrho_b,\varrho_c$, the quantity of $\rho_\phi$ is independent of the choice of ${\cal I}$ since
\be
\rho_\phi=\frac{p_\phi^2}{32\pi^2p_b^2p_c}
=\frac{{\bf V}^2\dot{\phi}^2}{{\bf V}^2}=\frac{\dot{\phi}^2}{2}.
\ee
Furthermore, \eqnref{eqn:qm' b} and \eqnref{eqn:qm' c} imply that the quantities $\mubar'_bb, \mubar'_cc$ depend only on $\dot{p_b}/p_b,\dot{p_c}/p_c$ and thus are independent of ${\cal I}$ (recall $p_b\propto L$, $p_c\propto L^0$). Consequently, \eqnref{eqn:qm' Kc}, \eqnref{eqn:qm' Kb} and \eqnref{eqn:qm' eom 7} tell us: $K_c$, $\bar{K}'_b$, $f$ and $K_\phi$ all scale as $\propto L$. Therefore, the phenomenological dynamics given by \eqnref{eqn:qm' diff eq for pc} and \eqnref{eqn:qm' diff eq for pb} is \emph{completely independent} of the choice of ${\cal I}$ as is the classical dynamics. In particular, the choice of ${\cal I}$ has no effect on the numerical factors $K_\phi^2/ F(K_c,K_{b\pm}), K_\phi^2/G(K_c,K_{b\pm})$ appearing in \eqnref{eqn:qm' crit density c} and \eqnref{eqn:qm' crit density b}. This is a desirable feature that the $\mubar$-scheme does not have.
[However, if we further impose the quantum corrections on the eigenvalue of the cotriad operator $\hat{\omega}_c$, this invariance is broken again.]

Even though the $\mubar'$-scheme is independent of ${\cal I}$, in case when the finite sized shell ${\cal I}\times S^2$ has a global meaning, the bounce conditions ($\rho_\phi\approx \rho_{c\pm,{\rm crit}},\rho_{b\pm,{\rm crit}}$) can be understood as: The volume of ${\cal I}\times S^2$ (i.e. ${\bf V}=4\pi p_b\sqrt{p_c}$) gets bounced when it undergoes the Planck regime (times numerical values $F^{1/2}(K_c,K_{b\pm})/K_\phi, G^{1/2}(K_c,K_{b\pm})/K_\phi$) measured by the reference of the momentum $p_\phi$.

\section{Scaling symmetry and relational measurements}\label{sec:scaling}

With LQC discreteness corrections, the phenomenological dynamics (both in the $\mubar$- and $\mubar'$-schemes) shows that both the big bang and big crunch singularities are resolved and replaced by the big bounces. The occurrence of bounces is indicated by the directional densities $\varrho_b$, $\varrho_c$ in the $\mubar$-scheme whereas it is signaled by the matter density $\rho_\phi$ in the $\mubar'$-scheme; the bounces take place when $\varrho_b$, $\varrho_c$ in the $\mubar$-scheme or $\rho_\phi$ in the $\mubar'$-scheme approaches the critical values.

It has also been noted that the classical dynamics and the phenomenological dynamics in the $\mubar'$-scheme are both completely independent of the choice of the finite sized interval ${\cal I}$, while the phenomenological dynamics in the $\mubar$-scheme reacts to the physical size of ${\cal I}$. This can be rephrased in terms of the scaling symmetry;\footnote{A dynamical system is said to be invariant under a certain scaling if for a given solution ($p_b(\tau)$, $p_c(\tau)$, $b(\tau)$, $c(\tau)$, $\phi(\tau)$ and $p_\phi$) to the dynamics, the rescaled functions also satisfy the equations of motion (i.e. Hamilton's equations and vanishing of Hamiltonian constraint). For the classical dynamics, the equations to be satisfied are \eqnref{eqn:cl eom 1}--\eqnref{eqn:cl eom 7}; for the $\mubar$-scheme, \eqnref{eqn:qm eom 1}--\eqnref{eqn:qm eom 7}; and for the $\mubar'$-scheme, \eqnref{eqn:qm' eom 1}--\eqnref{eqn:qm' eom 6} and \eqnref{eqn:qm' eom 7}.} that is, the classical dynamics and $\mubar'$-scheme phenomenological dynamics are invariant under the following scaling:
\ba\label{eqn:symmetry 1}
p_b,\ p_c&\longrightarrow& l p_b,\ p_c,\nn\\
b,\ c&\longrightarrow& b,\ lc,\nn\\
p_\phi,\ K_\phi&\longrightarrow&l p_\phi,\ l K_\phi,\nn\\
K_c&\longrightarrow&lK_c.
\ea
[Note that the scaling for $K_c$ should be accompanied by the same scaling on $K_b$ in classical dynamics and on $\bar{K}'_b$ as well $f$ in the $\mubar'$-scheme; that is $K_b,\,\bar{K}'_b,\,f\longrightarrow l K_b,\,l\bar{K}'_b,\,lf$.]
On the other hand, the $\mubar$-scheme does not respect this scaling.\footnote{Also note that the symmetry involving $\phi\longrightarrow\lambda\phi$ as stated in Equation (4.2) of \cite{Chiou:2007mg} is a mistake and should be dismissed.}

The fact that the symmetry in \eqnref{eqn:symmetry 1} only scales $p_b$ and $c$ but leaves $p_c$ and $b$ fixed seems to spoil the idea that length/area/volume is measurable only if the line/surface/bulk is \emph{coupled with the material reference} as suggested in \cite{Rovelli:1990ph,Rovelli:1992vv} and affirmed in \cite{Chiou:2007mg} for the Bianchi I model. This is because in the Kantowski-Sachs spacetime the area ${\bf A}_{\theta\phi}=\pi g_{\Omega\Omega}$ (contrary to ${\bf A}_{x\phi}$) has no ambiguity even in the absence of matter content; in a sense, ${\bf A}_{\theta\phi}$ is measurable with reference to the spherical curvature of $S^2$ and thus the reference to $p_\phi$ is unnecessary. In the $K_{b\pm}$-asymptotic phases, where the spherical curvature is negligible (compared to the anisotropy), we expect the same scaling symmetry as of the Bianchi I model (see Equation (4.1) in \cite{Chiou:2007mg} and recall \remref{rem:Bianchi I} in \secref{sec:classical solution}). That is, if we ignore $\gamma^2p_b^2$ in the bracket of \eqnref{eqn:cl rescaled Hamiltonian} and the corresponding term in the effective Hamiltonian, the classical dynamics and $\mubar'$-scheme dynamics in the $K_{b\pm}$-asymptotic phases will approximately respect the symmetry of scaling under:
\ba
p_b,\ p_c&\longrightarrow&\lambda_b\lambda_c p_b,\ \lambda_b^2 p_c\nn\\
b,\ c&\longrightarrow& \lambda_bb,\ \lambda_cc\nn\\
p_\phi,\ K_\phi&\longrightarrow&\lambda_b^2\lambda_c p_\phi,\ \lambda_b^2\lambda_c K_\phi,\nn\\
K_c&\longrightarrow&\lambda_b^2\lambda_c K_c.
\ea
The $\mubar$-scheme dynamics respects this approximate symmetry as well if we set $\lambda_b=\lambda_c$.

Additionally, the classical dynamics also admits the symmetries given by
\ba
\tau&\longrightarrow&\eta \tau,\nn\\
\gamma&\longrightarrow&\xi \gamma,\nn\\
p_b,\ p_c&\longrightarrow& \eta^2p_b,\ \eta^2p_c,\nn\\
b,\ c&\longrightarrow& \xi b,\ \xi c,\nn\\
p_\phi,\ K_\phi&\longrightarrow&\eta^{2} p_\phi,\ \eta^{2} K_\phi,\nn\\
K_c&\longrightarrow&\eta^{2} K_c.
\ea
The scaling symmetry regarding $\gamma\longrightarrow \xi \gamma$ is expected, since the Barbero-Immirzi parameter $\gamma$ has no effect on the classical dynamics. The scaling symmetry regarding $\tau\longrightarrow \eta \tau$ is also easy to understand, since there is no temporal scale introduced in the Hamiltonian.\footnote{For the Bianchi I cosmology studied in \cite{Chiou:2007mg}, a different scaling $p_I \longrightarrow p_I$ with $c_I\longrightarrow \eta^{-1}c_I$ is chosen to respect the symmetry regarding $\tau\longrightarrow \eta\tau$. This alternative scaling does not work in the case of Kantowski-Sachs spacetime, since it violates the Hamiltonian constraint \eqnref{eqn:cl rescaled Hamiltonian}. That is to say, the presence of the spatial curvature [i.e., the $\gamma^2p_b^2$ term in the bracket in \eqnref{eqn:cl rescaled Hamiltonian}] ties the temporal scale with the spatial scale; as a result, only the scaling $p_b,\,p_c \longrightarrow \eta^2p_b,\,\eta^2p_c$ (with gives the spatial direction the same scaling as in the temporal direction) with $b,\,c\longrightarrow b,\,c$ preserves the symmetry.} However, very surprisingly, the scaling symmetry involving $\tau\longrightarrow \eta \tau$ is violated for both the $\mubar$-scheme and $\mubar'$-scheme phenomenological dynamics. Curiously, this symmetry is restored if $\tau\longrightarrow \eta \tau$ is accompanied by $\gamma \longrightarrow\xi \gamma$ and one extra scaling is also imposed at the same time:
\be
\Delta\longrightarrow \xi^{-2}\eta^2\Delta.
\ee
This intriguing observation seems to suggest, albeit speculatively, that in the context of quantum gravity the fundamental scale (area gap) in spatial geometry gives rise to a temporal scale via the nonlocality of quantum gravity (i.e., using holonomies) and the Barbero-Immirzi parameter $\gamma$ somehow plays the role bridging the scalings in time and space. [This reminds us that, in LQG, the precise value of the area gap $\Delta$ is proportional to $\gamma$, and $\gamma$ is also the parameter which relates the \emph{intrinsic} geometry (encoded by spin connection $\Gamma^i_a$) with the \emph{extrinsic} curvature (${K_a}^i$) via ${A_a}^i=\Gamma^i_a-\gamma{K_a}^i$.]

Meanwhile, related to the above observations, the physical meaning of the directional factors $\varrho_I$ and matter density $\rho_\phi$ can be interpreted as the (inverse of) area and volume scales, \emph{measured by the reference of the matter content}. In this regard, we may say that the big bounces take place when one of the areas ${\bf A}_{\theta\phi}$, ${\bf A}_{x\phi}$ (in the $\mubar$-scheme) or the volume ${\bf V}$ (in the $\mubar'$-scheme) undergoes the Planck regime (up to a numerical factor) measured by the reference of the matter momentum $p_\phi$.
It is then tempting to regard not only $\phi$ as the ``internal clock'' (emergent time) but also $p_\phi$ as the ``internal rod'' --- namely, the measurement of both temporal and spatial geometries makes sense only in the presence of matter content.

The above observation for the scaling symmetries draws a close parallel to those in the Bianchi I model \cite{Chiou:2007mg} and seems again to support the ideas of the relational interpretation of quantum mechanics with real rods and clocks such as studied in \cite{Gambini:2006ph} (see also \cite{Rovelli:1990ph,Rovelli:1992vv}) with the caveat that the nonvanishing spherical curvature can make the reference to matter content unnecessary. The caveat however does not fail the relational interpretation immediately; rather, it suggests that we should put the spatial curvature on the equal footing as the matter content.\footnote{This reminds us that in the FRW model the curvature term for $k=\pm1$ can be regarded as ``matter'' with state parameter $w=-1/3$.} More precisely, apart from energy density of matters to be the metric reference, we should also take into account the energy densities of both curvature and anisotropic shear.\footnote{It has been shown in Appendix B of \cite{Chiou:2007sp} that the anisotropic shear behaves as a kind of anisotropic matter; in particular, the directional densities can be considered as the ``energy density carried from the classical anisotropic shear portioned to the specific direction''.}

Unfortunately, all the scaling symmetries break down in the detailed construction of LQC with the $\mubar$-scheme (the strategy to construct the fundamental theory of LQC in the $\mubar'$-scheme is still not clear) even for the isotropic model (where the $\mubar$- and $\mubar'$-schemes are identical). The fundamental LQC only respects the scaling symmetries at the leading order. This is due to the fact that the quantum evolution in the fundamental LQC is governed by a difference equation, in which the step size of difference introduces an additional scale in the deep Planck regime (see \cite{Ashtekar:2006wn} for the isotropic model and \cite{Chiou:2006qq} for Bianchi I model). In fact, already in the level of phenomenological dynamics, the scaling symmetries are violated if we further take into account the LQC corrections on the cotriad component $\omega_c$. For the fundamental theory of LQC, if we take the aforementioned symmetries seriously, we might be able to revise the detailed construction in the spirit of relational quantum theory such that the step size in the difference equation scales adaptively by the reference of the matter content.

\section{Summary and discussion}\label{sec:summary}

To summarize, we list the important facts for the classical dynamics, $\mubar$-scheme and $\mubar'$-scheme phenomenological dynamics in \tabref{tab:summary}. In the following, the main results are restated and their implications are discussed.


\begin{table}

\begin{tabular}{|c|c|c|}
\hline\hline
\rule[-0.15cm]{0cm}{0.5cm}
\textbf{Classical dynamics} & \textbf{Phenomenology in $\mubar$-scheme} & \textbf{Phenomenology in $\mubar'$-scheme}\\

\hline\hline
\rule[-0.2cm]{0cm}{0.6cm}
$p_\phi=\sqrt{4\pi G^{-1}}\,K_\phi={\bf V}\dot{\phi}$ & $p_\phi=\sqrt{4\pi G^{-1}}\,K_\phi={\bf V}\dot{\phi}$ & $p_\phi=\sqrt{4\pi G^{-1}}\,K_\phi={\bf V}\dot{\phi}$\\

\hline
\begin{tabular}{c}
\rule{0cm}{0.35cm}
$2\pi p_b=2\pi L\sqrt{g_{xx}g_{\Omega\Omega}}={\bf A}_{x\phi}={\bf A}_{x\theta}$\\
$\pi p_c=\pi g_{\Omega\Omega}={\bf A}_{\theta\phi}$\vspace{1mm}
\end{tabular}
&
\begin{tabular}{c}
$2\pi p_b=2\pi L\sqrt{g_{xx}g_{\Omega\Omega}}={\bf A}_{x\phi}={\bf A}_{x\theta}$\\
$\pi p_c=\pi g_{\Omega\Omega}={\bf A}_{\theta\phi}$
\end{tabular}
&
\begin{tabular}{c}
$2\pi p_b=2\pi L\sqrt{g_{xx}g_{\Omega\Omega}}={\bf A}_{x\phi}={\bf A}_{x\theta}$\\
$\pi p_c=\pi g_{\Omega\Omega}={\bf A}_{\theta\phi}$
\end{tabular}\\

\hline
\begin{tabular}{l}
$b=\gamma\frac{d}{d\tau}\sqrt{g_{\Omega\Omega}}=\frac{\gamma}{2p_c^{1/2}}\frac{dp_c}{d\tau}$\\
$c=\gamma\frac{d}{d\tau}(L\sqrt{g_{xx}})$\\
\quad$=\frac{\gamma}{p_c^{1/2}}\frac{dp_b}{d\tau}-\frac{\gamma p_b}{2p_c^{3/2}}\frac{dp_c}{d\tau}$
\end{tabular}
&
\begin{tabular}{l}
$\frac{\sin(\mubar_bb)}{\mubar_b}=\frac{1}{\cos(\mubar_cc)}
\frac{\gamma}{2p_c^{1/2}}\frac{dp_c}{d\tau}$\\
$\frac{\sin(\mubar_cc)}{\mubar_c}=\frac{1}{\cos(\mubar_bb)}
\frac{\gamma}{p_c^{1/2}}\frac{dp_p}{d\tau}$\\
$\qquad\qquad-\frac{1}{\cos(\mubar_cc)}
\frac{\gamma p_b}{2p_c^{3/2}}\frac{dp_c}{d\tau}$
\end{tabular}
&
\begin{tabular}{l}
$\frac{\sin(\mubar'_bb)}{\mubar'_b}=\frac{1}{\cos(\mubar'_cc)}
\frac{\gamma}{2p_c^{1/2}}\frac{dp_c}{d\tau}$\rule{0cm}{0.45cm}\\
$\frac{\sin(\mubar'_cc)}{\mubar'_c}=\frac{1}{\cos(\mubar'_bb)}
\frac{\gamma}{p_c^{1/2}}\frac{dp_p}{d\tau}$\\
$\qquad\qquad-\frac{1}{\cos(\mubar'_cc)}
\frac{\gamma p_b}{2p_c^{3/2}}\frac{dp_c}{d\tau}$\vspace{1mm}
\end{tabular}\\

\hline
\begin{tabular}{l}
$p_cc=\gamma K_c$\\
$p_bb=\gamma K_b(\phi)$
\end{tabular}
&
\begin{tabular}{l}
$p_c\frac{\sin(\mubar_cc)}{\mubar_cc}=\gamma K_c$\rule{0cm}{0.4cm}\vspace{0.5mm}\\
$p_b\frac{\sin(\mubar_bb)}{\mubar_bb}=\gamma \bar{K}_b(\phi)$\vspace{1mm}
\end{tabular}
&
\begin{tabular}{l}
$p_cc=\gamma\left[K_c+f(\phi)\right]$\\
$p_bb=\gamma\left[\bar{K}'_b(\phi)+f(\phi)\right]$
\end{tabular}\\

\hline
$K_\phi^2=2K_bK_c+K_b^2+p_b^2$
&
$K_\phi^2=2\bar{K}_bK_c+\bar{K}_b^2+p_b^2$
&
\begin{tabular}{l}
\rule{0cm}{0.55cm}
$\quad\frac{\Delta\gamma^2K_\phi^2}{p_b^2p_c}=$\vspace{0.75mm}\\
$2\sin\left(\sqrt{\frac{\Delta\gamma^2}{p_b^2p_c}}(\bar{K}'_b+f)\right)
\sin\left(\sqrt{\frac{\Delta\gamma^2}{p_b^2p_c}}(K_c+f)\right)$\vspace{0.5mm}\\
$\qquad\ +\sin^2\left(\sqrt{\frac{\Delta\gamma^2}{p_b^2p_c}}(\bar{K}'_b+f)\right)
+\frac{\Delta\gamma^2}{p_c}$\vspace{1mm}
\end{tabular}\\

\hline
\begin{tabular}{l}
$\frac{1}{p_c}\frac{dp_c}{d\phi}=2\frac{\sqrt{4\pi G}}{K_\phi}K_b(\phi)$\\ \\
$\frac{1}{p_b}\frac{dp_b}{d\phi}
=\frac{\sqrt{4\pi G}}{K_\phi}\left[K_b(\phi)+K_c\right]$\\ \\ \\
$\frac{K_\phi}{\sqrt{4\pi G}}\frac{dK_b}{d\phi}=-p_b^2$
\end{tabular}
&
\begin{tabular}{l}
$\frac{1}{p_c}\frac{dp_c}{d\phi}=2\frac{\sqrt{4\pi G}}{K_\phi}\cos(\mubar_cc)\bar{K}_b(\phi)$\\ \\
$\frac{1}{p_b}\frac{dp_b}{d\phi}=\frac{\sqrt{4\pi G}}{K_\phi}
\cos(\mubar_bb)\left[\bar{K}_b(\phi)+K_c\right]$\\ \\ \\
$\frac{K_\phi}{\sqrt{4\pi G}}\frac{d\bar{K}_b}{d\phi}=
-\cos(\mubar_bb)\,p_b^2$
\end{tabular}
&
\begin{tabular}{l}
$\frac{1}{p_c}\frac{dp_c}{d\phi}
=\frac{2\sqrt{4\pi G}}{K_\phi}\sqrt{\frac{p_b^2p_c}{\gamma^2\Delta}}
\cos\left(\sqrt{\frac{\gamma^2\Delta}{p_b^2p_c}}(K_c+f)\right)$\rule{0cm}{0.5cm}\vspace{0.5mm}\\
$\qquad\qquad\times
\sin\left(\sqrt{\frac{\gamma^2\Delta}{p_b^2p_c}}(\bar{K}'_b+f)\right)$\vspace{1mm}\\
$\frac{1}{p_b}\frac{dp_b}{d\phi}
=\frac{\sqrt{4\pi G}}{K_\phi}\sqrt{\frac{p_b^2p_c}{\gamma^2\Delta}}
\cos\left(\sqrt{\frac{\gamma^2\Delta}{p_b^2p_c}}(\bar{K}'_b+f)\right)$\vspace{1mm}\\
$\qquad\qquad\times\left[
\sin\left(\sqrt{\frac{\gamma^2\Delta}{p_b^2p_c}}(\bar{K}'_b+f)\right)\right.$\\
$\qquad\qquad\qquad\qquad+\left.\sin\left(\sqrt{\frac{\gamma^2\Delta}{p_b^2p_c}}(K_c+f)\right)
\right]$\\
$\frac{K_\phi}{\sqrt{4\pi G}}\frac{d\bar{K}'_b}{d\phi}=-p_b^2$\vspace{1mm}
\end{tabular}\\

\hline
\begin{tabular}{l}
$p_c,\,p_b\rightarrow 0$\\
toward both big bang and\\
big crunch singularities.\\ \\
$\mbox{}$
\end{tabular} &
\begin{tabular}{l}
$p_c$ bounces whenever\\
$\quad\varrho_c:=\frac{p_\phi^2}{32\pi^2p_c^3}
=\frac{K_\phi^2}{K_c^2}\rho_{\rm Pl}$;\\
$p_b$ bounces whenever\\
$\quad\varrho_b:=\frac{p_\phi^2}{32\pi^2p_b^3}
\approx\frac{K_\phi^2}{K_{b\pm}^2}\rho_{\rm Pl}$;\\
$p_b$ bounces periodically;\\
$p_c$ bounces only a few times.
\end{tabular} &
\begin{tabular}{l}
$p_c$ bounces at the moment when\\
$\quad\rho_\phi:=\frac{p_\phi^2}{32\pi^2p_b^2p_c}
\approx\frac{K_\phi^2}{F(K_c,K_{b\pm})}\rho_{\rm Pl}$;\\
$p_b$ bounces at the moment when\\
$\quad\rho_\phi:=\frac{p_\phi^2}{32\pi^2p_b^2p_c}
\approx\frac{K_\phi^2}{G(K_c,K_{b\pm})}\rho_{\rm Pl}$;\\
$p_b,\,p_c$ bounce roughly around\\
the same moments.\vspace{0.5mm}
\end{tabular}\\

\hline
\begin{tabular}{c}
No big bounce;\\
$K_c$ fixed
\end{tabular} &
\begin{tabular}{l}
Big bounces bridge classical cycles\\
with $K_c$ and $-K_c$ and give rise\\
to ``meta-classical'' phases.
\end{tabular}
&
\begin{tabular}{l}
Big bounces bridge classical cycles\\
with varying effective $K_c$:\\
$\cdots\ \utilde{\utilde{K}}_c\rightleftarrows\utilde{K}_c
\rightleftarrows K_c \rightleftarrows \widetilde{K}_c \rightleftarrows
\widetilde{\widetilde{K}}_c\ \cdots$\vspace{0.65mm}
\end{tabular}\\

\hline
\begin{tabular}{c}
Symmetry of scaling:\\
$\tau\longrightarrow\eta \tau$\\
$\gamma\longrightarrow\xi \gamma$\\
$p_b,\ p_c\longrightarrow l\eta^2 p_b,\ \eta^2p_c$\\
$b,\ c\longrightarrow \xi b,\ l\xi c$\\
$p_\phi,\ K_\phi\longrightarrow l\eta^{2}p_\phi,\ l\eta^{2}K_\phi$\\
$K_c\longrightarrow l\eta^{2}K_c$\\
$\mbox{}$\vspace{0.65mm}
\end{tabular}
&
\begin{tabular}{c}
Symmetry of scaling:\\
$\tau\longrightarrow\eta \tau$\\
$\gamma\longrightarrow\xi \gamma$\\
$p_b,\ p_c\longrightarrow \eta^2p_b,\ \eta^2p_c$\\
$b,\ c\longrightarrow \xi b,\ \xi c$\\
$p_\phi,\ K_\phi\longrightarrow \eta^{2}p_\phi,\ \eta^{2}K_\phi$\\
$K_c\longrightarrow \eta^{2}K_c$\\
$\Delta\longrightarrow\xi^{-2}\eta^2\Delta$
\vspace{0.65mm}
\end{tabular}
&
\begin{tabular}{c}
Symmetry of scaling:\\
$\tau\longrightarrow\eta \tau$\\
$\gamma\longrightarrow\xi \gamma$\\
$p_b,\ p_c\longrightarrow l\eta^2p_b,\ \eta^2p_c$\\
$b,\ c\longrightarrow \xi b,\ l\xi c$\\
$p_\phi,\ K_\phi\longrightarrow l\eta^{2}p_\phi,\ l\eta^{2}K_\phi$\\
$(K_c+f)\longrightarrow l\eta^{2}(K_c+f)$\\
$\Delta\longrightarrow\xi^{-2}\eta^2\Delta$
\vspace{0.65mm}
\end{tabular}\\

\hline
\begin{tabular}{c}
Approximate symmetry in\\
$K_{b\pm}$-asymptotic phases:\\
$p_b,\ p_c\longrightarrow \lambda_b\lambda_cp_b,\ \lambda_b^2p_c$\\
$b,\ c\longrightarrow \lambda_b b,\ \lambda_c c$\\
$p_\phi,\ K_\phi\longrightarrow \lambda_b^2\lambda_c p_\phi,\ \lambda_b^2\lambda_c K_\phi$\\
$K_c\longrightarrow \lambda_b^2\lambda_c K_c$\\
\end{tabular}
&
\begin{tabular}{c}
Approximate symmetry in\\
$K_{b\pm}$-asymptotic phases:\\
$p_b,\ p_c\longrightarrow \lambda^2 p_b,\ \lambda^2 p_c$\\
$b,\ c\longrightarrow \lambda b,\ \lambda c$\\
$p_\phi,\ K_\phi\longrightarrow \lambda^3p_\phi,\ \lambda^3K_\phi$\\
$K_c\longrightarrow \lambda^3K_c$
\end{tabular}
&
\begin{tabular}{c}
Approximate symmetry in\\
$K_{b\pm}$-asymptotic phases:\\
$p_b,\ p_c\longrightarrow \lambda_b\lambda_c p_b,\ \lambda_b^2p_c$\\
$b,\ c\longrightarrow \lambda_b b,\ \lambda_c c$\\
$p_\phi,\ K_\phi\longrightarrow \lambda_b^2\lambda_c p_\phi,\ \lambda_b^2\lambda_c K_\phi$\\
$(K_c+f)\longrightarrow \lambda_b^2\lambda_c (K_c+f)$\vspace{0.75mm}
\end{tabular}\\

\hline\hline
\end{tabular}
\caption{Summary of the classical dynamics, $\mubar$-scheme and $\mubar'$-scheme phenomenological dynamics.}\label{tab:summary}
\end{table}

With the LQC discreteness corrections, the phenomenological dynamics shows that the classical singularities (both big bang and big crunch) are resolved and \emph{replaced by the big bounces}. In the $\mubar$-scheme, it is the directional densities $\varrho_b$ and $\varrho_c$ that signal the occurrence of big bounces when $\varrho_b$ and $\varrho_c$ approach the critical values $\varrho_{b\pm,\,{\rm crit}}$ and $\varrho_{c,\,{\rm crit}}$ respectively. In the $\mubar'$-scheme, the indication of big bounces is the matter density $\rho_\phi$ and the big bounces take place around the moments when $\rho_\phi$ is close to the critical values $\rho_{b\pm,\,{\rm crit}}$ and $\rho_{c\pm,\,{\rm crit}}$.

The detailed evolution in the $\mubar$-scheme shows that the equations of motion (in terms of emergent time $\phi$) for $p_b$ and $\bar{K}_b$ are decoupled from $p_c$ and $c$ (the dependence on $p_c$ and $c$ is only through the constant $K_c$). As a result, the bouncing scenario of $p_b$ is unaffected by the varying of $p_c$ and perfectly periodic. On the other hand, $p_c$ only bounces a few times and grows up toward infinity in the far future and past.
By contrast, in the $\mubar'$-scheme, equations of motion for $p_b$ and $p_c$ are closely coupled through $\rho_\phi$ and thus $p_b$ and $p_c$ bounce roughly around the same moment.

The big bounces bridge different ``cycles'' of classical solutions. In the $\mubar$-scheme, the classical solutions with (effective) constants $K_c$ and $-K_c$ are bridged by the big bounces. For some periods of time, the phenomenological dynamics also yields ``meta-classical'' phases, which are absent in the ordinary classical evolution. (Further investigation is needed to know whether the occurrence of meta-classical phases indicates wrong semiclassical behavior.) On the other hand, in the $\mubar'$-scheme, starting with the constant $K_c$ in a given classical cycle, the evolution ends up with the new effective constant $\widetilde{K}_c$ or $\utilde{K}_c$ in the adjacent classical cycle across the big bounce. Furthermore, the bouncing behaviors of $p_b$ and $p_c$ oscillate more and more rapidly toward the future and past; eventually $p_b$ grows very huge (and $\rho_\phi$ subsides) while $p_c$ descends into the deep Planck regime, where the semiclassical analysis of phenomenological dynamics can no longer be trusted and the quantum corrections on the cotriad component $\omega_c$ become important.

In regard to the finite sized interval ${\cal I}$ chosen to make sense of the Hamiltonian formalism, the phenomenological dynamics in the $\mubar$-scheme depends on the choice of ${\cal I}$, and hence reacts to the macroscopic scales introduced by the boundary condition. (In terms of symmetry, it is said that the $\mubar$-scheme has no such scaling symmetry respected by the $\mubar'$-scheme and classical dynamics.) The phenomenological dynamics in the $\mubar'$-scheme, in contrast, is completely independent of ${\cal I}$ as is the classical dynamics. In case that the physical size of ${\cal I}$ has a global meaning (such as in the compactified Kantowski-Sachs model or in the lattice refining model of \cite{Bojowald:2007ra}), the condition for the bounce occurrence can be rephrased: In the $\mubar$/$\mubar'$-scheme (respectively), the physical \emph{area/volume} of the surfaces/volume of ${\bf A}_{x\theta}$, ${\bf A}_{\theta\phi}$ or ${\bf V}$ gets bounced when it undergoes the Planck regime (times a numerical value) measured by the reference of the momentum $p_\phi$.

While the $\mubar'$-scheme has the advantage that its phenomenological dynamics is independent of ${\cal I}$, the fundamental theory of LQC based on the $\mubar'$-scheme is difficult to construct. Both the $\mubar$- and $\mubar'$-schemes have desirable merits and it is still disputable which one (or yet another possibility) is more faithful to implement the underlying physics of quantum geometry.

In addition to the symmetry related to the choice of ${\cal I}$, both schemes admit additional symmetries of scaling, which are reminiscent of the relational interpretation of quantum mechanics, featuring the ideas of real rods and clocks. Furthermore, the symmetry involving the Barbero-Immirzi parameter is suggestive that the fundamental scale (area gap) in spatial geometry may give rise to a fundamental scale in temporal measurement. These symmetries however break down in the construction for the fundamental theory of LQC.

Most results obtained in \cite{Chiou:2007mg} for the Bianchi I model are analogously affirmed at the level of phenomenological dynamics for the Kantowski-Sachs spacetime and the close parallels between these two cases are well established. (Also see \remref{rem:Bianchi I} in\secref{sec:classical solution}.) Additionally, new features also arise due to the presence of the spherical curvature: The bouncing scenario exhibits (semi)-cyclic patterns; and in the $\mubar'$-scheme, $p_c$ eventually descends into the deep Planck regime, whereby the validity of the phenomenological dynamics could be questioned and more sophisticated treatment may be required to faithfully convey the quantum geometry of the full theory of LQG. At this stage, it is not clear what happens exactly when $p_c$ reaches Planck regime.

Meanwhile, it is noteworthy that the Kantowski-Sachs spacetime describes the interior of the Schwarzschild black hole (see \remref{rem:black hole} in \secref{sec:classical solution}). Although, by introducing the scalar field, we dismiss the issues for black holes and instead study a self-contained cosmological model, it is still instructive to compare the phenomenological dynamics of the Schwarzschild interior studied in \cite{Bohmer:2007wi} with the results obtained in this paper. It has been shown in \cite{Bohmer:2007wi} that in both the $\mu_o$- and $\mubar$-schemes (referred to as ``constant $\delta$ Hamiltonian'' and ``alternative quantum Hamiltonian'' in \cite{Bohmer:2007wi}), the phenomenological dynamics  bridges a classical black hole with a white hole through a bounce, whereas in the $\mubar'$-scheme phenomenological dynamics (referred to as ``improved quantum Hamiltonian'' in \cite{Bohmer:2007wi}), $p_c$ oscillates and eventually lands on a constant in deep Planck regime.\footnote{Therefore, it is claimed in \cite{Bohmer:2007wi} that the $\mubar$-scheme phenomenological dynamics extends a classical Schwarzschild black hole to a patch of a nonsingular charged Nariai universe, which gives constant $p_c$. However, a closer look suggests that the extended part is \emph{not} a patch of the classical Nariai universe but instead represents the quantum universe which \emph{formally} exhibits Nariai type metric, as the asymptotic constant value for $p_c$ is in the deep Planck regime ($\alt\Pl^2$). For this reason, some of the claims in \cite{Bohmer:2007wi} for the $\mubar'$-scheme may require further investigation.} This dichotomy is analogous to the qualitative difference in the bouncing cosmological scenarios between the $\mubar$- and $\mubar'$-schemes observed in this paper. This similarity also suggests that, by exploiting the strategy we used in this paper (in particular, to identify the constant $K_c$ and the functions $K_b$, $\bar{K}$, $\bar{K}'$), we could be able to reproduce the results of \cite{Bohmer:2007wi} for the black hole interior in greater detail (such as to pinpoint the bounce occurrence condition and the varying of the effective $K_c$); the quantum corrections on the inner side of horizon could also be studied.

Furthermore, as this paper focuses specifically on the model with a massless scalar field, it should be straightforward (with necessary approximation) to extend the results to the models with inclusion of generic matters. As studied in \cite{Chiou:2007sp} for the Bianchi I model, it is anticipated that there would be a competition among the matter density, anisotropy and spherical curvature to be the indication of the occurrence of big bounces. Studying the loop quantum geometry of Kantowski-Sachs spacetime with generic matters would further support or oppose our observations and help us to understand them in a broader context.


\acknowledgements{
The author would like to thank Birjoo Vaishnav for bringing this topic to his attention and having useful discussions. It is also greatly appreciated that Martin Bojowald spent his precious time carefully reading the draft of this paper and giving valuable comments. This work was supported in part by the NSF Grant No. PHY-0456913.
}

\vspace{0.5cm}

\appendix

\section{Phenomenological dynamics in the $\mu_o$-scheme}\label{sec:muzero dynamics}
One of the virtues of the improved strategy ($\mubar$- or $\mubar'$-scheme) in both the isotropic and Bianchi I models is to fix the serious drawback in the old precursor strategy ($\mu_o$-scheme) that the critical value of directional densities $\varrho_I$ (in the $\mubar$-scheme) or of matter density $\rho_\phi$ (in the $\mubar'$-scheme) at which the bounce occurs can be made arbitrarily small by increasing the momentum $p_\phi$, thereby leading to wrong semiclassical behavior.

Having learned from the isotropic and Bianchi I cases, we expect that the critical values of $\varrho_c$, $\varrho_b$ and $\rho_\phi$ at which the bounces occur can be made arbitrarily small by increasing the momentum $p_\phi$ in the $\mu_0$-scheme but are independent of $p_\phi$ in the $\mubar$- or $\mubar'$-scheme. The latter is what has been shown in the main text of this paper.\footnote{In the $\mubar$-scheme, the critical values $\varrho_{c,{\rm crit}}$ and $\varrho_{b\pm,{\rm crit}}$ depend on $p_\phi$ only through the ratios $K_\phi^2/K_c^2$ and $K_\phi^2/K_{b\pm}^2$; In the $\mubar'$-scheme, $\rho_{c\pm,{\rm crit}}$ and $\rho_{b\pm,{\rm crit}}$ depends on $p_\phi$ only through $K_\phi^2/F(K_c,K_{b\pm})$ and $K_\phi^2/G(K_c,K_{b\pm})$.} For comparison, the phenomenological dynamics in the $\mu_o$-scheme is presented here.

In the phenomenological theory of the $\mu_o$-scheme, we take the
prescription to replace $c$ and $b$ with $\sin(\muzero_cc)/\muzero_c$ and $\sin(\muzero_bb)/\muzero_b$ by introducing the \emph{fixed} numbers $\muzero_c$ and $\muzero_b$ for discreteness. Analogous to \eqnref{eqn:qm Hamiltonian},
we have the effective (rescaled) Hamiltonian constraint:
\ba\label{eqn:qm0 Hamiltonian}
H'_{\mu_o}&=&
-\frac{1}{2G\gamma}
\left\{
2\frac{\sin(\muzero_bb)}{\muzero_b}\frac{\sin(\muzero_cc)}{\muzero_c}\,p_bp_c
+\left(\frac{\sin(\muzero_bb)}{\muzero_b}\right)^2 p_b^2
+\gamma^2p_b^2
\right\}
+
\gamma\frac{p_\phi^2}{8\pi}.
\ea

Again, the equations of motion are given by the Hamiltonian constraint $H'_{\mu_o}=0$ and Hamilton's equations:
\ba\label{eqn:qm0 eom 1}
\frac{dp_\phi}{dt'}&=&\{p_\phi,H'_{\mu_o}\}=0\quad\Rightarrow\
p_\phi\ \text{is constant},\\
\label{eqn:qm0 eom 2}
\frac{d\phi}{dt'}&=&\{\phi,H'_{\mu_o}\}=\frac{\gamma}{4\pi}p_\phi,\\
\label{eqn:qm0 eom 3}
\frac{dc}{dt'}&=&\{c,H'_{\mu_o}\}=2G\gamma\,\frac{\partial\, H'_{\mu_o}}{\partial p_c}
=-2p_b\frac{\sin(\muzero_bb)}{\muzero_b}
\frac{\sin(\muzero_cc)}{\muzero_c},\\
\label{eqn:qm0 eom 4}
\frac{dp_c}{dt'}&=&\{p_c,H'_{\mu_o}\}=-2G\gamma\,
\frac{\partial\, H'_{\mu_o}}{\partial c}
=2p_bp_c\cos(\muzero_cc)
\frac{\sin(\muzero_bb)}{\muzero_b},\\
\label{eqn:qm0 eom 5}
\frac{db}{dt'}&=&\{b,H'_{\mu_o}\}=G\gamma\,\frac{\partial\, H'_{\mu_o}}{\partial p_b}
=-p_c\frac{\sin(\muzero_bb)}{\muzero_b}\frac{\sin(\muzero_cc)}{\muzero_c}
-p_b\left[\frac{\sin(\muzero_bb)}{\muzero_b}\right]^2
-\gamma^2p_b,\\
\label{eqn:qm0 eom 6}
\frac{dp_b}{dt'}&=&\{p_b,H'_{\mu_o}\}=-G\gamma\,
\frac{\partial\, H'_{\mu_o}}{\partial b}
=p_b\cos(\muzero_bb)
\left[p_b\frac{\sin(\muzero_bb)}{\muzero_b}
+p_c\frac{\sin(\muzero_cc)}{\muzero_c}\right],
\ea
which follow
\be\label{eqn:qm0 dKc/dt'}
\frac{d}{dt'}\left[p_c\frac{\sin(\muzero_cc)}{\muzero_c}\right]=0
\qquad\Rightarrow\qquad
p_c\frac{\sin(\muzero_cc)}{\muzero_c}=\gamma K_c
\ee
and
\be\label{eqn:qm0 dKb/dt'}
p_c\frac{\sin(\muzero_bb)}{\muzero_b}=:\gamma K_b^o(t'),
\qquad
\frac{d K_b^o}{dt'}
=-\gamma^2p_b^2\cos(\muzero_bb).
\ee
These are exactly the same as \eqnref{eqn:qm dKc/dt'}--\eqnref{eqn:qm Kb} except that the discreteness variables $\mubar_c$, $\mubar_b$ are now replaced by $\muzero_c$ and $\muzero_b$.

Therefore, exploiting the close resemblance between the $\mubar$-scheme and $\mu_o$-scheme, we can readily repeat the calculation we did in \secref{sec:mubar dynamics} and obtain the differential equations [cf. \eqnref{eqn:qm diff eq for Kb}--\eqnref{eqn:qm diff eq for pb}]:
\be\label{eqn:qm0 diff eq for Kb}
\gamma^{-1}\frac{d K^o_b}{dt'}=\frac{K_\phi}{\sqrt{4\pi G}}\frac{d K^o_b}{d\phi}
=\cos(\muzero_bb)\left(2K^o_bK^o_c+{K^o_b}^2-K_\phi^2\right),
\ee
\ba
\label{eqn:qm0 diff eq for pc}
\frac{1}{p_c}\frac{dp_c}{d\phi}&=&
2\frac{\sqrt{4\pi G}}{K_\phi}\,\cos(\muzero_cc)K^o_b,\\
\label{eqn:qm0 diff eq for pb}
\frac{1}{p_b}\frac{dp_b}{d\phi}&=&
\frac{\sqrt{4\pi G}}{K_\phi}\,\cos(\muzero_bb)\left[K^o_b+K_c\right],
\ea
where
\ba\label{eqn:qm0 cosc}
\cos(\muzero_cc)&=&\pm\sqrt{1-\sin^2\muzero_cc}
=\pm\sqrt{1-\left(\frac{\gamma\muzero_cK_c}{p_c}\right)^2}
=\pm\sqrt{1-
\bigg(\frac{\varrho_c}{\varrho^{\mu_o}_{c,\,{\rm crit}}}\bigg)^{2/3}}\,,\\
\label{eqn:qm0 cosb}
\cos(\muzero_bb)&=&\pm\sqrt{1-\sin^2\muzero_bb}
\approx\pm\sqrt{1-\left(\frac{\gamma\muzero_cK_{b\pm}}{p_b}\right)^2}
=\pm\sqrt{1-
\bigg(\frac{\varrho_b}{\varrho^{\mu_o}_{b\pm,\,{\rm crit}}}\bigg)^{2/3}}\,,
\ea
which give the bouncing solutions similar to those given in the $\mubar$-scheme phenomenological dynamics except that the critical values at which the big bounce takes place are given by
\ba
\varrho^{\mu_o}_{c,\,{\rm crit}}&:=&
\left[
\frac{K_\phi^2}{K_c^2}\,
\frac{\rho_{\rm Pl}\Delta}{{\muzero_c}^2}
\right]^{3/2}
\frac{1}{p_\phi},\\
\varrho^{\mu_o}_{b\pm,\,{\rm crit}}&:=&
\left[
\frac{K_\phi^2}{K_{b\pm}^2}\,
\frac{\rho_{\rm Pl}\Delta}{{\muzero_b}^2}
\right]^{3/2}
\frac{1}{p_\phi},
\ea
which can be made arbitrarily small by increasing the value of $p_\phi$. As a result, the $\mu_o$-scheme gives wrong semiclassical behavior and should be improved by the $\mubar$- or $\mubar'$-scheme to fix the problem.

\section{Effective Hamiltonian in the $\mubar$-schemes}\label{sec:mubar schemes}

In this appendix, starting from the Hamiltonian constraint of LQG, we derive the gravitational part of the Hamiltonian with Kantowski-Sachs symmetry and give a heuristic argument for the prescription given by \eqnref{eqn:sin prescription}. The motivations for both the $\mubar$- and $\mubar'$-schemes are addressed in detail. The advantages and drawbacks of both schemes are also remarked.

The gravitational part of the classical Hamiltonian constraint in the full (unreduced) theory is given by
\be\label{eqn:Hamiltonian full}
H_{\rm grav}=\frac{1}{8\pi G}\int d^3x N e^{-1}
\left\{{\epsilon_i}^{jk}F^i_{ab}\tE{a}{j}\tE{b}{k}
-2(1+\gamma^2){K_{[a}}^i{K_{b]}}^j\tE{a}{i}\tE{b}{j}\right\},
\ee
where $e:=|\det\tilde{E}|^{1/2}=\sqrt{q}$.

With the Kantowski-Sachs symmetry, the connection potential is given by \eqnref{eqn:symmetric A}, which leads to the field strength:
\ba\label{eqn:F}
F&=&\frac{1}{2}F_{ab}dx^a\wedge dx^b = dA+A\wedge A\nn\\
&=&(\tilde{b}^2-1)\tau_3\sin\theta\, d\theta\wedge d\phi
+2\tilde{b}\tilde{c}\tau_2\sin\theta\, d\phi\wedge dx
-2\tilde{b}\tilde{c}\tau_1 dx\wedge d\theta.
\ea
On the other hand, the densitized triad given by \eqnref{eqn:symmetric E} gives $e=\sqrt{q}=\tilde{p}_b\sqrt{\tilde{p}_c}\sin\theta$ and the corresponding cotriad:
\be
\omega=\omega_a^i\tau_idx^a=\frac{\tilde{p}_b}{\sqrt{\tilde{p}_c}}\tau_3 dx
+\sqrt{\tilde{p}_c}\tau_2 d\theta
-\sqrt{\tilde{p}_c}\tau_1 \sin\theta d\phi
\ee
via $\omega_a^i\tE{a}{j}=\sqrt{q}\,\delta^i_j$.
The compatibility relation $d\omega+\Gamma\wedge\omega=0$ then yields the spin connection:
\be
\Gamma=\tau_3\cos\theta d\phi.
\ee
Consequently, the extrinsic curvature $K$ is given by
\be
\gamma K:=A-\Gamma=\tilde{c}\tau_3dx+\tilde{b}\tau_2d\theta-\tilde{b}\tau_1\sin\theta d\phi,
\ee
which follows
\be\label{eqn:K K}
\gamma^2 K\wedge K=\tilde{b}^2\tau_3\sin\theta\, d\theta\wedge d\phi
+\tilde{b}\tilde{c}\tau_2\sin\theta\, d\phi\wedge dx
-\tilde{b}\tilde{c}\tau_1 dx\wedge d\theta.
\ee
Therefore, we have
\be\label{eqn:part1}
{\epsilon_i}^{jk}F^i_{ab}\tE{a}{j}\tE{b}{k}
=(\tilde{b}^2-1)\tilde{p}_b^2\sin^2\theta
+2\tilde{b}\tilde{c}\tilde{p}_b\tilde{p}_c\sin^2\theta
\ee
and
\be\label{eqn:part2}
\gamma^2{K_{[a}}^i{K_{b]}}^j\tE{a}{i}\tE{b}{j}
=\frac{1}{2}\tilde{b}^2\tilde{p}_b^2\sin^2\theta
+\tilde{b}\tilde{c}\tilde{p}_b\tilde{p}_c\sin^2\theta.
\ee
Putting \eqnref{eqn:part1} and \eqnref{eqn:part2} into \eqnref{eqn:Hamiltonian full} and restricting the integral to the finite sized shell ${\cal I}\times S^2$ as prescribed in \eqnref{eqn:Omega tilde}, we then have the gravitational part of the classical Hamiltonian in terms of the reduced variables:
\ba\label{eqn:H grav}
H_{\rm grav}&=&-\frac{N}{8\pi G\gamma^2}\int_{{\cal I}\times S^2}
\!\!d^3x\, \frac{\sin\theta}{\tilde{p}_b\sqrt{\tilde{p}_c}}
\left[
2\tilde{b}\tilde{c}\tilde{p}_b\tilde{p}_c
+(\tilde{b}^2+\gamma^2)\tilde{p}+b^2
\right]\nn\\
&=&-\frac{N}{2G\gamma^2}\left[
2bc\sqrt{p_c}+(b^2+\gamma^2)\frac{p_b}{\sqrt{p_c}}
\right],
\ea
which is the $H_{\rm grav}$ given in \eqnref{eqn:cl H grav}.

When the quantization is performed in the context of LQC, there are two loop quantum corrections. The first is the modification on the cotriad operator $\hat{\omega}_c$, which is negligible and ignored in this paper.
The second is due to the fact that the connections ${A_a}^i$ (or $b,\,c$) do not exist and should be replaced by holonomies (or exponentials of $b,\,c$).

Following the standard techniques in gauge theories, the curvature component $F^i_{ab}$ can be expressed in terms of holonomies (i.e. Wilson loops). Given a small surface $\alpha$ center in $\vec{x}$, Stokes' theorem allows us to write the curvature component as
\be\label{eqn:curvature}
\tau_i F^i_{ab}(\vec{x})\approx\frac{1}{\epsilon^\alpha_{ab}}\int_\alpha \tau_i F^i_{cd}\, dx^c\wedge dx^d
\approx\frac{1}{\epsilon^\alpha_{ab}}\left[{\cal P}\exp\left(\oint_{\partial\alpha} \tau_i{A_c}^i\, dx^c\right)-1\right],\quad
\ee
where $\partial\alpha$ is the boundary loop of $\alpha$ and $\epsilon^\alpha_{ab}=\int_{\alpha}dx^a\wedge dx^b$ is the coordinate area of $\alpha$ projected in ``$ab$-direction''.
This is a good approximation provided $\epsilon^\alpha_{ab}$ is small enough and in fact it becomes exact in the continuous limit $\epsilon^\alpha_{ab}\rightarrow 0$.

\begin{figure}
\begin{picture}(510,100)(0,0)

\put(140,90){\textbf{(a)}}
\put(340,90){\textbf{(b)}}

\put(20,5){
\scalebox{0.7}
{\includegraphics{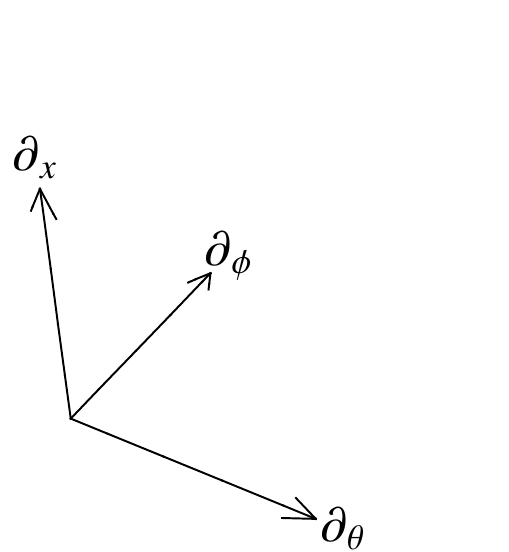}}
}

\put(130,-42){
\scalebox{0.72}
{\includegraphics{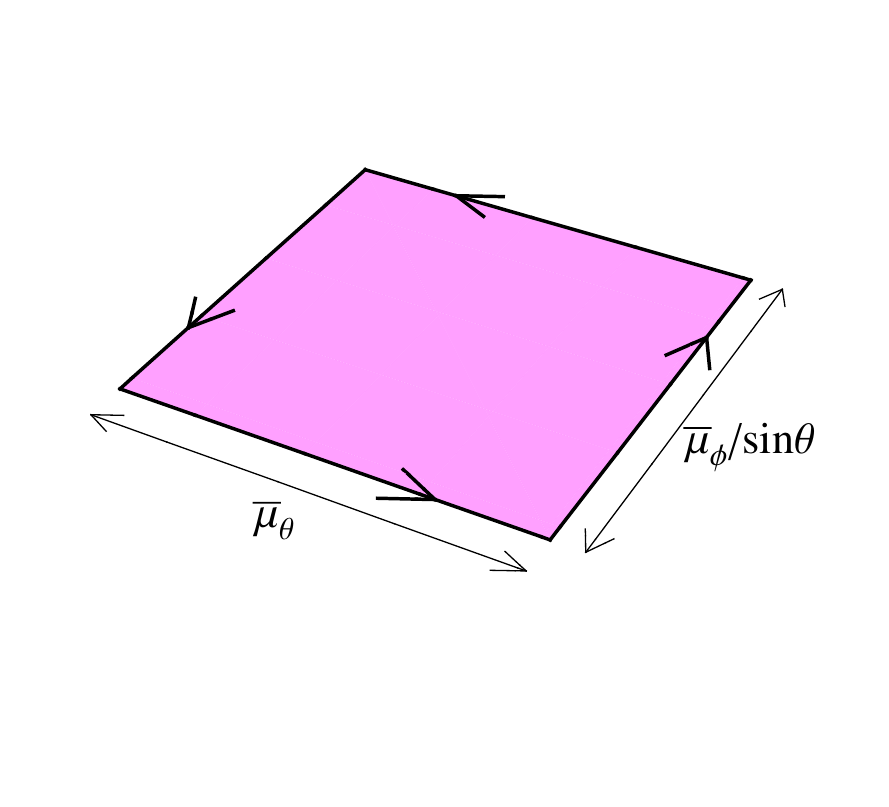}}
}

\put(320,-37){
\scalebox{0.72}
{\includegraphics{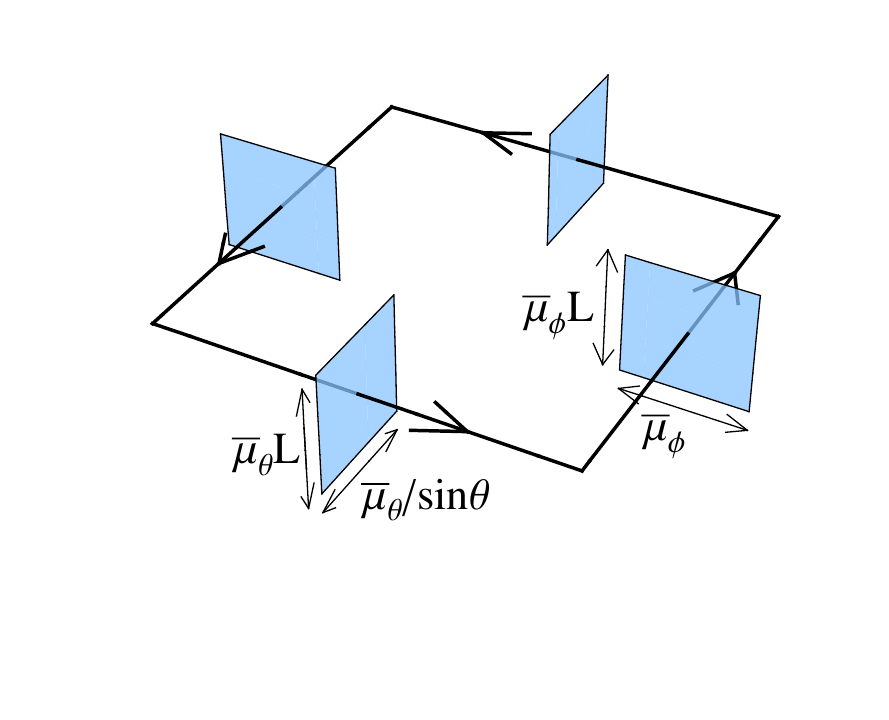}}
}

\end{picture}
\caption{\textbf{(a)} The surface in pink is $\Box_{\theta\phi}$, the physical area of which is to be shrunk to $\Delta$ in the $\mubar'$-scheme. \textbf{(b)} The surfaces in blue are $\boxdot_\theta$ and $\boxdot_\phi$, the physical areas of which are to be shrunk to $\Delta$ in the $\mubar$-scheme.}\label{fig:holonomy}
\end{figure}

With the Kantowski-Sachs symmetry, we choose $\alpha$ to be small rectangular surfaces $\Box_{\theta\phi}$, $\Box_{\phi x}$ and $\Box_{x\theta}$ normal to the vectors $\partial_x$, $\partial_\theta$ and $\partial_\phi$ respectively. The \emph{coordinate} lengths of the edges of $\alpha$ along the directions $\partial_x$, $\partial_\theta$ and $\partial_\phi$ are denoted as $\mubar_x L$, $\mubar_\theta$ and $\mubar_\phi/\sin\theta$ with the discreteness parameters introduced. (See \figref{fig:holonomy}(a).)
We then read off from \eqnref{eqn:curvature} that
\ba
\label{eqn:Fab1}
F^1_{x\theta}&\approx&
-\frac{2}{\mubar_x\mubar_\theta}{\rm Tr}
\left[\tau_1 \left(h_x^{(\mubar_x)}h_\theta^{(\mubar_\theta)}
(h_x^{(\mubar_x)})^{-1}(h_\theta^{(\mubar_\theta)})^{-1}-1\right)
\right],\\
\label{eqn:Fab2}
F^2_{\phi\theta}&\approx&
-\frac{2\sin\theta}{\mubar_\phi\mubar_x}{\rm Tr}
\left[\tau_2 \left(h_\phi^{(\mubar_\phi)}h_x^{(\mubar_x)}
(h_\phi^{(\mubar_\phi)})^{-1}(h_x^{(\mubar_x)})^{-1}-1\right)
\right],\\
\label{eqn:Fab3}
F^3_{\theta\phi}&\approx&
-\frac{2\sin\theta}{\mubar_\theta\mubar_\phi}{\rm Tr}
\left[\tau_3 \left(h_\theta^{(\mubar_\theta)}h_\phi^{(\mubar_\phi)}
(h_\theta^{(\mubar_\theta)})^{-1}(h_\phi^{(\mubar_\phi)})^{-1}-1\right)
\right],
\ea
with $h_x^{(\mubar_x)}$, $h_\theta^{(\mubar_th)}$ and $h_\phi^{(\mubar_\phi)}$ being the holonomies along the individual edge of $\alpha$:
\ba
h_x^{(\mubar_x)}(A)&:=&{\cal P}\exp\left(\int_x^{x+\mubar_x L}\!\!\tau_i{A_x}^{i} \,dx\right)
=e^{\,\mubar_x c\tau_3}
=\cos\left(\frac{\mubar_x c}{2}\right)+2\sin\left(\frac{\mubar_x c}{2}\right)\,\tau_3,\\
h_\theta^{(\mubar_\theta)}(A)&:=&{\cal P}\exp\left(\int_\theta^{\theta+\mubar_\theta}\!\!\tau_i{A_\theta}^{i} \,d\theta\right)
=e^{\,\mubar_\theta b\tau_2}
=\cos\left(\frac{\mubar_\theta b}{2}\right)+2\sin\left(\frac{\mubar_\theta b}{2}\right)\,\tau_2,\\
h_\phi^{(\mubar_\phi)}(A)&:=&{\cal P}\exp\left(\int_\phi^{\phi+\frac{\mubar_\phi}{\sin\theta}}\!\!\tau_i{A_\phi}^{i} \,d\phi\right)
=\exp\left(\mubar_\phi(-b\tau_1+\tau_3\cot\theta)\right).
\ea

Essentially, this is to replace the components of the connection $A$ in the form of \eqnref{eqn:symmetric A} with the holonomies $h_x^{(2\mubar_x)}$, $h_\theta^{(2\mubar_\theta)}$, and $h_\phi^{(2\mubar_\phi)}$ via:
\ba
\gamma\tau_i{K_x}^i&=&\tau_i{A_x}^i=\tilde{c}\tau_3=\frac{c\tau_3}{L}\nn\\
&\approx&\frac{1}{2\mubar_x L}
\left[h_x^{(2\mubar_x)}(A)-1\right]
=
\frac{1}{2\mubar_x L}\left[
\cos\left(\mubar_x c\right)+
2\sin\left(\mubar_x c\right)\tau_3-1
\right],\\
\gamma\tau_i{K_\theta}^i&=&\tau_i{A_\theta}^i=\tilde{b}\tau_2=b \tau_2\nn\\
&\approx&\frac{1}{2\mubar_\theta}
\left[h_\theta^{(2\mubar_\theta)}(A)-1\right]
=\frac{1}{2\mubar_\theta}\left[
\cos\left(\mubar_\theta b\right)+
2\sin\left(\mubar_\theta b\right)\tau_2-1
\right],\\
\gamma\tau_i{K_\phi}^i&=&\tau_i\left({A_\phi}^i-{\Gamma_\phi}^i\right)
=-\tilde{b}\tau_1\sin\theta=-b \tau_1\sin\theta\nn\\
&\approx&\frac{\sin\theta}{\mubar_\phi}
\left[h_\phi^{(2\mubar_\phi)}(A-\Gamma)-1\right]
=\frac{\sin\theta}{2\mubar_\phi}
\left[{\cal P}\exp\left(\int_\phi^{\phi+\frac{2\mubar_\phi}{\sin\theta}}\!\!\tau_i
\left({A_\phi}^{i}-{\Gamma_\phi}^i\right) \,d\phi\right)-1\right]\nn\\
&=&\frac{\sin\theta}{2\mubar_\phi}
\left[\exp\left(-2\mubar_\phi b\tau_1\right)-1\right]
=\frac{\sin\theta}{2\mubar_\phi}\left[
\cos\left(\mubar_\phi b\right)-
2\sin\left(\mubar_\phi b\right)\tau_1-1
\right],
\ea
where the extra factor 2 in $h_x^{(2\mubar_x)}$, $h_\theta^{(2\mubar_\theta)}$, and $h_\phi^{(2\mubar_\phi)}$ is adopted to be consistent with \eqnref{eqn:Fab1}--\eqnref{eqn:Fab3}.
According to \eqnref{eqn:Hamiltonian full}, \eqnref{eqn:F} and \eqnref{eqn:K K}, the relevant components of $F$ and $K\wedge K$ appearing in $H_{\rm grav}$ are $\tilde{b}^2\tau_3\sin\theta$, $\tilde{b}\tilde{c}\tau_2\sin\theta$ and $-\tilde{b}\tilde{c}\tau_1$, which now can be expressed in terms of holonomies as
\ba
\tilde{b}^2\tau_3\sin\theta&=&[\tilde{b}\tau_2,-\tilde{b}\tau_1\sin\theta]
\approx\frac{\sin\theta}{4\mubar_\theta\mubar_\phi}
[h_\theta^{(2\mubar_\theta)},h_\phi^{(2\mubar_\phi)}]
=\frac{\sin\theta}{\mubar_\theta\mubar_\phi}
\sin\left(\mubar_\theta b\right)\sin\left(\mubar_\phi b\right)\tau_3,\\
\tilde{b}\tilde{c}\tau_2\sin\theta&=&[-\tilde{b}\tau_1\sin\theta,\tilde{c}\tau_3]
\approx\frac{\sin\theta}{4\mubar_\phi\mubar_x L}
[h_\phi^{(2\mubar_\phi)},h_x^{(2\mubar_x)}]
=\frac{\sin\theta}{\mubar_x\mubar_\phi L}
\sin\left(\mubar_\phi b\right)\sin\left(\mubar_x c\right)\tau_2,\qquad\\
-\tilde{b}\tilde{c}\tau_1&=&[\tilde{c}\tau_3,\tilde{b}\tau_2]
\approx\frac{1}{4\mubar_x\mubar_\theta L}
[h_x^{(2\mubar_x)},h_\theta^{(2\mubar_\theta)}]
=\frac{1}{\mubar_x\mubar_\theta L}
\sin\left(\mubar_x c\right)\sin\left(\mubar_\theta b\right)\tau_1.
\ea
Equivalently, to take into account the discreteness corrections of LQG, we make the following prescription:
\ba\label{eqn:b c to sin}
b^2\tau_3&\longrightarrow& \frac{\sin\left(\mubar_\theta b\right)\sin\left(\mubar_\phi b\right)}{\mubar_\theta\mubar_\phi}\tau_3,\nn\\
bc\tau_2&\longrightarrow& \frac{\sin\left(\mubar_\phi b\right)\sin\left(\mubar_x c\right)}{\mubar_\phi\mubar_x}\tau_2,\nn\\
bc\tau_1&\longrightarrow& \frac{\sin\left(\mubar_x c\right)\sin\left(\mubar_\theta b\right)}{\mubar_x\mubar_\theta}\tau_1.
\ea

The continuous limit with $\mubar_x,\mubar_\theta,\mubar_\phi\rightarrow0$ recovers the classical Hamiltonian constraint in \eqnref{eqn:H grav}. However, the very feature of LGC is that the continuous limit does not exist and the failure of the limit to exist is intimately related with the underlying quantum geometry of LQG, where eigenvalues of the area operator are \emph{discrete} and has an \emph{area gap} $\Delta$. In LQC, to implement the discreteness as imprint from the full theory of LQG, we have to set the discreteness parameters $\mubar_x$, $\mubar_\theta$ and $\mubar_\phi$ to be finite. There are many possibilities to fix the discreteness parameters, but as in the Bianchi I model (see Appendix B of \cite{Chiou:2007mg}), two well-motivated strategies ($\mubar$- and $\mubar'$-schemes) are of particular interest and presented in the following.

The first strategy to impose the discreteness of LQG is to set the \emph{physical} areas of $\Box_{\theta\phi}$, $\Box_{\phi x}$ and $\Box_{x\theta}$ to be $\Delta$ (depicted in \figref{fig:holonomy}(a)). That is
\ba
\left(\sqrt{g_{\theta\theta}}\,\mubar_\theta\right)
\left(\sqrt{g_{\phi\phi}}\,\frac{\mubar_\phi}{\sin\theta}\right)
&=&p_c\mubar_\theta\mubar_\phi=\Delta,\nn\\
\left(\sqrt{g_{\phi\phi}}\,\frac{\mubar_\phi}{\sin\theta}\right)
\left(\sqrt{g_{xx}}\,\mubar_x L\right)
&=&p_b\mubar_\phi\mubar_x=\Delta,\nn\\
\left(\sqrt{g_{xx}}\,\mubar_x L\right)
\left(\sqrt{g_{\theta\theta}}\,\mubar_\theta\right)
&=&p_b\mubar_x\mubar_\theta=\Delta,
\ea
and consequently
\be
\mubar_\theta=\mubar_\phi\equiv\mubar'_b=\sqrt{\frac{\Delta}{p_c}}\,,
\qquad
\mubar_x\equiv\mubar'_c=\frac{\sqrt{p_c\Delta}}{p_b},
\ee
which, along with \eqnref{eqn:b c to sin}, gives the ``$\mubar'$-scheme'' in \eqnref{eqn:mubar'}.

Instead of shrinking the areas of $\Box_{\theta\phi}$, $\Box_{\phi x}$ and $\Box_{x\theta}$ to $\Delta$, the second strategy is to associate each \emph{edge} of $\Box_{\theta\phi}$, $\Box_{\phi x}$ and $\Box_{x\theta}$ with an area and then shrink the associated areas to $\Delta$. (See \figref{fig:holonomy}(b).) For instance,
The edge of $\Box_{\theta\phi}$ in $\partial_\theta$-direction is of \emph{coordinate} length $\mubar_\theta$, with which, most naturally, we associate a rectangle $\boxdot_\theta$ normal to $\partial_\theta$-direction of \emph{coordinate} lengths $\mubar_\theta L$ and $\mubar_\theta/\sin\theta$ on its edges. We then set the \emph{physical} area of $\boxdot_\theta$ (and similarly of $\boxdot_\phi$ and $\boxdot_x$ as well) to $\Delta$. That is
\ba
\left(\sqrt{g_{\phi\phi}}\,\frac{\mubar_\theta}{\sin\theta}\right)
\left(\sqrt{g_{xx}}\,\mubar_\theta L\right)
&=&p_b\mubar_\theta^2=\Delta,\nn\\
\left(\sqrt{g_{xx}}\,\mubar_\phi L\right)
\left(\sqrt{g_{\theta\theta}}\,\mubar_\phi\right)
&=&p_b\mubar_\phi^2=\Delta,\nn\\
\left(\sqrt{g_{\theta\theta}}\,\mubar_x\right)
\left(\sqrt{g_{\phi\phi}}\,\frac{\mubar_x}{\sin\theta}\right)
&=&p_c\mubar_x^2=\Delta,
\ea
and consequently
\be
\mubar_\theta=\mubar_\phi\equiv\mubar_b=\sqrt{\frac{\Delta}{p_b}}\,,
\qquad
\mubar_x\equiv\mubar_c=\sqrt{\frac{\Delta}{p_c}}\,,
\ee
which, along with \eqnref{eqn:b c to sin}, gives the ``$\mubar$-scheme'' in \eqnref{eqn:mubar}.

[Note that in both pictures of the $\mubar'$- and $\mubar$-schemes, if we rather set the \emph{coordinate} areas of $\Box_{\theta\phi}, \cdots$ or $\boxdot_\theta, \cdots$ to be $\Delta$, we end up with constant $\mubar_b$ and $\mubar_c$. This is the ``$\mu_o$-scheme'' used in the precursor strategy.
The wrong semiclassical behavior of the $\mu_o$-scheme partly originates from the problem that the \emph{coordinate} area (both of $\Box_{\theta\phi}, \cdots$ and of $\boxdot_{\theta}, \cdots$) is not totally physically relevant.]

In the full theory of LQG, when the Hamiltonian acts on an spin network state, it adds a new link of spin-$1/2$ and the coloring of the links on which the new link is attached is increased or decreased by $1/2$. The resulting spin network state is equivalent to the original state superimposed with a triangular loop of spin-$1/2$ links. This triangular loop is essentially the Wilson loop discussed above.
In the context of spin network states, the coloring of a link corresponds to the area of the surface which the link penetrates and the smallest coloring spin-$1/2$ gives rise to the \emph{area gap} $\Delta$. Therefore, it is in this sense that the $\mubar$-scheme is a more direct implementation of the underlying discreteness of quantum geometry than the $\mubar'$-scheme. (See FIG. 6 and the pertinent text in \cite{Chiou:2007mg} for more comments.)

Furthermore, in the fundamental quantum theory of LQC, the $\mubar$-scheme has the important virtue that we can define the affine variables $v_b$, $v_c$ via
\be
\frac{\partial}{\partial v_b}:=4\pi\gamma\Pl^2\,\mubar_b\frac{\partial}{\partial p_b},
\qquad
\frac{\partial}{\partial v_c}:=4\pi\gamma\Pl^2\,\mubar_c\frac{\partial}{\partial p_c}
\ee
such that Hamiltonian constraint of the fundamental quantum theory gives the evolution as a difference equation in terms of $v_b$, $v_c$ and therefore the methodology used for the isotropic model \cite{{Ashtekar:2006wn}} and Bianchi I model \cite{Chiou:2006qq} should be easily applied. This strategy fails in the $\mubar'$-scheme since
\be
\left[\mubar'_b\frac{\partial}{\partial p_b},\ \mubar'_c\frac{\partial}{\partial p_c}\right]\neq 0
\ee
and hence the corresponding affine variables do not exist. This makes it difficult to construct the fundamental quantum theory in the $\mubar'$-scheme.

On the other hand, as studied in \secref{sec:phenomenological dynamics}, the $\mubar'$-scheme has the advantage over the $\mubar$-scheme that the phenomenological dynamics in the $\mubar'$-scheme is independent of the choice of ${\cal I}$. The difference for this point between the $\mubar$- and $\mubar'$-schemes can be understood, heuristically but instructively, by estimating the quantities $\mubar_bb$ and $\mubar_cc$ with the classical formulae; that is, substituting \eqnref{eqn:cl b} and \eqnref{eqn:cl c} for $b$ and $c$, we have
\ba
\label{eqn:mubar b}
\mubar_bb&\approx&\gamma\Delta^{1/2}\left(\frac{2{\bf A}_{\theta\phi}}{{\bf A}_{x\phi}}\right)^{1/2}
\left(\frac{1}{\sqrt{g_{\Omega\Omega}}}\frac{d\sqrt{g_{\Omega\Omega}}}{d\tau}\right),\\
\label{eqn:mubar c}
\mubar_cc&\approx&\gamma\Delta^{1/2}\left(\frac{{\bf A}_{x\phi}}{2{\bf A}_{\theta\phi}}\right)
\left(\frac{1}{\sqrt{g_{xx}}}\frac{d\sqrt{g_{xx}}}{d\tau}\right),\\
\label{eqn:mubar' b}
\mubar'_bb&\approx&\gamma\Delta^{1/2}
\left(\frac{1}{\sqrt{g_{\Omega\Omega}}}\frac{d\sqrt{g_{\Omega\Omega}}}{d\tau}\right),\\
\label{eqn:mubar' c}
\mubar'_cc&\approx&\gamma\Delta^{1/2}
\left(\frac{1}{\sqrt{g_{xx}}}\frac{d\sqrt{g_{xx}}}{d\tau}\right).
\ea
Since the quantities $\mubar_bb$ and $\mubar_cc$ indicate how significant the quantum corrections are (quantum corrections are negligible if $\mubar_bb,\mubar_cc\ll 1$ ), \eqnref{eqn:mubar b} and \eqnref{eqn:mubar c} tell that in the $\mubar$-scheme, the place at which the quantum effects become appreciable is tied up with not only the ``Hubble rates'' for $\sqrt{g_{\Omega\Omega}}$ and $\sqrt{g_{xx}}$ but also the physical geometry of ${\cal I}\times S^2$. On the other hand, \eqnref{eqn:mubar' b} and \eqnref{eqn:mubar' c} show that $\mubar'_bb$ and $\mubar'_cc$ are proportional only to the ``Hubble rates'' in the classical regime and the different choice of ${\cal I}$ is irrelevant in the $\mubar'$-scheme.

Finally, it should be noted that, in the language of the lattice refining model in \cite{Bojowald:2007ra}, the $\mubar$-scheme corresponds to the refinement pattern that the number of lattice vertices is proportional to the transverse area, reminiscent of the idea depicted in \figref{fig:holonomy}(b) with the transverse surfaces shrunk to $\Delta$. A stability analysis however suggests that the $\mubar$-scheme leads to an unstable difference equation of evolution in the fundamental theory of LQC and thus, in this regard, could be problematic as a good quantization scheme.

Both the $\mubar$- and $\mubar'$-schemes have desirable and undesirable features of their own. In order to understood them more deeply, in the main text we study both schemes and their ramifications at the level of phenomenological dynamics.


\end{document}